\documentclass[aps,reprint,superscriptaddress,amsmath]{revtex4-1}

\usepackage{siunitx}
\usepackage{graphicx}
\usepackage[colorlinks=true,allcolors=blue]{hyperref}

\newcommand{\CaRuO}{Ca\textsubscript{3}Ru\textsubscript{2}O\textsubscript{7}}
\newcommand{\AFMa}{AFM$_a$}
\newcommand{\AFMb}{AFM$_b$}

\begin{document}

\title{Strain control of a bandwidth-driven spin reorientation in \CaRuO{}}

\author{C.~D.~Dashwood}
\email{cameron.dashwood.17@ucl.ac.uk}
\thanks{These authors contributed equally to this work.}
\author{A.~H.~Walker}
\email{a.walker.17@ucl.ac.uk}
\thanks{These authors contributed equally to this work.}
\affiliation{London Centre for Nanotechnology and Department of Physics and Astronomy, University College London, London, WC1E 6BT, United Kingdom}

\author{M.~P.~Kwasigroch}
\affiliation{Department of Mathematics, University College London, London, WC1H 0AY, United Kingdom}
\affiliation{Trinity College, Cambridge, CB2 1TQ, United Kingdom}

\author{L.~S.~I.~Veiga}
\affiliation{London Centre for Nanotechnology and Department of Physics and Astronomy, University College London, London, WC1E 6BT, United Kingdom}
\affiliation{Diamond Light Source, Harwell Science and Innovation Campus, Didcot, Oxfordshire, OX11 0DE, United Kingdom}

\author{Q.~Faure}
\affiliation{London Centre for Nanotechnology and Department of Physics and Astronomy, University College London, London, WC1E 6BT, United Kingdom}
\affiliation{Laboratoire Le\'{o}n Brillouin, CEA, CNRS, Universit\'{e} Paris-Saclay, CEA-Saclay, 91191 Gif-sur-Yvette, France}

\author{J.~G.~Vale}
\affiliation{London Centre for Nanotechnology and Department of Physics and Astronomy, University College London, London, WC1E 6BT, United Kingdom}

\author{D.~G.~Porter}
\affiliation{Diamond Light Source, Harwell Science and Innovation Campus, Didcot, Oxfordshire, OX11 0DE, United Kingdom}

\author{P.~Manuel}
\author{D.~D.~Khalyavin}
\author{F.~Orlandi}
\affiliation{ISIS Neutron and Muon Source, STFC Rutherford Appleton Laboratory, Didcot, Oxfordshire, OX11 0QX, United Kingdom}

\author{C.~V.~Colin}
\affiliation{Universit\'{e} Grenoble Alpes, CNRS, Institut N\'{e}el, 38000 Grenoble, France}

\author{O.~Fabelo}
\affiliation{Institut Laue-Langevin, 71 Avenue des Martyrs, CS 20156, 38042 Grenoble, France}

\author{F.~Kr\"{u}ger}
\affiliation{London Centre for Nanotechnology and Department of Physics and Astronomy, University College London, London, WC1E 6BT, United Kingdom}
\affiliation{ISIS Neutron and Muon Source, STFC Rutherford Appleton Laboratory, Didcot, Oxfordshire, OX11 0QX, United Kingdom}

\author{R.~S.~Perry}
\affiliation{London Centre for Nanotechnology and Department of Physics and Astronomy, University College London, London, WC1E 6BT, United Kingdom}

\author{R.~D.~Johnson}
\affiliation{Department of Physics and Astronomy, University College London, London, WC1E 6BT, United Kingdom}

\author{A.~G.~Green}
\author{D.~F.~McMorrow}
\affiliation{London Centre for Nanotechnology and Department of Physics and Astronomy, University College London, London, WC1E 6BT, United Kingdom}

\begin{abstract}
The layered-ruthenate family of materials possess an intricate interplay of structural, electronic and magnetic degrees of freedom that yields a plethora of delicately balanced ground states. This is exemplified by \CaRuO{}, which hosts a coupled transition in which the lattice parameters jump, the Fermi surface partially gaps and the spins undergo a \ang{90} in-plane reorientation. Here, we show how the transition is driven by a lattice strain that tunes the electronic bandwidth. We apply uniaxial stress to single crystals of \CaRuO{}, using neutron and resonant x-ray scattering to simultaneously probe the structural and magnetic responses. These measurements demonstrate that the transition can be driven by externally induced strain, stimulating the development of a theoretical model in which an internal strain is generated self-consistently to lower the electronic energy. We understand the strain to act by modifying tilts and rotations of the RuO\textsubscript{6} octahedra, which directly influences the nearest-neighbour hopping. Our results offer a blueprint for uncovering the driving force behind coupled phase transitions, as well as a route to controlling them.
\end{abstract}

\maketitle

\section*{Introduction}

The coupling between structural and electronic degrees of freedom in quantum materials generates a variety of ground states and drives transitions between them. Textbook examples include the Peierls transition in 1D materials, in which a periodic lattice deformation leads to a metal-insulator transition and the formation of a charge-density wave \cite{Peierls1955, Frohlich1954}. Similarly, the cooperative Jahn-Teller effect describes the spontaneous distortion of a crystalline lattice to lower the electronic degeneracy and give rise to orbital ordering \cite{Kanamori1960}. Recently, there has been considerable interest in twisted bilayer materials, which host a spectrum of electronic phases---from Mott insulators \cite{Cao2018a} to unconventional superconductors \cite{Cao2018b}---as the bandwidth is tuned by the twist angle. The key role of the lattice in many quantum materials offers a powerful set of control parameters with which to tune their phases, but also presents a considerable challenge in developing a comprehensive understanding of the interactions that give rise to these phases.

In this context, the application of stress has arisen as a powerful method to tune the electronic properties of quantum materials, including superconducting \cite{Hicks2014, Steppke2017}, charge/spin-density wave \cite{Kim2018, Kim2021, Wang2022, Choi2022, Simutis2022}, nematic \cite{Malinowski2020, Bartlett2021, Worasaran2021, Sanchez2021} and topological \cite{Nicholson2021, Lin2021} phases. Here, we use uniaxial stress to drive a coupled spin reorientation and Fermi surface reconstruction in \CaRuO{}, uncovering the central role of the lattice and facilitating a microscopic understanding of the transition.

\begin{figure*}
	\includegraphics[width=\linewidth]{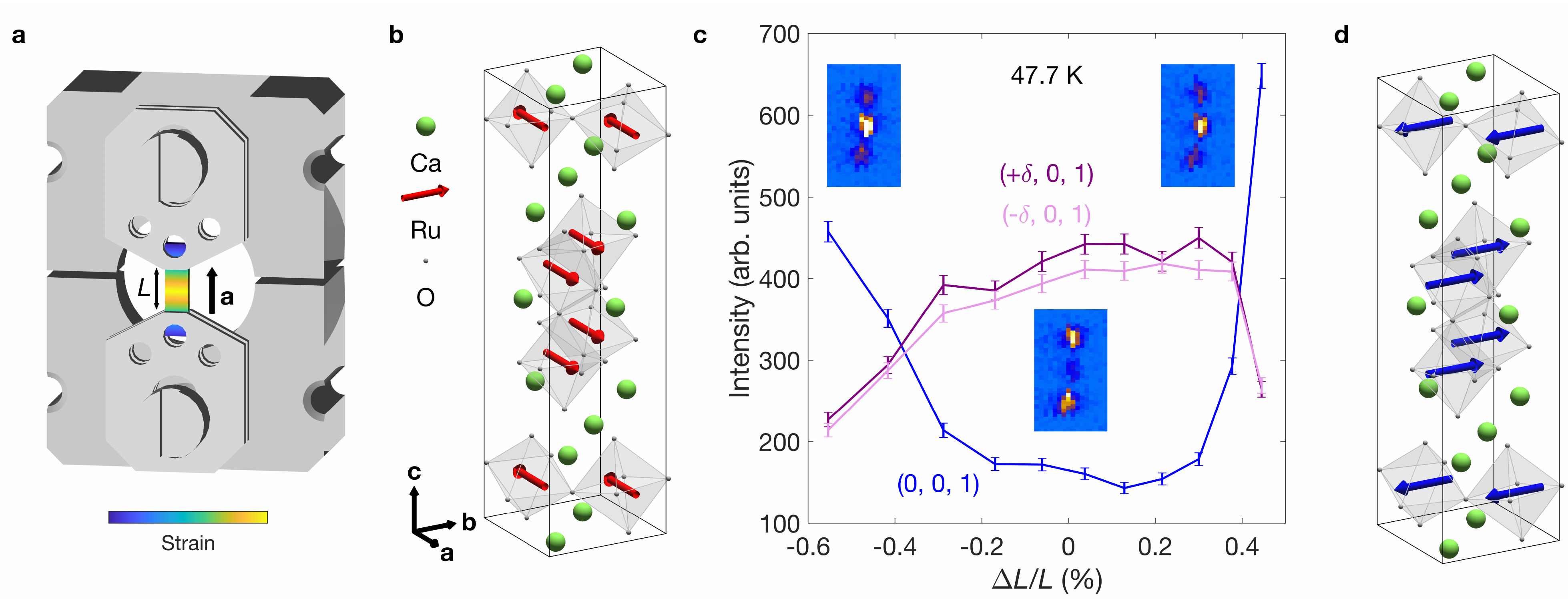}
	\caption{\label{Ca327_strain_neutron}\textbf{Neutron scattering under stress. a} Schematic of the strain cell used in the neutron scattering experiment. The bridges of the cell and sample plates are shown in grey, and the sample is shown with a blue-to-yellow colourmap to indicate the strain gradient induced along its length. Neutrons are scattered in transmission through the central strained region of the sample. \textbf{b} Magnetic structure of \CaRuO{} in the \AFMa{} phase \cite{Bohnenbuck2008}. \textbf{c} Integrated intensity of the commensurate $(0, 0, 1)$ and incommensurate $(\pm\delta, 0, 1)$ peaks as a function of applied strain, $\Delta L / L$. The background signal comes from scattering from the strain cell, and errors are standard deviations. The insets show detector images at zero and maximum compressive/tensile applied strain. \textbf{d} Magnetic structure of \CaRuO{} in the \AFMb{} phase \cite{Bohnenbuck2008}.}
\end{figure*}

\CaRuO{} is a bilayer member of the Ruddlesden-Popper ruthenates, A$_{n + 1}$Ru$_n$O$_{3n + 1}$, across which varying structural distortions lead to a diversity of ground states. In the monolayer compounds these range from superconductivity in Sr\textsubscript{2}RuO\textsubscript{4} \cite{Maeno1994, Hicks2014} to a Mott insulating state in Ca\textsubscript{2}RuO\textsubscript{4} \cite{Nakatsuji2000, Ricco2018}, while bilayer Sr\textsubscript{3}Ru\textsubscript{2}O\textsubscript{7} displays an electronic nematic phase with spin-density wave order \cite{Yamase2007, Berridge2009, Lester2015, Lindquist2022}. \CaRuO{} crystallises in the $Bb2_1m$ space group ($a \approx$ \SI{5.3}{\angstrom}, $b \approx$ \SI{5.5}{\angstrom}, $c \approx$ \SI{19.5}{\angstrom}), in which a combination of octahedral tilts around $\mathbf{b}$ ($X_3^-$, $a^- a^- c^0$ in Glazer notation) and rotations around $\mathbf{c}$ ($X_2^+$, $a^0 a^0 c^+$) combine to unlock polar lattice displacements \cite{Lei2018}. It undergoes a coupled structural, electronic and magnetic transition, where the lack of inversion symmetry causes a magnetic cycloid to form and mediate a spin-reorientation transition (SRT). Below $T_N \approx$ \SI{60}{\kelvin}, the spins align along $\mathbf{a}$, coupled ferromagnetically within the bilayers and antiferromangetically between them, in the \AFMa{} phase (see Fig.~\ref{Ca327_strain_neutron}b) \cite{Bohnenbuck2008}. On cooling through \SI{49}{\kelvin}, an incommensurate cycloid (ICC) develops with $\mathbf{q} = (\delta, 0, 1)$ ($\delta \approx 0.023$) and the spins rotating in the $\mathbf{a}$--$\mathbf{b}$ plane \cite{Dashwood2020, Faure2023}. The envelope of the cycloid evolves from elongated along $\mathbf{a}$, to circular, to elongated along $\mathbf{b}$. The SRT concludes at \SI{46}{\kelvin} with the cycloid collapsing into the collinear \AFMb{} phase, where the spins are globally rotated by \ang{90} from \AFMa{} (Fig.~\ref{Ca327_strain_neutron}d). The SRT is accompanied by a rapid change in the lattice parameters \cite{Yoshida2005} and a partial gapping of the Fermi surface \cite{Baumberger2006, Xing2018, Horio2019, Markovic2020}.

Although there have been various proposals for the mechanism behind the SRT, including recent work that attributes it to the energy gain associated with a Rashba-based hybridisation of bands at the Fermi level \cite{Markovic2020}, these proposals have generally neglected the structural degrees of freedom. We use piezoelectric cells to apply continuously tuneable uniaxial stresses to \CaRuO{} single crystals. A combination of neutron and resonant x-ray scattering enables us to directly probe the magnetic structure and offer insight into the response of the lattice to applied stress. Our measurements reveal that the SRT can be fully driven by strain at fixed temperature, and allow the construction of temperature-strain phase diagrams for orthogonal in-plane stresses. This motivates the development of a minimal theoretical model in which strain tunes the electronic hopping. As well as reproducing our strain data, a self-consistent solution of the model (with realistic values for the electronic parameters) reveals that the transition with temperature is driven by an internal strain that arises to lower the electronic energy. We interpret this strain as acting via the tilts and rotations of the RuO\textsubscript{6} octahedra, as corroborated by temperature-dependent diffraction measurements. Our results therefore offer a promising mechanism to drive electronic and magnetic phase transitions in correlated perovskite materials.

\section*{Results}

\subsection*{Driving the spin reorientation with strain}

As a well-established bulk probe of magnetic order, neutron scattering is a natural choice for our experiments. The large sample size generally needed is, however, at odds with the inverse scaling of achievable strain on sample length -- an issue that is compounded by the increased background scattering from the strain cell. To overcome these issues, we used the latest generation of piezoelectric strain cells from Razorbill Instruments \cite{Razorbill}, which can apply large strains to samples measured in a transmission scattering geometry with minimal background from the cell. Combined with the high count-rate of the WISH instrument at the ISIS Neutron and Muon Source \cite{Chapon2011}, this enabled us to achieve sufficient signal at strains of up to $0.5\%$. A schematic of the strain cell is shown in Fig.~\ref{Ca327_strain_neutron}a, with a \CaRuO{} sample spanning a distance $L$ between the sample plates. Applying a voltage across the piezoelectric stacks changes the distance between the plates by $\Delta L$, producing an applied strain $\Delta L / L$ along the $a$-axis of the sample. Further experimental details can be found in the Methods.

Figure \ref{Ca327_strain_neutron}c shows the result of a strain sweep at a fixed temperature in the centre of the ICC phase. We tracked the intensities of the commensurate $(0, 0, 1)$ magnetic peak, which arises from both the \AFMa{} and \AFMb{} structures, and satellite $(\pm\delta, 0, 1)$ peaks that arise from the mediating cycloidal phase. At $\Delta L / L = 0$, we see intense satellite peaks and a small remnant commensurate peak due to temperature and strain gradients in the sample. Under both tensile ($\Delta L / L > 0$) and compressive ($\Delta L / L < 0$) strain the satellite peaks are suppressed and the commensurate peak is simultaneously enhanced. The commensurate intensity is lower under compression, consistent with entering the \AFMa{} phase, while the higher commensurate intensity under tension is consistent with the \AFMb{} phase \cite{Dashwood2020}. This assignment is confirmed by strain dependences at temperatures just above and below the SRT (see the Supplementary Information). The slight change in incommensurate intensity at low strains is consistent with the variation of moment size through the ICC phase. Our neutron data therefore provide strong evidence that we can continuously drive the cycloid-mediated SRT with strain at fixed temperature.

Despite this, our neutron setup has a number of drawbacks. The transitions between the magnetic phases can be seen to occur over a finite width of around $0.3\%$ due to the neutron beam illuminating a large region of the sample across which there is a significant strain gradient (see Fig.~\ref{Ca327_strain_neutron}a). Further, despite the large displacements that can be generated by the cell, the maximum strains of $|\Delta L / L| \approx 0.5\%$ are not large enough to fully enter the collinear phases. Finally, and most significantly, the scattering geometry necessitated by the strain cell blocks access to any nuclear Bragg peaks with a finite component along the stress direction. This precludes a determination of the lattice parameters, and therefore the true strain generated in the sample, forcing us to rely on the displacement of the sample plates as a measure of the applied strain. As we will see in the following section, the fact that the epoxy is significantly softer than the sample leads to only a fraction of the applied strain being transmitted to the sample.

\subsection*{Response of the lattice to applied stress}

\begin{figure}
	\includegraphics[width=\linewidth]{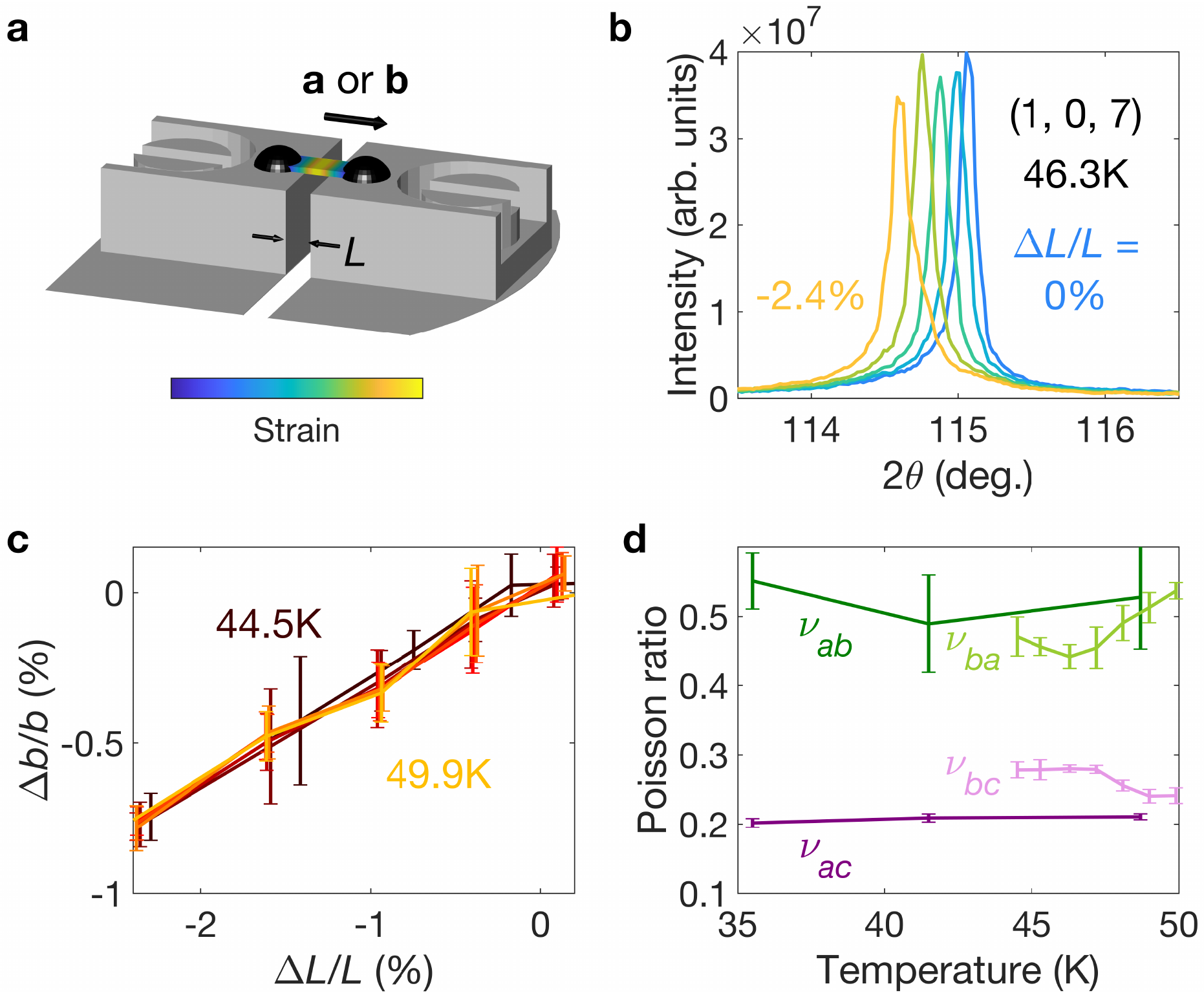}
	\caption{\label{Ca327_strain_xray}\textbf{X-ray scattering under stress. a} Schematic of the strain cell used for x-ray scattering. The principles of operation are the same as the neutron setup, except that the sample is mounted on top of raised sample plates to give a large sphere of access for the incident and scattered beams. \textbf{b} $2\theta$ scans of the structural $(1, 0, 7)$ Bragg peak at various applied strains. \textbf{c} True strain, $\Delta b / b$, as a function of applied strain, $\Delta L/ L$, at a range of temperatures through the SRT. \textbf{d} Temperature dependences of the Poisson ratios determined from the relative strains along orthogonal axes. All errors are standard deviations.}
\end{figure}

A full characterisation of the response of the lattice to applied stress is crucial to understanding the magnetic response, as highlighted by previous contradictory reports of the electronic and magnetic changes under uniaxial pressure in \CaRuO{} \cite{Cao2003, Yoshida2008}. To achieve this, we turned to synchrotron x-ray scattering at beamline I16 of the Diamond Light Source. Size limitations of the closed-cycle cryostat required use of a smaller strain cell than the neutron measurements. The lower stresses are mitigated, however, by the high flux and small focus of the x-ray beam, allowing the use of smaller samples to achieve applied strains over $2\%$. To maximise beam access we mounted the samples on top of raised sample plates, as shown in Fig.~\ref{Ca327_strain_xray}a (further details can be found in the Methods). Two samples were measured, one with stress applied along $\mathbf{a}$ and the other along $\mathbf{b}$.

We tracked the positions of multiple structural Bragg peaks while applying compressive stresses. Figure \ref{Ca327_strain_xray}b shows representative $2\theta$ scans of the $(1, 0, 7)$ peak for stress applied along $\mathbf{b}$. The shifting of the peak arises from the changing lattice parameters, from which we can determine the true strain, $\Delta b / b$. This is shown in Fig.~\ref{Ca327_strain_xray}c, where we see a linear, temperature-insensitive dependence on $\Delta L / L$, with around 40\% of the applied strain transmitted to the sample. We were unable to apply tensile strains due to the asymmetric sample mounting, which also induced a slight bending of the sample under compression (see the Supplementary Information). The bending did not induce significant strain gradients through the probed region, however, as evidenced by the minimal broadening of the peaks in Fig.~\ref{Ca327_strain_xray}b.

\begin{figure*}
	\includegraphics[width=\linewidth]{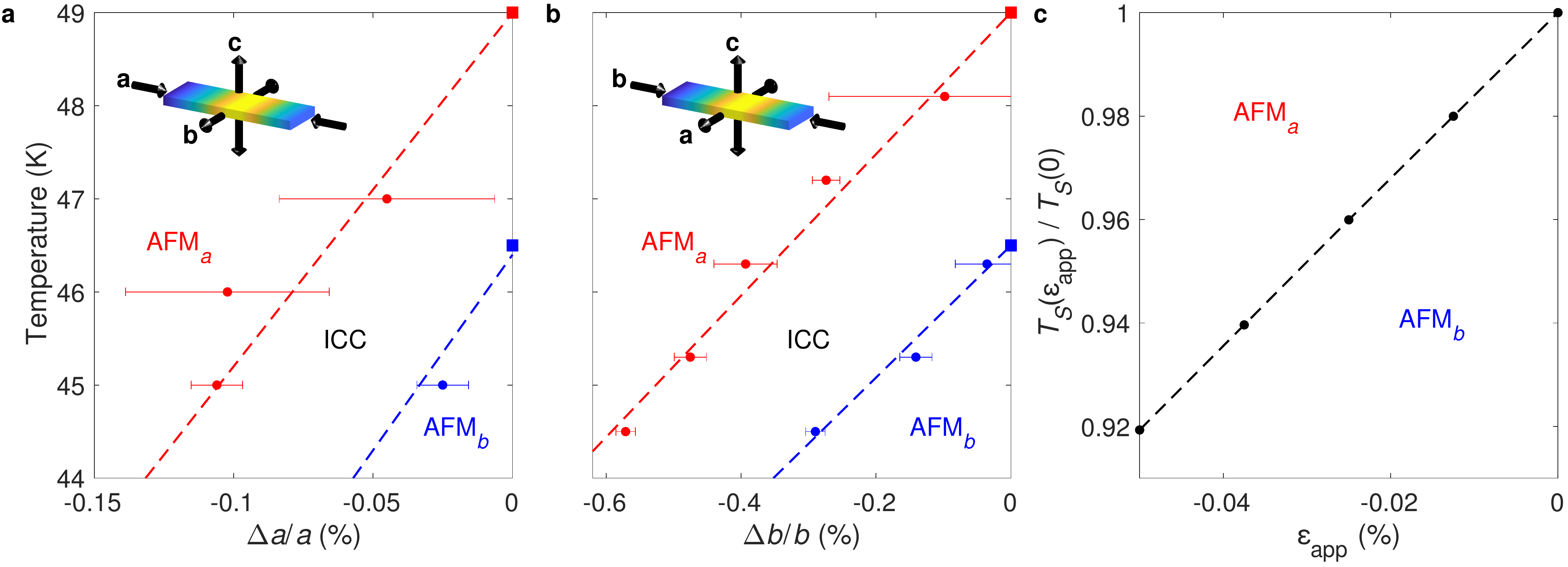}
	\caption{\label{Ca327_strain_phase_diagrams} \textbf{Temperature-strain phase diagrams. a, b} Experimental phase diagrams for stress applied along the $a$- and $b$-axes respectively. The points with errors (standard deviations) are transition temperatures extracted from the strain measurements, and the squares indicate the zero-strain transition temperatures from ref.~\cite{Dashwood2020}. The dashed lines are guides to the eye. The insets indicate the tensile strains induced along the orthogonal directions according to the Poisson ratios. \textbf{c} Theoretical phase diagram under applied strain $\varepsilon_\text{app}$, calculated by self-consistently minimising the free energy in Eq.~(\ref{eq:free_energy}) using the parameter values $U = 8$, $\lambda = 0.5$,  $\theta = \SI{15}{\degree}$, $\mu = 8.5$, $\kappa =  400$, $\nu =  12$, $t_0 =  1.12$ and $\varepsilon_0 = 0.066$.}
\end{figure*}

Our x-ray measurements also enable us to quantify the strains along the other crystalline axes, $\varepsilon_x = \Delta x / x$, that arise from the uniaxial stresses applied by the cell, and thereby calculate the Poisson ratios $\nu_{xy} = -\varepsilon_y / \varepsilon_x$. The results are plotted as a function of temperature in Fig.~\ref{Ca327_strain_xray}d. The anisotropy of the in-plane ($\nu_{ab} \approx \nu_{ba} \approx 0.5$) and out-of-plane ($\nu_{ac} \approx \nu_{bc} \approx 0.2$) Poisson ratios is as expected for a layered material like \CaRuO{}.

We can also use x-ray scattering to probe the magnetic phases by exploiting the resonant scattering enhancement on tuning the incident x-ray energy to the Ru $L_2$ edge (\SI{2.967}{\kilo\electronvolt}). Alongside the positions of the structural peaks, we monitored the intensities of the $(\delta, 0, 5)$ and $(0, 0, 5)$ magnetic peaks at each strain value, the latter at two orthogonal azimuthal angles to differentiate between the \AFMa{} and \AFMb{} phases \cite{Dashwood2020}. Repeating these strain sweeps at a range of temperatures enabled us to construct the phase diagrams shown in Fig.~\ref{Ca327_strain_phase_diagrams}a, b for stress applied along $\mathbf{a}$ and $\mathbf{b}$ respectively (further details of how the phase diagrams were constructed can be found in the Supplementary Information).

The phase diagrams show the transition temperatures varying linearly with compressive strain along both $\mathbf{a}$ and $\mathbf{b}$. The data quality is worse for the sample with stress applied along $\mathbf{a}$, most likely due to the poorer crystalline quality of the sample causing a heterogeneous distribution of strain (see the Supplementary Information). Despite this, we can still determine a rate of change of the transition temperatures of approximately \SI{40}{\kelvin} per percent strain along $\mathbf{a}$. The transition temperatures change at a slower rate for the sample stressed along $\mathbf{b}$, at around \SI{7}{\kelvin} per percent strain along $\mathbf{b}$. Although we could not access the tensile sides of the phase diagrams in our x-ray experiments, our neutron results suggest that the trends should continue unchanged. It is interesting to note that the phase boundaries move in the same direction in both phase diagrams, despite the opposite in-plane deformations of the lattice (as depicted in the insets of Fig.~\ref{Ca327_strain_phase_diagrams}a, b).

\subsection*{A strain-coupled electronic model}

To understand the role of strain in the SRT, we developed a phenomenological model in which strain tunes the electronic hopping between nearest-neighbour Ru orbitals. In the spirit of developing a minimal model that captures the physics under study, we included only the Ru $d_{xz}$ and $d_{yz}$ orbitals in a perovskite monolayer. Such a two-band model is consistent with ARPES studies that show that the $d_{xy}$ orbital does not contribute to the Fermi surface \cite{Horio2019}. We consider a Hamiltonian $\hat{H} = \hat{H}_t + \hat{H}_U + \hat{H}_\lambda - \mu \hat N$. $\hat{H}_t$ is a tight-binding Hamiltonian fitted to ARPES data \cite{Horio2019} (see the Methods). $\hat{H}_U = U \sum_{i \tau \alpha} \hat{n}_{i \tau \alpha \uparrow} \hat{n}_{i \tau \alpha \downarrow}$ is the on-site intra-orbital Hubbard interaction, where the lattice site is labelled $i$, the orbital is labelled $\alpha \in \{xz, yz\}$, and $\tau \in \{A,B\}$ is a sublattice index that is introduced to allow for the staggered octahedral tilting. The interaction is treated in the Hartree-Fock mean-field approximation, and all other local electron-electron interactions are neglected for simplicity. The Hamiltonian is then a function of site-averaged charge, $\rho_{\tau \alpha}$, and magnetisation, $\mathbf{M}_{\tau \alpha}$, fields for each orbital and sublattice. $\hat{H}_\lambda$ is the spin-orbit interaction \cite{Mohapatra2020},
\begin{equation}
	\hat{H}_\lambda = -\frac{i \lambda}{2} \sum_{i, \tau} \left(\underline{c}_{i \tau, xz}^\dagger \sigma_z^\tau \underline{c}_{i \tau, yz} + \text{h.c.}\right),
\end{equation}
where $\underline{c}_{i \tau \alpha}^\dagger = (c_{i \tau \alpha \uparrow}^\dagger, c_{i \tau \alpha \downarrow}^\dagger)$. This is an on-site term that couples the two orbitals. Octahedral tilting affects the spin-orbit term by rotating the spin quantisation axes on the sublattice sites by $\pm\theta$ about the $b$-axis, as $\sigma_z^{A(B)} = \cos(\theta) \sigma_z \pm \sin(\theta) (\sigma_x - \sigma_y) / \sqrt{2}$. The octahedral rotation (as well as the polar distortion that leads to Rashba-like spin-orbit effects) does not enter the Hamiltonian explicitly, although it does influence the hopping as described in the next section.

Strain enters the model via a phenomenological field, $\varepsilon$, that couples to the nearest-neighbour hopping, $t$, of the tight-binding model (see the Methods for further details). We take a minimal coupling of $t = t_0 + \nu \varepsilon$, where $t_0$ and $\nu$ are parameters, which is valid when the strain magnitude is small. The coupling of strain to further-neighbour hopping is neglected. We also introduce a strain cost set by a parameter, $\kappa$, in the free energy
\begin{multline}
	F(\mathbf{M}, \rho, \varepsilon, \varepsilon_\text{app}) = -T \sum_{n, \mathbf{k}} \ln\left(1 + e^{-(\epsilon_{n \mathbf{k}} - \mu) / T}\right) \\
    + U \sum_{\tau, \alpha} \left(\mathbf{M}_{\tau \alpha}^2 - \rho_{\tau \alpha}^2\right) + \frac{1}{2} \kappa \left(\varepsilon_0 + \varepsilon - \varepsilon_\text{app}\right)^2, \label{eq:free_energy}
\end{multline}
where $\epsilon_{n \mathbf{k}}$ are the eigenvalues of the electronic Hamiltonian, $\mu$ is the chemical potential, and $\varepsilon_0$ and $\varepsilon_\text{app}$ are parameters. The last term is the contribution of the non-electronic degrees of freedom to quadratic order. Within this parameterisation, $\varepsilon$ is the strain measured relative to the unstrained system which we define, via our choice of value of $\varepsilon_0$ and without loss of generality, as the system at the N\'{e}el temperature. This is a natural reference point for studying the magnetic phase diagram. As well as the internal strain $\varepsilon$, we include an externally applied strain $\varepsilon_\text{app}$ that moves the minimum of the strain cost function away from zero. The electronic state is determined at a particular temperature and externally applied strain by minimising the free energy with respect to the electronic degrees of freedom, $\mathbf{M}_{\tau\alpha}$ and $\rho_{\tau\alpha}$, and the strain field, $\varepsilon$, in a self-consistent manner.

In the absence of strain, $\kappa = \nu = 0$, the solution of this model exhibits easy $a$- and easy $b$-axis ferromagnetism that corresponds to the \AFMa{} and \AFMb{} phases in the full structure. This results from the interplay of the on-site interaction and spin-orbit coupling, which generate the magnetic anisotropy, and the octahedral tilting that breaks its symmetry in the $\mathbf{a}$--$\mathbf{b}$ plane. The anisotropy is strongly dependent on the nearest-neighbour hopping, and an $\mathbf{a} \rightarrow \mathbf{b}$ reorientation of the ferromagnetic order is achieved by increasing $t$ and therefore the electronic bandwidth. Although our minimal model cannot reproduce the cycloid that mediates the SRT, we would expect this to appear naturally when the full bilayer is considered, due to a uniform Dzyaloshinskii-Moriya interaction that becomes dominant close to the reorientation transition \cite{Dashwood2020}.

\begin{figure}
	\includegraphics[width=\linewidth]{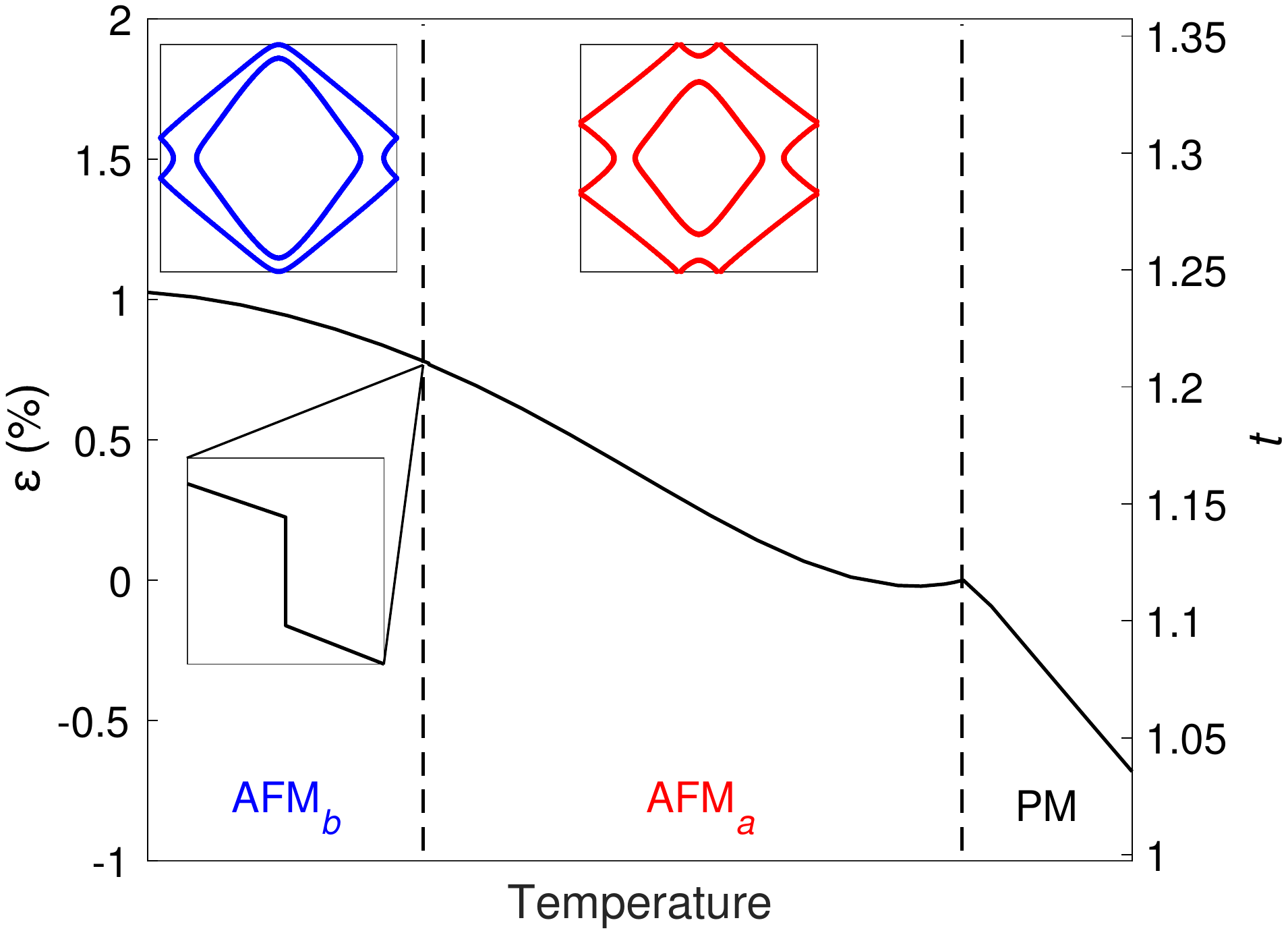}
	\caption{\label{Ca327_strain_temp_dep} \textbf{Self-consistent internal strain.} Strain field, $\varepsilon$, as a function of temperature, with the magnetic phase transitions indicated by vertical dashed lines. The corresponding change in the effective nearest-neighbour hopping parameter, $t$, is measured on the right-hand axis. The zoomed region shows a small discontinuity in $\varepsilon$ across the SRT of magnitude $|\Delta \varepsilon| \sim 0.005\%$. The insets show Fermi surfaces calculated in the two magnetic phases. The parameter values are the same as in Fig.~\ref{Ca327_strain_phase_diagrams}c. The calculated value of the hopping parameter at the transition corresponds to $U/t \sim 6.6$, which is consistent with values of $t$ \cite{Horio2019} and $U$ \cite{Puggioni2020} previously fitted to the observed electronic structure of \CaRuO{}.}
\end{figure}

With a finite $\kappa$ and $\nu$, we find an internal strain being generated self-consistently even with zero applied strain, $\varepsilon_\text{app} = 0$, as shown in Fig.~\ref{Ca327_strain_temp_dep}. We see $\varepsilon$ increase as the temperature is reduced through the \AFMa{} phase, lowering the energy of the electronic system at the expense of an elastic energy cost. When the hopping reaches a threshold value the SRT is triggered. This transition is first order and results in a small positive jump in the strain (see zoomed region in Fig.~\ref{Ca327_strain_temp_dep}) as well as a jump in the magnetisation. We also capture some of the changes in the Fermi surface that are seen in ARPES studies. We find a continuous Fermi surface reconstruction (insets in Fig.~\ref{Ca327_strain_temp_dep}) that is not directly driven by the transition, but instead by the increase in magnetisation (and Stoner gap) as temperature is reduced. We emphasise that in this self-consistent calculation, the SRT and Fermi surface reconstruction are achieved as a function of temperature without varying the band filling. The model does not capture the partial gapping of the Fermi surface that is also seen by ARPES \cite{Baumberger2006, Horio2019}.

To make contact with our strain experiments, we also solve the model with a negative applied strain, $\varepsilon_\text{app} < 0$. This reduces the self-consistent strain, $\varepsilon$, and thus the nearest-neighbour hopping, resulting in a reduction in the critical temperature. As shown in Fig.~\ref{Ca327_strain_phase_diagrams}c, the linear suppression of the transition temperature matches the experimental phase diagrams for compressive stress applied along $\mathbf{a}$ and $\mathbf{b}$ in Fig.~\ref{Ca327_strain_phase_diagrams}a, b. As well as reproducing the main sequence of magnetic phases with temperature, our minimal model is therefore able to capture the qualitative features of our strain measurements.

\subsection*{Microscopic understanding of the transition}

\begin{figure*}
	\includegraphics[width=\linewidth]{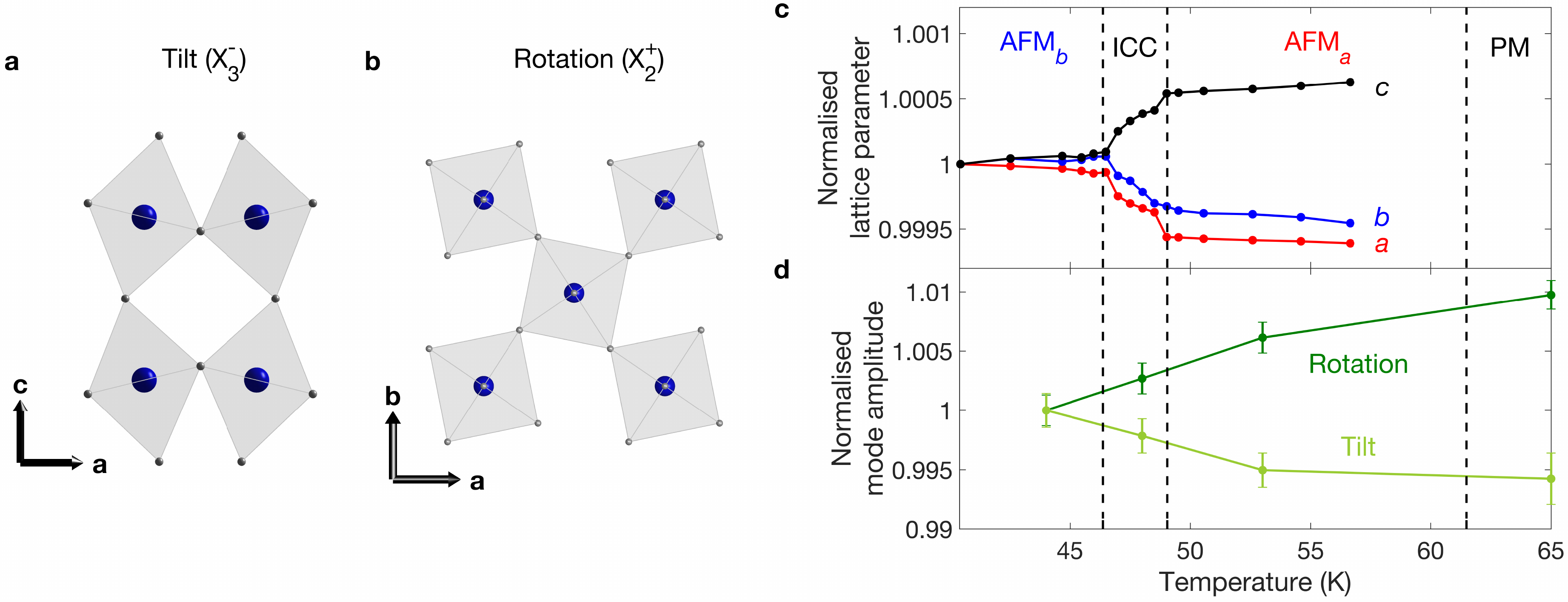}
	\caption{\label{Ca327_strain_tilt_rotation}\textbf{Octahedral tilt and rotation modes. a} Schematic of the tilt mode, consisting of staggered tilts of the RuO\textsubscript{6} octahedra around the $b$-axis. \textbf{b} Schematic of the rotation mode, consisting of staggered rotations of the octahedra around the $c$-axis. \textbf{c} Temperature dependence of the lattice parameters of \CaRuO{} from x-ray scattering, normalised to the values at \SI{40.4}{\kelvin}. \textbf{d} Temperature dependence of the octahedral rotation and tilt amplitudes determined from a symmetry analysis of single-crystal neutron diffraction data, normalised to the values at \SI{44}{\kelvin}. Errors are standard deviations.}
\end{figure*}

Our theoretical model is able to reproduce the phenomenology of the SRT by introducing a strain field that couples to the nearest-neighbour hopping. For a full understanding of the mechanism behind the transition, however, we need to connect this strain field to microscopic distortions of the \CaRuO{} lattice. Given their strong influence on the ground states across the Ruddlesden-Popper series of ruthenates, the octahedral tilt and rotation degrees of freedom (depicted in Fig.~\ref{Ca327_strain_tilt_rotation}a, b) are the most obvious candidates.

To be able to identify the tilt and rotation modes with the strain field of the model, we must first show that they directly influence the nearest-neighbour hopping. This can be achieved with a heuristic model of the overlap between neighbouring Ru $d$ orbitals, assuming perfectly rigid RuO\textsubscript{6} octahedra. Considering the tilt and rotation modes separately (valid for small distortions), we associate them with angles $\theta$ and $\phi$ respectively. In a similar manner to the two-centre method of Slater and Koster \cite{Slater1954}, the orbitals on a distorted Ru--O--Ru bond are expanded in a basis of orbitals quantised with respect to an undistorted bond axis (further details can be found in the Methods). We find that the intra-orbital hopping is equal along both nearest-neighbour bonds, with angle-dependent hopping parameters
\begin{equation}
	\begin{aligned}
		t(\theta) &= \frac{1}{4} \left[\cos(2 \sqrt{2} \theta) + 4 \cos(\sqrt{2} \theta) - 1\right] \ (dpd\pi) \\
		&+ \ \cos^3(\sqrt{2} \theta) \ (dd\pi) \ + \ \frac{3}{4} \sin^2(\sqrt{2} \theta) \ (dd\sigma) \\
		&+ \ \frac{1}{4} \sin^2(\sqrt{2} \theta) \cos(2 \sqrt{2} \theta) \ (dd\delta) \\\\
		t(\phi) &= (dpd\pi) + \cos^2(\phi) \ (dd\pi) + \sin^2(\phi) \ (dd\delta)
	\label{eq:effective_hoppings}
	\end{aligned}
\end{equation}
where $(l l^\prime m)$ are Slater-Koster integrals due to direct hopping between the $d$ orbitals, and $(dpd\pi)$ is the $\pi$-bonding integral due to indirect hopping via an oxygen $p$ orbital on an undistorted bond. We see that increasing the rotation reduces the contribution from the direct $\pi$-bonding while allowing a weaker $\delta$-bond to develop, therefore reducing the effective hopping. The effect of tilt on the hopping depends on the precise hierarchy of bonding integrals, but we expect from the behaviour of the (Sr$_{1 - x}$Ca$_x$)\textsubscript{3}Ru\textsubscript{2}O\textsubscript{7} series that increasing tilt will also reduce the hopping \cite{Peng2010, Rivero2017, Rivero2018}.

The importance of the octahedral tilts and rotations in the transition is further supported by (zero-stress) measurements of the \CaRuO{} structure as a function of temperature. Figure \ref{Ca327_strain_tilt_rotation}c shows the lattice parameters determined by resonant x-ray scattering. The in-plane lattice parameters are seen to increase and $c$ to decrease on cooling through the SRT, all by around 0.05\%. The lack of significant changes either side of the SRT points to the low phonon populations at these temperatures, and therefore the importance of the electronic coupling in the structural changes. We also directly determined the symmetry-adapted tilt ($X_3^-$) and rotation ($X_2^+$) mode amplitudes from neutron diffraction measurements, the results of which are shown in Fig.~\ref{Ca327_strain_tilt_rotation}d. On cooling through the SRT, the tilts increase by around 0.6\% while the rotations decrease by around 1\%. These changes are an order of magnitude larger than the changes in bond lengths \cite{Yoshida2005}, supporting our earlier assumption that the octahedra remain rigid. Based on our orbital-overlap argument above, we would expect the rotations to increase the hopping on cooling, slightly offset by the increase in tilt. This is in agreement with our model, in which the self-consistent strain and thus the hopping increase with decreasing temperature. We therefore have a clear picture of how the transition proceeds with temperature: an internal strain, corresponding to changes in the octahedral tilts and rotations, arises through feedback with the electronic system to modulate the hopping and thereby alter the magnetic anisotropy.

Finally, we can apply our understanding of the effects of the tilts and rotations to our strain experiments. While the precise changes in tilt and rotation angle under applied stress are unknown, we can make some general deductions from our knowledge of the Poisson ratios (ignoring strain along $\mathbf{c}$ which is likely to be mostly accommodated within the CaO rock-salt layers). When compressing along $\mathbf{a}$, the dominant effect is likely to be an increase in the tilt angle to accomodate the increased orthorhombicity, together with a smaller reduction in the rotation angle to account for the accompanying expansion along $\mathbf{b}$. This leads to a reduction in hopping (offset by the reduced rotation) which suppresses the transition temperature in agreement with Fig.~\ref{Ca327_strain_phase_diagrams}a. By contrast, the dominant effect when compressing along $\mathbf{b}$ is likely to be an increase in the rotation angle. This alone would cause an equal contraction along $\mathbf{a}$, such that a significant decrease in the tilt is needed to allow an expansion along $\mathbf{a}$ according to the Poisson ratio. This could explain the slope of the transition temperature in Fig.~\ref{Ca327_strain_phase_diagrams}b if the net effect is still a reduction in the hopping due to the increased rotation, but one that is largely offset by the reduction in tilt.

\section*{Discussion}

In this work, we have demonstrated that the spin orientation in \CaRuO{} can be controlled with applied stress. By isolating the lattice degrees of freedom, our measurements have provided a unique window into the role of strain in a coupled structural, electronic and magnetic phase transition. Neutron scattering gave us access to the bulk magnetic phases, while resonant x-ray scattering enabled us to probe in detail how strain is transmitted to the lattice, and thereby construct temperature-strain phase diagrams. These experiments motivated a theoretical model in which a strain field couples to the electronic hopping, and drives a spin reorientation when it reaches a critical value.

Our combination of experiment and theory has therefore uncovered a mechanism whereby strain, either externally induced or thermally generated through feedback with the electronic system, drives a global reorientation of the spins and concomitant reconstruction of the Fermi surface in \CaRuO{}. Given the ubiquity of the octahedral structural unit, we expect this mechanism to be relevant to other transition-metal oxides. For the Ruddlesden-Popper ruthenates in particular, octahedral tilts and rotations have a profound effect on the ground state, whether that is a Mott insulator or unconventional superconductor. Strain then offers the possibility of continuously tuning between these phases without the disorder introduced by chemical doping, and possibly uncovering phases not seen in the ambient structures.

\section*{Methods}

\subsection*{Samples}

Single crystals of \CaRuO{} were grown using the floating zone method and characterised by wavelength-dispersive x-ray spectrometry, resistivity and magnetization measurements, and x-ray and neutron diffraction \cite{Dashwood2020, Faure2023}. The strain axis was aligned with Laue diffraction. A polarised-light microscope was used to identify orthorhombic twin domains, and single-domain pieces were then cut out using a wire saw. The single-domain nature of the samples was checked during the scattering measurements. The samples were found to naturally cleave during cutting, producing long, thin bars with clean $(0, 0, 1)$ faces for scattering (and precluding measurements with stress applied along $\mathbf{c}$). The dimensions of the sample used for neutron scattering were $L \times W \times T \approx 1 \times 0.3 \times$ \SI{0.05}{\milli\metre}, for x-ray scattering with stress along $\mathbf{a}$ were $0.3 \times 0.1 \times$ \SI{0.04}{\milli\metre}, and for x-ray scattering with stress along $\mathbf{b}$ were $0.2 \times 0.1 \times$ \SI{0.03}{\milli\metre}.

\subsection*{Neutron scattering under stress}

Neutron scattering measurements under stress were performed at the WISH instrument of the ISIS Neutron and Muon Source \cite{Chapon2011}. We used the CS200T strain cell from Razorbill Instruments \cite{Razorbill}. The cell incorporates a \ang{90} access cone allowing transmission measurements with minimal background signal, and a capacitive displacement sensor using which we can calculate the applied strain. The cell was mounted on a custom cryostat stick with feedthroughs for the cables to power the cell and measure the capacitance of the displacement sensor. Voltage was applied to the cell with an RP100 power supply from Razorbill Instruments, and the capacitance was measured with a Keysight E4980AL LCR meter, both of which were integrated into the instrument control software to enable automated control and data acquisition. A temperature sensor was thermally contacted with the body of the cell to give accurate readings of the sample temperature. Cadmium shielding was used to reduce background scattering from the bridges of the cell and from unstrained regions of the sample near the sample plates, while still allowing full beam access in the horizontal plane. The sample was sandwiched between two pairs of sample plates and secured with Stycast 2850FT epoxy, such that $\mathbf{a}$ lies along the stress direction and vertical detector banks. The sample was cooled without any applied stress, and held at fixed temperatures during the strain sweeps. Data analysis was performed with \textsc{Mantid} \cite{Arnold2014}. All data are normalised to the cumulative current and a beam monitor. The intensities were obtained by diffraction focussing a small area on the detector around the peak, and then integrating the resulting time-of-flight spectra.

\subsection*{X-ray scattering under stress}

X-ray scattering measurements were performed at beamline I16 of the Diamond Light Source, with the incident energy tuned to the Ru $L_2$ edge (\SI{2.967}{\kilo\electronvolt}) by performing 4 bounces on the monochromator and using an extended helium-filled beam pipe. Here, we used the CS100 strain cell from Razorbill Instruments \cite{Razorbill}, which is smaller than the CS200T and necessitates a reflection scattering geometry. A custom mount was designed to hold the cell in the closed-cycle cryocooler, with feedthroughs added for the power and capacitance cables. As at WISH, the RP100 power supply and E4980AL LCR meter were interfaced with the beamline software to enable remote control and scripting of the measurements. A Cernox temperature sensor was thermally contacted with the body of the cell. The sample was mounted to the raised sample plates with Stycast 2850FT epoxy, with either $\mathbf{a}$ or $\mathbf{b}$ along the stress direction. No top-plates were used in order to maximise beam access, but the asymmetric mounting resulted in stress being transmitted mostly through the lower surface of the sample causing it to bend (see the Supplementary Information). This, along with the difference in elastic moduli of the sample ($\sim$\SI{100}{\giga\pascal}) and epoxy ($\sim$\SI{4}{\giga\pascal}), leads to the true strain being significantly lower than the applied strain. The diffractometer was operated with a vertical scattering geometry in fixed-azimuth mode, and the scattered intensity was measured with an in-vacuum Pilatus 100K area detector in ultrahigh gain mode. All temperature changes were conducted with zero applied stress, and the temperature was kept constant during strain measurements. Data analysis was carried out using the \textsc{Py16} program \cite{Porter2020}, and reciprocal-space maps were reconstructed from rocking scans using the \textsc{MillerSpaceMapper} software \cite{Basham2015}. Intensities were obtained by summing over a region-of-interest on the area detector and fitting the resulting rocking curves with pseudo-Voigt profiles plus a constant background.

\subsection*{Neutron diffraction}

The neutron diffraction data shown in Fig.~\ref{Ca327_strain_tilt_rotation}d were taken on the four-circle diffractometer D9 at the Institut Laue Langevin. A wavelength of \SI{0.836}{\angstrom} was obtained from a Cu(220) monochromator giving access to 800 reflections (up to $h = 10$, $k = 10$, $l = 15$). The symmetry mode analysis was performed using the \textsc{Amplimodes} software \cite{Orobengoa2009} from the Bilbao Crystallographic Server, considering the high-symmetry parent phase $I4/mmm$ (139) and the low symmetry phase $Bb2_1m$ (36). The mode amplitudes were directly refined at each temperature by the least-squares method using \textsc{Fullprof} \cite{Rodriguez-Carvajal1993}.

\subsection*{Strain-coupled electronic model}

We apply the tight-binding model for \CaRuO~that has been fitted to ARPES data \cite{Horio2019}. The tight-binding Hamiltonian in a crystal momentum basis is
\begin{equation}
	\begin{aligned}
		\hat H_t(\mathbf{k}) = \sum_\tau \Big[&\epsilon_x(\mathbf{k}) \ \underline{c}_{xz, \tau}^\dagger(\mathbf{k}) \ \sigma_0 \ \underline{c}_{xz, \bar{\tau}}(\mathbf{k}) \\
		+ \ &\epsilon_y(\mathbf{k}) \ \underline{c}_{yz, \tau}^\dagger(\mathbf{k}) \ \sigma_0 \ \underline{c}_{yz, \bar{\tau}}(\mathbf{k})\Big] \\
		+ \sum_{\alpha, \tau} \Big[&\epsilon_{ab}(\mathbf{k}) \ \underline{c}_{\alpha, \tau}^\dagger(\mathbf{k}) \ \sigma_0 \ \underline{c}_{\alpha, \tau}(\mathbf{k}) \\
		+ \ &\epsilon_{ab}^\prime(\mathbf{k}) \ \underline{c}_{\alpha, \tau}^\dagger(\mathbf{k}) \ \sigma_0 \ \underline{c}_{\bar{\alpha}, \tau}(\mathbf{k})\Big]
		\label{eq:Horio_TB}
	\end{aligned}
\end{equation}
where $\underline{c}_{\alpha \tau}^\dagger(\mathbf{k}) = (c_{\alpha \tau \uparrow}^\dagger(\mathbf{k}),  c_{\alpha \tau \downarrow}^\dagger(\mathbf{k}))$, $\sigma_0$ is the $2 \times 2$ identity matrix in spin space, $\alpha \in \{xz, yz\}$ and $\tau \in \{A, B\}$. Bars are used to denote flipped orbital and sublattice indices, i.e.~$\bar{xz} = yz$, $\bar{A} = B$, and vice versa. In the tetragonal crystal momentum basis, $\mathbf{k} = (k_x, k_y)$, the dispersions are
\begin{equation}
	\begin{split}
		\epsilon_x(\mathbf{k}) &= -2 t \cos(k_x a) \\
		\epsilon_y(\mathbf{k}) &= -2 t \cos(k_y a) \\
		\epsilon_{ab}(\mathbf{k}) &= -2 t_a \cos\left((k_x - k_y) a\right) - 2 t_b \cos\left((k_x + k_y) a\right) \\
		\epsilon_{ab}^\prime(\mathbf{k}) &= -2 t_a^\prime \cos\left((k_x - k_y) a\right) - 2 t_b^\prime \cos\left((k_x + k_y) a\right) \\
		\label{eq:Horio_dispersions}
	\end{split}
\end{equation}
where $k_{x, y} \in [-\frac{\pi}{a}, \frac{\pi}{a}]$ with lattice constant $a$. $\epsilon_{x(y)}$ describes the nearest-neighbour hopping between $xz$($yz$) orbitals along the $x$($y$) direction, and $\epsilon_{ab}$ and $\epsilon_{ab}^\prime$ describe the intra- and inter-orbital next-nearest-neighbour hopping along the $a$ and $b$ orthorhombic directions respectively. We use $t_a = 0.04 \tilde{t}$, $t_b = 0.16 \tilde{t}$ and $t_a^\prime = - t_b^\prime = 0.079 \tilde{t}$, with $\tilde{t} = 1.20$. The coefficients were chosen by fitting the Fermi surfaces in both magnetic phases to those measured using ARPES \cite{Horio2019}. Only the intra-orbital nearest-neighbour hopping parameter couples to the strain field, as $t = t_0 + \nu \varepsilon$. Strictly, the octahedral tilting and rotation also induce inter-orbital nearest-neighbour hopping between the $d_{xz}$ and $d_{yz}$ orbitals \cite{Mohapatra2020}. This hopping mode is neglected in ref.~\cite{Horio2019}, and we also neglect it as a sub-leading effect of the lattice distortions.

In our minimal model, we consider only the on-site intra-orbital repulsion, $\hat{H}_U = U \sum_{i \tau \alpha} \hat{n}_{i \tau \alpha \uparrow} \hat{n}_{i \tau \alpha \downarrow}$, and neglect the inter-orbital repulsion and Hund's coupling for simplicity. $\hat{H}_U$ is treated in the Hartree-Fock mean-field approximation by decoupling in the charge, $\hat{\rho}_{i \tau \alpha} = \frac{1}{2} \ \underline{c}_{i \tau \alpha}^\dagger \ \sigma_0 \ \underline{c}_{i \tau \alpha}$, and spin, $\hat{\mathbf{S}}_{i \tau \alpha} = \frac{1}{2} \ \underline{c}_{i \tau \alpha}^\dagger \ \boldsymbol{\sigma} \ \underline{c}_{i \tau \alpha}$, channels, where $\boldsymbol{\sigma} = (\sigma_x, \sigma_y, \sigma_z)$ is a vector of Pauli matrices. Defining $\rho_{\tau \alpha}$ and $\mathbf{M}_{\tau \alpha}$ as the site-averaged charge and spin densities, the interaction decouples to
\begin{equation}
	\begin{aligned}
		\hat{H}_U^\text{MF} &= U \sum_{i, \tau, \alpha} \underline{c}_{i \tau \alpha}^\dagger \left(\rho_{\tau \alpha} \sigma_0 - \mathbf{M}_{\tau \alpha} \cdot \boldsymbol{\sigma}\right) \underline{c}_{i \tau \alpha} \\
		&+ U \sum_{ \tau, \alpha} \left(\mathbf{M}_{\tau \alpha}^2 - \rho_{\tau \alpha}^2\right).
	\end{aligned}
\end{equation}
The system is described in the mean-field and at fixed chemical potential, $\mu$, by the Hamiltonian, $\hat{H}^\text{MF} = \hat{H}_t + \hat{H}_\lambda + \hat{H}_U^\text{MF} - \mu \hat{N} = \sum_{n \mathbf{k}} \left(\epsilon_{n \mathbf{k}} - \mu\right) \hat{n}_{n \mathbf{k}} + U \sum_{ \tau \alpha} \left(\mathbf{M}_{\tau \alpha}^2 - \rho_{\tau \alpha}^2\right)$. The free energy is evaluated via the partition function to give Eq.~(\ref{eq:free_energy}) (in the absence of coupling to strain, $\nu = \kappa = 0$) in terms of the energy eigenvalues, $\epsilon_{n \mathbf{k}}$, in a band and crystal momentum basis. The strain field is then introduced into the model as described in the Results.

The values of  strain parameters ($\nu$, $\kappa$) are chosen such that the calculated strain is of the same order of magnitude as seen in experiment (see Figs.~\ref{Ca327_strain_temp_dep} and \ref{Ca327_strain_tilt_rotation}c), and the magnitude of strain-hopping coupling, $\nu$, is comparable to a value calculated from first principles for a related material \cite{Li2022}. It is worth noting that the solution of this model is invariant under the scaling of the strain parameters as $\nu \rightarrow x \nu$, $\kappa \rightarrow x^2 \kappa$, $\varepsilon_0 \rightarrow \varepsilon_0 / x$ and $\varepsilon_\text{app} \rightarrow \varepsilon_\text{app} / x$. This scales the self-consistent strain solution as $\varepsilon \rightarrow \varepsilon / x$ but leaves the electronic solution unchanged.

\subsection*{\label{Hopping_details}Effective hopping on a distorted lattice}

\begin{figure*}
	\includegraphics[width=\linewidth]{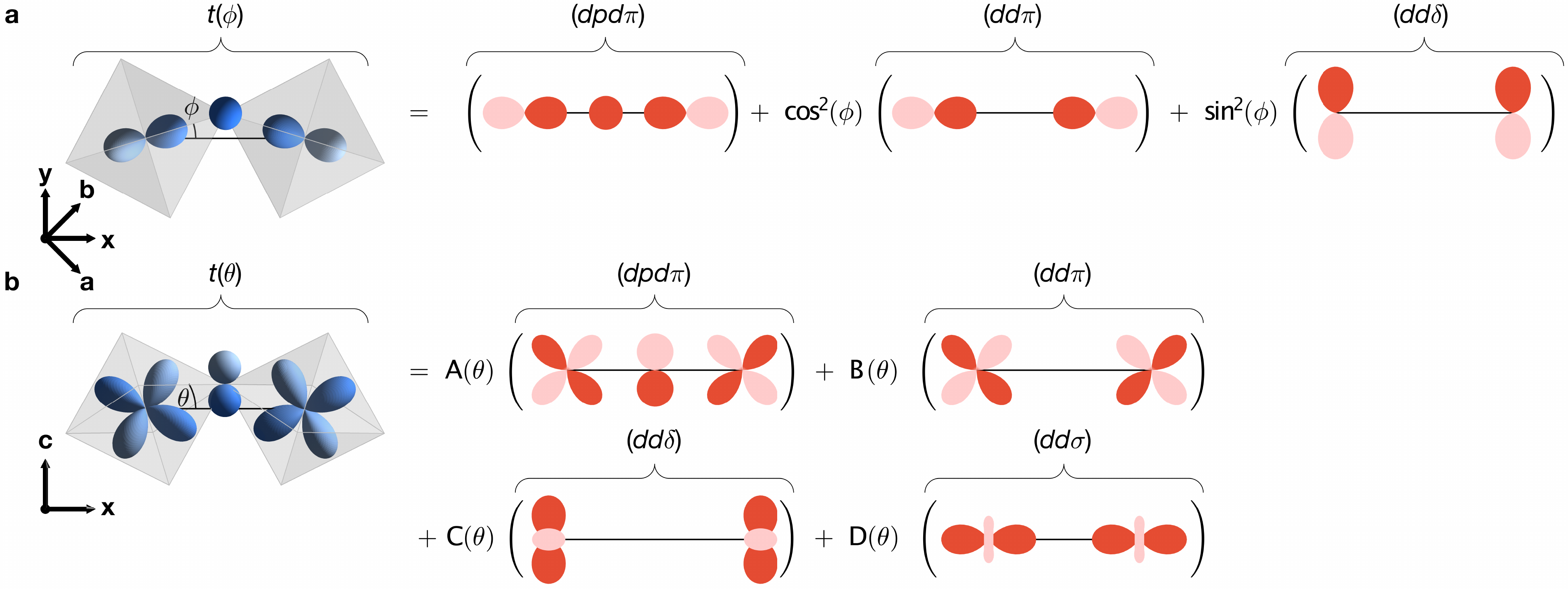}
	\caption{\label{Ca327_strain_hopping}\textbf{Rotation- and tilt-dependence of hopping. a, b} Dependence of the in-plane nearest-neighbour hopping, $t$, between Ru $d_{xz}$ orbitals along $\mathbf{x}$ on the octahedral rotation angle, $\phi$, and octahedral tilt angle, $\theta$, respectively. The orbitals on the distorted Ru--O--Ru bond are shown in blue on the left side of each equation. Direct hopping is expressed in terms of Slater-Koster integrals, $(l l^\prime m)$, illustrated in red on the right side of each equation. The indirect hopping, due to $\pi$-bonding with oxygen $p$ orbitals, is expressed in terms of $(dpd\pi)$, the $d$--$p$--$d$ $\pi$-bonding integral on a straight Ru--O--Ru bond (also shown in red). All oxygen $p$ orbitals are considered but only the $p_z$ orbital is shown for clarity. The coefficients in (\textbf{b}) are $A(\theta) = \frac{1}{4} \left[\cos(2 \sqrt{2} \theta) + 4 \cos(\sqrt{2} \theta) - 1\right]$, $B(\theta) = \cos^3(\sqrt{2} \theta)$, $C(\theta) = \frac{1}{4} \sin^2(\sqrt{2} \theta) \cos(2 \sqrt{2} \theta)$ and $D(\theta) = \frac{3}{4} \sin^2(\sqrt{2} \theta)$. Identical results are obtained for the hopping between $d_{yz}$ orbitals along $\mathbf{y}$.}
\end{figure*}

We calculate the effective nearest-neighbour hopping between ruthenium $d$ orbitals in the presence of octahedral tilting and rotational distortions (separately), considering the direct orbital overlap and the indirect overlap via oxygen $p$ orbitals.

The hopping is first defined in the absence of lattice distortions, i.e.~for a straight Ru--O--Ru bond. In the case of hopping due to direct $d$ orbital overlap, we apply the Slater-Koster two-centre approximation,
\begin{equation}
	t_{l m, l^\prime m^\prime}^\text{Ru--Ru} = \langle \psi_{l m}^\text{Ru}(\mathbf{r}) | \hat{H}^\text{Ru--Ru} | \psi_{l^\prime m^\prime}^\text{Ru}(\mathbf{r} + R \hat{\mathbf{x}}) \rangle
\end{equation}
where $\psi_{l m}^\text{Ru}$ is an atomic orbital centred on a ruthenium ion whose angular momentum quantisation axis is dictated by the crystal field, and $\hat{H}^\text{Ru--Ru}$ is the sum of the kinetic energy operator and ionic potentials on the bond. In the case of indirect hopping, the effective hopping is calculated as a second order process in which an electron or hole hops between ruthenium $d$ orbitals via an oxygen $p$ orbital. The hopping matrix element for a straight Ru--O--Ru bond is 
\begin{widetext}
	\begin{equation}
		t_{l m, l^\prime m^\prime}^\text{Ru--O--Ru} = -\sum_{l^\prime, m^\prime} \frac{\langle \psi_{l m}^\text{Ru}(\mathbf{r}) | \hat{H}^\text{Ru--O} | \psi_{l^\prime m^\prime}^\text{O}(\mathbf{r} + \frac{R}{2}\hat{\mathbf{x}}) \rangle \langle \psi_{l^\prime m^\prime}^\text{O}(\mathbf{r} + \frac{R}{2}\hat{\mathbf{x}}) | \hat{H}^\text{O--Ru} | \psi_{l m}^\text{Ru}(\mathbf{r} + R \hat{\mathbf{x}}) \rangle}{E_{l m} - E_{l^\prime m^\prime}^\prime}
	\end{equation}
\end{widetext}
where $E_{l m}$ and $E_{l^\prime m^\prime}^\prime$ are the total energies of the system when the electron/hole occupies a ruthenium orbital, $\psi_{l m}^\text{Ru}$, or an oxygen orbital,  $\psi_{l^\prime m^\prime}^\text{O}$, respectively. The numerator is a product of matrix elements for Ru--O and O--Ru hopping processes. We make a key assumption that the lattice distortions are a weak perturbation on the oxygen $p$ orbitals, which remain degenerate. Every indirect hopping process then has the same intermediate energy, $E^\prime$, and the effect of lattice distortions is only to modify the two-centre integrals in the numerator.

The lattice distortions are introduced by rotating all orbitals in space accordingly. Since the orbital orientation is dictated most strongly by the local (octahedral) crystal field, we assume that the ruthenium $d$ orbitals are rotated/tilted along with the octahedra. Note that we decompose the (small) tilt about $\mathbf{b}$ into tilts of $\theta / \sqrt 2$ about the $x$- and $y$-axes. We also assume that the change in Ru--Ru distance due to the distortions has a sub-leading effect on the hopping and as such is neglected.

The rotated orbitals are then expanded in a basis of atomic orbitals that are quantised with respect an appropriate bond axis. In the direct case this is the Ru--Ru bond. The effective hopping can then be written in terms of matrix elements for the hopping of an electron between the basis orbitals, which are the Slater-Koster integrals
\begin{equation}
	(l l^\prime m) \delta_{m m^\prime} = \langle \phi_{l m}^\text{Ru}(\mathbf{r}) | \hat{H}^\text{Ru--Ru} | \phi_{l^\prime m^\prime}^\text{Ru}(\mathbf{r} + R \hat{\mathbf{x}}) \rangle
\end{equation}
where $\phi_{l m}$ is an orbital that is quantised with respect to the bond axis. As is convention, we use the labels $l, l^\prime \in \{s, p, d\}$ and $m \in \{\sigma, \pi, \delta \}$ on the left-hand side. In the indirect case we expand in bases of orbitals that are quantised with respect to the Ru--O and O--Ru bonds. We find that the effective hopping can be written as a function of the hopping along a straight bond,
\begin{widetext}
	\begin{equation}
		(l l^\prime l m) \delta_{m m^\prime} = - \frac{\langle \phi_{l m}^\text{Ru}(\mathbf{r}) | \hat{H}^\text{Ru--O} | \phi_{l^\prime m^\prime}^\text{O}(\mathbf{r} + R \hat{\mathbf{x}}) \rangle \langle \phi_{l^\prime m^\prime}^\text{O}(\mathbf{r} + R \hat{\mathbf{x}}) | \hat{H}^\text{O--Ru} | \phi_{l m}^\text{Ru}(\mathbf{r} + 2 R \hat{\mathbf{x}}) \rangle}{E_{l m} - E_{l^\prime m^\prime}^\prime}.
	\end{equation}
\end{widetext}
On the LHS we have introduced a similar notation for the indirect hopping matrix element to that of the Slater-Koster integrals. The numerator is a product of two Slater-Koster integrals. 

By summing the direct and indirect hoppings in each case, we find the effective nearest-neighbour hoppings, Eq.~(\ref{eq:effective_hoppings}). We find that increasing the rotation angle reduces the contribution from the direct $\pi$-bonding while allowing a weaker $\delta$-bond to develop, therefore reducing the effective hopping. Increasing the tilt angle reduces the contribution from both $\pi$-bonding modes, while allowing $\sigma$- and $\delta$-bonding modes to develop. Since the indirect $\pi$-bonding, $(dpd\pi)$, likely dominates, we expect tilting to cause a net reduction in the effective hopping.

\section*{Data availability}

Raw data from the neutron scattering measurements under stress are available at \url{https://doi.org/10.5286/ISIS.E.RB1920210}. Raw data from the neutron diffraction measurements are available at \url{https://doi.org/10.5291/ILL-DATA.EASY-951}. Source data used to plot the figures are available at \url{https://doi.org/10.6084/m9.figshare.23508354}.

\bibliography{Ca327_strain}

\begin{thebibliography}{50}%
\makeatletter
\providecommand \@ifxundefined [1]{%
 \@ifx{#1\undefined}
}%
\providecommand \@ifnum [1]{%
 \ifnum #1\expandafter \@firstoftwo
 \else \expandafter \@secondoftwo
 \fi
}%
\providecommand \@ifx [1]{%
 \ifx #1\expandafter \@firstoftwo
 \else \expandafter \@secondoftwo
 \fi
}%
\providecommand \natexlab [1]{#1}%
\providecommand \enquote  [1]{``#1''}%
\providecommand \bibnamefont  [1]{#1}%
\providecommand \bibfnamefont [1]{#1}%
\providecommand \citenamefont [1]{#1}%
\providecommand \href@noop [0]{\@secondoftwo}%
\providecommand \href [0]{\begingroup \@sanitize@url \@href}%
\providecommand \@href[1]{\@@startlink{#1}\@@href}%
\providecommand \@@href[1]{\endgroup#1\@@endlink}%
\providecommand \@sanitize@url [0]{\catcode `\\12\catcode `\$12\catcode
  `\&12\catcode `\#12\catcode `\^12\catcode `\_12\catcode `\%12\relax}%
\providecommand \@@startlink[1]{}%
\providecommand \@@endlink[0]{}%
\providecommand \url  [0]{\begingroup\@sanitize@url \@url }%
\providecommand \@url [1]{\endgroup\@href {#1}{\urlprefix }}%
\providecommand \urlprefix  [0]{URL }%
\providecommand \Eprint [0]{\href }%
\providecommand \doibase [0]{http://dx.doi.org/}%
\providecommand \selectlanguage [0]{\@gobble}%
\providecommand \bibinfo  [0]{\@secondoftwo}%
\providecommand \bibfield  [0]{\@secondoftwo}%
\providecommand \translation [1]{[#1]}%
\providecommand \BibitemOpen [0]{}%
\providecommand \bibitemStop [0]{}%
\providecommand \bibitemNoStop [0]{.\EOS\space}%
\providecommand \EOS [0]{\spacefactor3000\relax}%
\providecommand \BibitemShut  [1]{\csname bibitem#1\endcsname}%
\let\auto@bib@innerbib\@empty
\bibitem [{\citenamefont {Peierls}(1955)}]{Peierls1955}%
  \BibitemOpen
  \bibfield  {author} {\bibinfo {author} {\bibfnamefont {R.~E.}\ \bibnamefont
  {Peierls}},\ }\href
  {https://global.oup.com/academic/product/quantum-theory-of-solids-9780192670175}
  {\emph {\bibinfo {title} {Quantum theory of solids}}},\ International series
  of monographs on physics\ (\bibinfo  {publisher} {Clarendon Press, Oxford},\
  \bibinfo {year} {1955})\BibitemShut {NoStop}%
\bibitem [{\citenamefont {Fr\"{o}hlich}(1954)}]{Frohlich1954}%
  \BibitemOpen
  \bibfield  {author} {\bibinfo {author} {\bibfnamefont {H.}~\bibnamefont
  {Fr\"{o}hlich}},\ }\href {\doibase 10.1098/rspa.1954.0116} {\bibfield
  {journal} {\bibinfo  {journal} {Proc. R. Soc. Lond. A}\ }\textbf {\bibinfo
  {volume} {223}},\ \bibinfo {pages} {296} (\bibinfo {year}
  {1954})}\BibitemShut {NoStop}%
\bibitem [{\citenamefont {Kanamori}(1960)}]{Kanamori1960}%
  \BibitemOpen
  \bibfield  {author} {\bibinfo {author} {\bibfnamefont {J.}~\bibnamefont
  {Kanamori}},\ }\href {\doibase 10.1063/1.1984590} {\bibfield  {journal}
  {\bibinfo  {journal} {J. App. Phys.}\ }\textbf {\bibinfo {volume} {31}},\
  \bibinfo {pages} {S14} (\bibinfo {year} {1960})}\BibitemShut {NoStop}%
\bibitem [{\citenamefont {Cao}\ \emph {et~al.}(2018{\natexlab{a}})\citenamefont
  {Cao}, \citenamefont {Fatemi}, \citenamefont {Demir}, \citenamefont {Fang},
  \citenamefont {Tomarken}, \citenamefont {Luo}, \citenamefont
  {Sanchez-Yamagishi}, \citenamefont {Watanabe}, \citenamefont {Taniguchi},
  \citenamefont {Kaxiras}, \citenamefont {Ashoori},\ and\ \citenamefont
  {Jarillo-Herrero}}]{Cao2018a}%
  \BibitemOpen
  \bibfield  {author} {\bibinfo {author} {\bibfnamefont {Y.}~\bibnamefont
  {Cao}}, \bibinfo {author} {\bibfnamefont {V.}~\bibnamefont {Fatemi}},
  \bibinfo {author} {\bibfnamefont {A.}~\bibnamefont {Demir}}, \bibinfo
  {author} {\bibfnamefont {S.}~\bibnamefont {Fang}}, \bibinfo {author}
  {\bibfnamefont {S.~L.}\ \bibnamefont {Tomarken}}, \bibinfo {author}
  {\bibfnamefont {J.~Y.}\ \bibnamefont {Luo}}, \bibinfo {author} {\bibfnamefont
  {J.~D.}\ \bibnamefont {Sanchez-Yamagishi}}, \bibinfo {author} {\bibfnamefont
  {K.}~\bibnamefont {Watanabe}}, \bibinfo {author} {\bibfnamefont
  {T.}~\bibnamefont {Taniguchi}}, \bibinfo {author} {\bibfnamefont
  {E.}~\bibnamefont {Kaxiras}}, \bibinfo {author} {\bibfnamefont {R.~C.}\
  \bibnamefont {Ashoori}}, \ and\ \bibinfo {author} {\bibfnamefont
  {P.}~\bibnamefont {Jarillo-Herrero}},\ }\href {\doibase 10.1038/nature26154}
  {\bibfield  {journal} {\bibinfo  {journal} {Nature}\ }\textbf {\bibinfo
  {volume} {556}},\ \bibinfo {pages} {80} (\bibinfo {year}
  {2018}{\natexlab{a}})}\BibitemShut {NoStop}%
\bibitem [{\citenamefont {Cao}\ \emph {et~al.}(2018{\natexlab{b}})\citenamefont
  {Cao}, \citenamefont {Fatemi}, \citenamefont {Fang}, \citenamefont
  {Watanabe}, \citenamefont {Taniguchi}, \citenamefont {Kaxiras},\ and\
  \citenamefont {Jarillo-Herrero}}]{Cao2018b}%
  \BibitemOpen
  \bibfield  {author} {\bibinfo {author} {\bibfnamefont {Y.}~\bibnamefont
  {Cao}}, \bibinfo {author} {\bibfnamefont {V.}~\bibnamefont {Fatemi}},
  \bibinfo {author} {\bibfnamefont {S.}~\bibnamefont {Fang}}, \bibinfo {author}
  {\bibfnamefont {K.}~\bibnamefont {Watanabe}}, \bibinfo {author}
  {\bibfnamefont {T.}~\bibnamefont {Taniguchi}}, \bibinfo {author}
  {\bibfnamefont {E.}~\bibnamefont {Kaxiras}}, \ and\ \bibinfo {author}
  {\bibfnamefont {P.}~\bibnamefont {Jarillo-Herrero}},\ }\href {\doibase
  10.1038/nature26160} {\bibfield  {journal} {\bibinfo  {journal} {Nature}\
  }\textbf {\bibinfo {volume} {556}},\ \bibinfo {pages} {43} (\bibinfo {year}
  {2018}{\natexlab{b}})}\BibitemShut {NoStop}%
\bibitem [{\citenamefont {Hicks}\ \emph {et~al.}(2014)\citenamefont {Hicks},
  \citenamefont {Brodsky}, \citenamefont {Yelland}, \citenamefont {Gibbs},
  \citenamefont {Bruin}, \citenamefont {Barber}, \citenamefont {Edkins},
  \citenamefont {Nishimura}, \citenamefont {Yonezawa}, \citenamefont {Maeno},\
  and\ \citenamefont {Mackenzie}}]{Hicks2014}%
  \BibitemOpen
  \bibfield  {author} {\bibinfo {author} {\bibfnamefont {C.~W.}\ \bibnamefont
  {Hicks}}, \bibinfo {author} {\bibfnamefont {D.~O.}\ \bibnamefont {Brodsky}},
  \bibinfo {author} {\bibfnamefont {E.~A.}\ \bibnamefont {Yelland}}, \bibinfo
  {author} {\bibfnamefont {A.~S.}\ \bibnamefont {Gibbs}}, \bibinfo {author}
  {\bibfnamefont {J.~A.~N.}\ \bibnamefont {Bruin}}, \bibinfo {author}
  {\bibfnamefont {M.~E.}\ \bibnamefont {Barber}}, \bibinfo {author}
  {\bibfnamefont {S.~D.}\ \bibnamefont {Edkins}}, \bibinfo {author}
  {\bibfnamefont {K.}~\bibnamefont {Nishimura}}, \bibinfo {author}
  {\bibfnamefont {S.}~\bibnamefont {Yonezawa}}, \bibinfo {author}
  {\bibfnamefont {Y.}~\bibnamefont {Maeno}}, \ and\ \bibinfo {author}
  {\bibfnamefont {A.~P.}\ \bibnamefont {Mackenzie}},\ }\href {\doibase
  10.1126/science.1248292} {\bibfield  {journal} {\bibinfo  {journal}
  {Science}\ }\textbf {\bibinfo {volume} {344}},\ \bibinfo {pages} {283}
  (\bibinfo {year} {2014})}\BibitemShut {NoStop}%
\bibitem [{\citenamefont {Steppke}\ \emph {et~al.}(2017)\citenamefont
  {Steppke}, \citenamefont {Zhao}, \citenamefont {Barber}, \citenamefont
  {Scaffidi}, \citenamefont {Jerzembeck}, \citenamefont {Rosner}, \citenamefont
  {Gibbs}, \citenamefont {Maeno}, \citenamefont {Simon}, \citenamefont
  {Mackenzie},\ and\ \citenamefont {Hicks}}]{Steppke2017}%
  \BibitemOpen
  \bibfield  {author} {\bibinfo {author} {\bibfnamefont {A.}~\bibnamefont
  {Steppke}}, \bibinfo {author} {\bibfnamefont {L.}~\bibnamefont {Zhao}},
  \bibinfo {author} {\bibfnamefont {M.~E.}\ \bibnamefont {Barber}}, \bibinfo
  {author} {\bibfnamefont {T.}~\bibnamefont {Scaffidi}}, \bibinfo {author}
  {\bibfnamefont {F.}~\bibnamefont {Jerzembeck}}, \bibinfo {author}
  {\bibfnamefont {H.}~\bibnamefont {Rosner}}, \bibinfo {author} {\bibfnamefont
  {A.~S.}\ \bibnamefont {Gibbs}}, \bibinfo {author} {\bibfnamefont
  {Y.}~\bibnamefont {Maeno}}, \bibinfo {author} {\bibfnamefont {S.~H.}\
  \bibnamefont {Simon}}, \bibinfo {author} {\bibfnamefont {A.~P.}\ \bibnamefont
  {Mackenzie}}, \ and\ \bibinfo {author} {\bibfnamefont {C.~W.}\ \bibnamefont
  {Hicks}},\ }\href {\doibase 10.1126/science.aaf9398} {\bibfield  {journal}
  {\bibinfo  {journal} {Science}\ }\textbf {\bibinfo {volume} {355}},\ \bibinfo
  {pages} {eaaf9398} (\bibinfo {year} {2017})}\BibitemShut {NoStop}%
\bibitem [{\citenamefont {Kim}\ \emph {et~al.}(2018)\citenamefont {Kim},
  \citenamefont {Souliou}, \citenamefont {Barber}, \citenamefont
  {Lefran\c{c}ois}, \citenamefont {Minola}, \citenamefont {Tortora},
  \citenamefont {Heid}, \citenamefont {Nandi}, \citenamefont {Borzi},
  \citenamefont {Garbarino}, \citenamefont {Bosak}, \citenamefont {Porras},
  \citenamefont {Loew}, \citenamefont {K\"{o}nig}, \citenamefont {Moll},
  \citenamefont {Mackenzie}, \citenamefont {Keimer}, \citenamefont {Hicks},\
  and\ \citenamefont {Le~Tacon}}]{Kim2018}%
  \BibitemOpen
  \bibfield  {author} {\bibinfo {author} {\bibfnamefont {H.-H.}\ \bibnamefont
  {Kim}}, \bibinfo {author} {\bibfnamefont {S.~M.}\ \bibnamefont {Souliou}},
  \bibinfo {author} {\bibfnamefont {M.~E.}\ \bibnamefont {Barber}}, \bibinfo
  {author} {\bibfnamefont {E.}~\bibnamefont {Lefran\c{c}ois}}, \bibinfo
  {author} {\bibfnamefont {M.}~\bibnamefont {Minola}}, \bibinfo {author}
  {\bibfnamefont {M.}~\bibnamefont {Tortora}}, \bibinfo {author} {\bibfnamefont
  {R.}~\bibnamefont {Heid}}, \bibinfo {author} {\bibfnamefont {N.}~\bibnamefont
  {Nandi}}, \bibinfo {author} {\bibfnamefont {R.~A.}\ \bibnamefont {Borzi}},
  \bibinfo {author} {\bibfnamefont {G.}~\bibnamefont {Garbarino}}, \bibinfo
  {author} {\bibfnamefont {A.}~\bibnamefont {Bosak}}, \bibinfo {author}
  {\bibfnamefont {J.}~\bibnamefont {Porras}}, \bibinfo {author} {\bibfnamefont
  {T.}~\bibnamefont {Loew}}, \bibinfo {author} {\bibfnamefont {M.}~\bibnamefont
  {K\"{o}nig}}, \bibinfo {author} {\bibfnamefont {P.~J.~W.}\ \bibnamefont
  {Moll}}, \bibinfo {author} {\bibfnamefont {A.~P.}\ \bibnamefont {Mackenzie}},
  \bibinfo {author} {\bibfnamefont {B.}~\bibnamefont {Keimer}}, \bibinfo
  {author} {\bibfnamefont {C.~W.}\ \bibnamefont {Hicks}}, \ and\ \bibinfo
  {author} {\bibfnamefont {M.}~\bibnamefont {Le~Tacon}},\ }\href {\doibase
  10.1126/science.aat4708} {\bibfield  {journal} {\bibinfo  {journal}
  {Science}\ }\textbf {\bibinfo {volume} {362}},\ \bibinfo {pages} {1040}
  (\bibinfo {year} {2018})}\BibitemShut {NoStop}%
\bibitem [{\citenamefont {Kim}\ \emph {et~al.}(2021)\citenamefont {Kim},
  \citenamefont {Lefran\c{c}ois}, \citenamefont {Kummer}, \citenamefont
  {Fumagalli}, \citenamefont {Brookes}, \citenamefont {Betto}, \citenamefont
  {Nakata}, \citenamefont {Tortora}, \citenamefont {Porras}, \citenamefont
  {Loew}, \citenamefont {Barber}, \citenamefont {Braicovich}, \citenamefont
  {Mackenzie}, \citenamefont {Hicks}, \citenamefont {Keimer}, \citenamefont
  {Minola},\ and\ \citenamefont {Le~Tacon}}]{Kim2021}%
  \BibitemOpen
  \bibfield  {author} {\bibinfo {author} {\bibfnamefont {H.-H.}\ \bibnamefont
  {Kim}}, \bibinfo {author} {\bibfnamefont {E.}~\bibnamefont {Lefran\c{c}ois}},
  \bibinfo {author} {\bibfnamefont {K.}~\bibnamefont {Kummer}}, \bibinfo
  {author} {\bibfnamefont {R.}~\bibnamefont {Fumagalli}}, \bibinfo {author}
  {\bibfnamefont {N.~B.}\ \bibnamefont {Brookes}}, \bibinfo {author}
  {\bibfnamefont {D.}~\bibnamefont {Betto}}, \bibinfo {author} {\bibfnamefont
  {S.}~\bibnamefont {Nakata}}, \bibinfo {author} {\bibfnamefont
  {M.}~\bibnamefont {Tortora}}, \bibinfo {author} {\bibfnamefont
  {J.}~\bibnamefont {Porras}}, \bibinfo {author} {\bibfnamefont
  {T.}~\bibnamefont {Loew}}, \bibinfo {author} {\bibfnamefont {M.~E.}\
  \bibnamefont {Barber}}, \bibinfo {author} {\bibfnamefont {L.}~\bibnamefont
  {Braicovich}}, \bibinfo {author} {\bibfnamefont {A.~P.}\ \bibnamefont
  {Mackenzie}}, \bibinfo {author} {\bibfnamefont {C.~W.}\ \bibnamefont
  {Hicks}}, \bibinfo {author} {\bibfnamefont {B.}~\bibnamefont {Keimer}},
  \bibinfo {author} {\bibfnamefont {M.}~\bibnamefont {Minola}}, \ and\ \bibinfo
  {author} {\bibfnamefont {M.}~\bibnamefont {Le~Tacon}},\ }\href {\doibase
  10.1103/PhysRevLett.126.037002} {\bibfield  {journal} {\bibinfo  {journal}
  {Phys. Rev. Lett.}\ }\textbf {\bibinfo {volume} {126}},\ \bibinfo {pages}
  {037002} (\bibinfo {year} {2021})}\BibitemShut {NoStop}%
\bibitem [{\citenamefont {Wang}\ \emph {et~al.}(2022)\citenamefont {Wang},
  \citenamefont {von Arx}, \citenamefont {Mazzone}, \citenamefont {Mustafi},
  \citenamefont {Horio}, \citenamefont {K\"{u}spert}, \citenamefont {Choi},
  \citenamefont {Bucher}, \citenamefont {Wo}, \citenamefont {Zhao},
  \citenamefont {Zhang}, \citenamefont {Asmara}, \citenamefont {Sassa},
  \citenamefont {M\r{a}nsson}, \citenamefont {Christensen}, \citenamefont
  {Janoschek}, \citenamefont {Kurosawa}, \citenamefont {Momono}, \citenamefont
  {Oda}, \citenamefont {Fischer}, \citenamefont {Schmitt},\ and\ \citenamefont
  {Chang}}]{Wang2022}%
  \BibitemOpen
  \bibfield  {author} {\bibinfo {author} {\bibfnamefont {Q.}~\bibnamefont
  {Wang}}, \bibinfo {author} {\bibfnamefont {K.}~\bibnamefont {von Arx}},
  \bibinfo {author} {\bibfnamefont {D.~G.}\ \bibnamefont {Mazzone}}, \bibinfo
  {author} {\bibfnamefont {S.}~\bibnamefont {Mustafi}}, \bibinfo {author}
  {\bibfnamefont {M.}~\bibnamefont {Horio}}, \bibinfo {author} {\bibfnamefont
  {J.}~\bibnamefont {K\"{u}spert}}, \bibinfo {author} {\bibfnamefont
  {J.}~\bibnamefont {Choi}}, \bibinfo {author} {\bibfnamefont {D.}~\bibnamefont
  {Bucher}}, \bibinfo {author} {\bibfnamefont {H.}~\bibnamefont {Wo}}, \bibinfo
  {author} {\bibfnamefont {J.}~\bibnamefont {Zhao}}, \bibinfo {author}
  {\bibfnamefont {W.}~\bibnamefont {Zhang}}, \bibinfo {author} {\bibfnamefont
  {T.~C.}\ \bibnamefont {Asmara}}, \bibinfo {author} {\bibfnamefont
  {Y.}~\bibnamefont {Sassa}}, \bibinfo {author} {\bibfnamefont
  {M.}~\bibnamefont {M\r{a}nsson}}, \bibinfo {author} {\bibfnamefont {N.~B.}\
  \bibnamefont {Christensen}}, \bibinfo {author} {\bibfnamefont
  {M.}~\bibnamefont {Janoschek}}, \bibinfo {author} {\bibfnamefont
  {T.}~\bibnamefont {Kurosawa}}, \bibinfo {author} {\bibfnamefont
  {N.}~\bibnamefont {Momono}}, \bibinfo {author} {\bibfnamefont
  {M.}~\bibnamefont {Oda}}, \bibinfo {author} {\bibfnamefont {M.~H.}\
  \bibnamefont {Fischer}}, \bibinfo {author} {\bibfnamefont {T.}~\bibnamefont
  {Schmitt}}, \ and\ \bibinfo {author} {\bibfnamefont {J.}~\bibnamefont
  {Chang}},\ }\href {\doibase 10.1038/s41467-022-29465-4} {\bibfield  {journal}
  {\bibinfo  {journal} {Nat. Commun.}\ }\textbf {\bibinfo {volume} {13}},\
  \bibinfo {pages} {1795} (\bibinfo {year} {2022})}\BibitemShut {NoStop}%
\bibitem [{\citenamefont {Choi}\ \emph {et~al.}(2022)\citenamefont {Choi},
  \citenamefont {Wang}, \citenamefont {J\"{o}hr}, \citenamefont {Christensen},
  \citenamefont {K\"{u}spert}, \citenamefont {Bucher}, \citenamefont
  {Biscette}, \citenamefont {Fischer}, \citenamefont {H\"{u}cker},
  \citenamefont {Kurosawa}, \citenamefont {Momono}, \citenamefont {Oda},
  \citenamefont {Ivashko}, \citenamefont {Zimmermann}, \citenamefont
  {Janoschek},\ and\ \citenamefont {Chang}}]{Choi2022}%
  \BibitemOpen
  \bibfield  {author} {\bibinfo {author} {\bibfnamefont {J.}~\bibnamefont
  {Choi}}, \bibinfo {author} {\bibfnamefont {Q.}~\bibnamefont {Wang}}, \bibinfo
  {author} {\bibfnamefont {S.}~\bibnamefont {J\"{o}hr}}, \bibinfo {author}
  {\bibfnamefont {N.~B.}\ \bibnamefont {Christensen}}, \bibinfo {author}
  {\bibfnamefont {J.}~\bibnamefont {K\"{u}spert}}, \bibinfo {author}
  {\bibfnamefont {D.}~\bibnamefont {Bucher}}, \bibinfo {author} {\bibfnamefont
  {D.}~\bibnamefont {Biscette}}, \bibinfo {author} {\bibfnamefont {M.~H.}\
  \bibnamefont {Fischer}}, \bibinfo {author} {\bibfnamefont {M.}~\bibnamefont
  {H\"{u}cker}}, \bibinfo {author} {\bibfnamefont {T.}~\bibnamefont
  {Kurosawa}}, \bibinfo {author} {\bibfnamefont {N.}~\bibnamefont {Momono}},
  \bibinfo {author} {\bibfnamefont {M.}~\bibnamefont {Oda}}, \bibinfo {author}
  {\bibfnamefont {O.}~\bibnamefont {Ivashko}}, \bibinfo {author} {\bibfnamefont
  {M.~v.}\ \bibnamefont {Zimmermann}}, \bibinfo {author} {\bibfnamefont
  {M.}~\bibnamefont {Janoschek}}, \ and\ \bibinfo {author} {\bibfnamefont
  {J.}~\bibnamefont {Chang}},\ }\href {\doibase 10.1103/PhysRevLett.128.207002}
  {\bibfield  {journal} {\bibinfo  {journal} {Phys. Rev. Lett.}\ }\textbf
  {\bibinfo {volume} {128}},\ \bibinfo {pages} {207002} (\bibinfo {year}
  {2022})}\BibitemShut {NoStop}%
\bibitem [{\citenamefont {Simutis}\ \emph {et~al.}(2022)\citenamefont
  {Simutis}, \citenamefont {K\"{u}spert}, \citenamefont {Wang}, \citenamefont
  {Choi}, \citenamefont {Bucher}, \citenamefont {Boehm}, \citenamefont
  {Bourdarot}, \citenamefont {Bertelsen}, \citenamefont {Wang}, \citenamefont
  {Kurosawa}, \citenamefont {Momono}, \citenamefont {Oda}, \citenamefont
  {M\r{a}nsson}, \citenamefont {Sassa}, \citenamefont {Janoschek},
  \citenamefont {Christensen}, \citenamefont {Chang},\ and\ \citenamefont
  {Mazzone}}]{Simutis2022}%
  \BibitemOpen
  \bibfield  {author} {\bibinfo {author} {\bibfnamefont {G.}~\bibnamefont
  {Simutis}}, \bibinfo {author} {\bibfnamefont {J.}~\bibnamefont
  {K\"{u}spert}}, \bibinfo {author} {\bibfnamefont {Q.}~\bibnamefont {Wang}},
  \bibinfo {author} {\bibfnamefont {J.}~\bibnamefont {Choi}}, \bibinfo {author}
  {\bibfnamefont {D.}~\bibnamefont {Bucher}}, \bibinfo {author} {\bibfnamefont
  {M.}~\bibnamefont {Boehm}}, \bibinfo {author} {\bibfnamefont
  {F.}~\bibnamefont {Bourdarot}}, \bibinfo {author} {\bibfnamefont
  {M.}~\bibnamefont {Bertelsen}}, \bibinfo {author} {\bibfnamefont {C.~N.}\
  \bibnamefont {Wang}}, \bibinfo {author} {\bibfnamefont {T.}~\bibnamefont
  {Kurosawa}}, \bibinfo {author} {\bibfnamefont {N.}~\bibnamefont {Momono}},
  \bibinfo {author} {\bibfnamefont {M.}~\bibnamefont {Oda}}, \bibinfo {author}
  {\bibfnamefont {M.}~\bibnamefont {M\r{a}nsson}}, \bibinfo {author}
  {\bibfnamefont {Y.}~\bibnamefont {Sassa}}, \bibinfo {author} {\bibfnamefont
  {M.}~\bibnamefont {Janoschek}}, \bibinfo {author} {\bibfnamefont {N.~B.}\
  \bibnamefont {Christensen}}, \bibinfo {author} {\bibfnamefont
  {J.}~\bibnamefont {Chang}}, \ and\ \bibinfo {author} {\bibfnamefont
  {D.}~\bibnamefont {Mazzone}},\ }\href {\doibase 10.1038/s42005-022-01061-4}
  {\bibfield  {journal} {\bibinfo  {journal} {Commun. Phys.}\ }\textbf
  {\bibinfo {volume} {5}},\ \bibinfo {pages} {296} (\bibinfo {year}
  {2022})}\BibitemShut {NoStop}%
\bibitem [{\citenamefont {Malinowski}\ \emph {et~al.}(2020)\citenamefont
  {Malinowski}, \citenamefont {Jiang}, \citenamefont {Sanchez}, \citenamefont
  {Mutch}, \citenamefont {Liu}, \citenamefont {Went}, \citenamefont {Liu},
  \citenamefont {Ryan}, \citenamefont {Kim},\ and\ \citenamefont
  {Chu}}]{Malinowski2020}%
  \BibitemOpen
  \bibfield  {author} {\bibinfo {author} {\bibfnamefont {P.}~\bibnamefont
  {Malinowski}}, \bibinfo {author} {\bibfnamefont {Q.}~\bibnamefont {Jiang}},
  \bibinfo {author} {\bibfnamefont {J.~J.}\ \bibnamefont {Sanchez}}, \bibinfo
  {author} {\bibfnamefont {J.}~\bibnamefont {Mutch}}, \bibinfo {author}
  {\bibfnamefont {Z.}~\bibnamefont {Liu}}, \bibinfo {author} {\bibfnamefont
  {P.}~\bibnamefont {Went}}, \bibinfo {author} {\bibfnamefont {J.}~\bibnamefont
  {Liu}}, \bibinfo {author} {\bibfnamefont {P.~J.}\ \bibnamefont {Ryan}},
  \bibinfo {author} {\bibfnamefont {J.-W.}\ \bibnamefont {Kim}}, \ and\
  \bibinfo {author} {\bibfnamefont {J.-H.}\ \bibnamefont {Chu}},\ }\href
  {\doibase 10.1038/s41567-020-0983-9} {\bibfield  {journal} {\bibinfo
  {journal} {Nat. Phys.}\ }\textbf {\bibinfo {volume} {16}},\ \bibinfo {pages}
  {1189} (\bibinfo {year} {2020})}\BibitemShut {NoStop}%
\bibitem [{\citenamefont {Bartlett}\ \emph {et~al.}(2021)\citenamefont
  {Bartlett}, \citenamefont {Steppke}, \citenamefont {Hosoi}, \citenamefont
  {Noad}, \citenamefont {Park}, \citenamefont {Timm}, \citenamefont
  {Shibauchi}, \citenamefont {Mackenzie},\ and\ \citenamefont
  {Hicks}}]{Bartlett2021}%
  \BibitemOpen
  \bibfield  {author} {\bibinfo {author} {\bibfnamefont {J.~M.}\ \bibnamefont
  {Bartlett}}, \bibinfo {author} {\bibfnamefont {A.}~\bibnamefont {Steppke}},
  \bibinfo {author} {\bibfnamefont {S.}~\bibnamefont {Hosoi}}, \bibinfo
  {author} {\bibfnamefont {H.}~\bibnamefont {Noad}}, \bibinfo {author}
  {\bibfnamefont {J.}~\bibnamefont {Park}}, \bibinfo {author} {\bibfnamefont
  {C.}~\bibnamefont {Timm}}, \bibinfo {author} {\bibfnamefont {T.}~\bibnamefont
  {Shibauchi}}, \bibinfo {author} {\bibfnamefont {A.~P.}\ \bibnamefont
  {Mackenzie}}, \ and\ \bibinfo {author} {\bibfnamefont {C.~W.}\ \bibnamefont
  {Hicks}},\ }\href {\doibase 10.1103/PhysRevX.11.021038} {\bibfield  {journal}
  {\bibinfo  {journal} {Phys. Rev. X}\ }\textbf {\bibinfo {volume} {11}},\
  \bibinfo {pages} {021038} (\bibinfo {year} {2021})}\BibitemShut {NoStop}%
\bibitem [{\citenamefont {Worasaran}\ \emph {et~al.}(2021)\citenamefont
  {Worasaran}, \citenamefont {Ikeda}, \citenamefont {Palmstrom}, \citenamefont
  {Straquadine}, \citenamefont {Kivelson},\ and\ \citenamefont
  {Fisher}}]{Worasaran2021}%
  \BibitemOpen
  \bibfield  {author} {\bibinfo {author} {\bibfnamefont {T.}~\bibnamefont
  {Worasaran}}, \bibinfo {author} {\bibfnamefont {M.~S.}\ \bibnamefont
  {Ikeda}}, \bibinfo {author} {\bibfnamefont {J.~C.}\ \bibnamefont
  {Palmstrom}}, \bibinfo {author} {\bibfnamefont {J.~A.~W.}\ \bibnamefont
  {Straquadine}}, \bibinfo {author} {\bibfnamefont {S.~A.}\ \bibnamefont
  {Kivelson}}, \ and\ \bibinfo {author} {\bibfnamefont {I.~R.}\ \bibnamefont
  {Fisher}},\ }\href {\doibase 10.1126/science.abb9280} {\bibfield  {journal}
  {\bibinfo  {journal} {Science}\ }\textbf {\bibinfo {volume} {372}},\ \bibinfo
  {pages} {973} (\bibinfo {year} {2021})}\BibitemShut {NoStop}%
\bibitem [{\citenamefont {Sanchez}\ \emph {et~al.}(2021)\citenamefont
  {Sanchez}, \citenamefont {Malinowski}, \citenamefont {Mutch}, \citenamefont
  {Liu}, \citenamefont {Kim}, \citenamefont {Ryan},\ and\ \citenamefont
  {Chu}}]{Sanchez2021}%
  \BibitemOpen
  \bibfield  {author} {\bibinfo {author} {\bibfnamefont {J.~J.}\ \bibnamefont
  {Sanchez}}, \bibinfo {author} {\bibfnamefont {P.}~\bibnamefont {Malinowski}},
  \bibinfo {author} {\bibfnamefont {J.}~\bibnamefont {Mutch}}, \bibinfo
  {author} {\bibfnamefont {J.}~\bibnamefont {Liu}}, \bibinfo {author}
  {\bibfnamefont {J.-W.}\ \bibnamefont {Kim}}, \bibinfo {author} {\bibfnamefont
  {P.~J.}\ \bibnamefont {Ryan}}, \ and\ \bibinfo {author} {\bibfnamefont
  {J.-H.}\ \bibnamefont {Chu}},\ }\href {\doibase 10.1038/s41563-021-01082-4}
  {\bibfield  {journal} {\bibinfo  {journal} {Nat. Mater.}\ }\textbf {\bibinfo
  {volume} {20}},\ \bibinfo {pages} {1519} (\bibinfo {year}
  {2021})}\BibitemShut {NoStop}%
\bibitem [{\citenamefont {Nicholson}\ \emph {et~al.}(2021)\citenamefont
  {Nicholson}, \citenamefont {Rumo}, \citenamefont {Pulkkinen}, \citenamefont
  {Kremer}, \citenamefont {Salzmann}, \citenamefont {Mottas}, \citenamefont
  {Hildebrand}, \citenamefont {Jaouen}, \citenamefont {Kim}, \citenamefont
  {Mukherjee}, \citenamefont {Ma}, \citenamefont {Muntwiler}, \citenamefont
  {von Rohr}, \citenamefont {Cacho},\ and\ \citenamefont
  {Monney}}]{Nicholson2021}%
  \BibitemOpen
  \bibfield  {author} {\bibinfo {author} {\bibfnamefont {C.~W.}\ \bibnamefont
  {Nicholson}}, \bibinfo {author} {\bibfnamefont {M.}~\bibnamefont {Rumo}},
  \bibinfo {author} {\bibfnamefont {A.}~\bibnamefont {Pulkkinen}}, \bibinfo
  {author} {\bibfnamefont {G.}~\bibnamefont {Kremer}}, \bibinfo {author}
  {\bibfnamefont {B.}~\bibnamefont {Salzmann}}, \bibinfo {author}
  {\bibfnamefont {M.}~\bibnamefont {Mottas}}, \bibinfo {author} {\bibfnamefont
  {B.}~\bibnamefont {Hildebrand}}, \bibinfo {author} {\bibfnamefont
  {T.}~\bibnamefont {Jaouen}}, \bibinfo {author} {\bibfnamefont {T.~K.}\
  \bibnamefont {Kim}}, \bibinfo {author} {\bibfnamefont {S.}~\bibnamefont
  {Mukherjee}}, \bibinfo {author} {\bibfnamefont {K.-Y.}\ \bibnamefont {Ma}},
  \bibinfo {author} {\bibfnamefont {M.}~\bibnamefont {Muntwiler}}, \bibinfo
  {author} {\bibfnamefont {F.~O.}\ \bibnamefont {von Rohr}}, \bibinfo {author}
  {\bibfnamefont {C.}~\bibnamefont {Cacho}}, \ and\ \bibinfo {author}
  {\bibfnamefont {C.}~\bibnamefont {Monney}},\ }\href {\doibase
  10.1038/s43246-021-00130-5} {\bibfield  {journal} {\bibinfo  {journal}
  {Commun. Mater.}\ }\textbf {\bibinfo {volume} {2}},\ \bibinfo {pages} {25}
  (\bibinfo {year} {2021})}\BibitemShut {NoStop}%
\bibitem [{\citenamefont {Lin}\ \emph {et~al.}(2021)\citenamefont {Lin},
  \citenamefont {Ochi}, \citenamefont {Noguchi}, \citenamefont {Kuroda},
  \citenamefont {Sakoda}, \citenamefont {Nomura}, \citenamefont {Tsubota},
  \citenamefont {Zhang}, \citenamefont {Bareille}, \citenamefont {Kurokawa},
  \citenamefont {Arai}, \citenamefont {Kawaguchi}, \citenamefont {Tanaka},
  \citenamefont {Yaji}, \citenamefont {Harasawa}, \citenamefont {Hashimoto},
  \citenamefont {Lu}, \citenamefont {Shin}, \citenamefont {Arita},
  \citenamefont {Tanda},\ and\ \citenamefont {Kondo}}]{Lin2021}%
  \BibitemOpen
  \bibfield  {author} {\bibinfo {author} {\bibfnamefont {C.}~\bibnamefont
  {Lin}}, \bibinfo {author} {\bibfnamefont {M.}~\bibnamefont {Ochi}}, \bibinfo
  {author} {\bibfnamefont {R.}~\bibnamefont {Noguchi}}, \bibinfo {author}
  {\bibfnamefont {K.}~\bibnamefont {Kuroda}}, \bibinfo {author} {\bibfnamefont
  {M.}~\bibnamefont {Sakoda}}, \bibinfo {author} {\bibfnamefont
  {A.}~\bibnamefont {Nomura}}, \bibinfo {author} {\bibfnamefont
  {M.}~\bibnamefont {Tsubota}}, \bibinfo {author} {\bibfnamefont
  {P.}~\bibnamefont {Zhang}}, \bibinfo {author} {\bibfnamefont
  {C.}~\bibnamefont {Bareille}}, \bibinfo {author} {\bibfnamefont
  {K.}~\bibnamefont {Kurokawa}}, \bibinfo {author} {\bibfnamefont
  {Y.}~\bibnamefont {Arai}}, \bibinfo {author} {\bibfnamefont {K.}~\bibnamefont
  {Kawaguchi}}, \bibinfo {author} {\bibfnamefont {H.}~\bibnamefont {Tanaka}},
  \bibinfo {author} {\bibfnamefont {K.}~\bibnamefont {Yaji}}, \bibinfo {author}
  {\bibfnamefont {A.}~\bibnamefont {Harasawa}}, \bibinfo {author}
  {\bibfnamefont {M.}~\bibnamefont {Hashimoto}}, \bibinfo {author}
  {\bibfnamefont {D.}~\bibnamefont {Lu}}, \bibinfo {author} {\bibfnamefont
  {S.}~\bibnamefont {Shin}}, \bibinfo {author} {\bibfnamefont {R.}~\bibnamefont
  {Arita}}, \bibinfo {author} {\bibfnamefont {S.}~\bibnamefont {Tanda}}, \ and\
  \bibinfo {author} {\bibfnamefont {T.}~\bibnamefont {Kondo}},\ }\href
  {\doibase 10.1038/s41563-021-01004-4} {\bibfield  {journal} {\bibinfo
  {journal} {Nat. Mater.}\ }\textbf {\bibinfo {volume} {20}},\ \bibinfo {pages}
  {1093} (\bibinfo {year} {2021})}\BibitemShut {NoStop}%
\bibitem [{\citenamefont {Bohnenbuck}\ \emph {et~al.}(2008)\citenamefont
  {Bohnenbuck}, \citenamefont {Zegkinoglou}, \citenamefont {Strempfer},
  \citenamefont {Sch\"{u}\ss{}ler-Langeheine}, \citenamefont {Nelson},
  \citenamefont {Leininger}, \citenamefont {Wu}, \citenamefont {Schierle},
  \citenamefont {Lang}, \citenamefont {Srajer}, \citenamefont {Ikeda},
  \citenamefont {Yoshida}, \citenamefont {Iwata}, \citenamefont {Katano},
  \citenamefont {Kikugawa},\ and\ \citenamefont {Keimer}}]{Bohnenbuck2008}%
  \BibitemOpen
  \bibfield  {author} {\bibinfo {author} {\bibfnamefont {B.}~\bibnamefont
  {Bohnenbuck}}, \bibinfo {author} {\bibfnamefont {I.}~\bibnamefont
  {Zegkinoglou}}, \bibinfo {author} {\bibfnamefont {J.}~\bibnamefont
  {Strempfer}}, \bibinfo {author} {\bibfnamefont {C.}~\bibnamefont
  {Sch\"{u}\ss{}ler-Langeheine}}, \bibinfo {author} {\bibfnamefont {C.~S.}\
  \bibnamefont {Nelson}}, \bibinfo {author} {\bibfnamefont {P.}~\bibnamefont
  {Leininger}}, \bibinfo {author} {\bibfnamefont {H.-H.}\ \bibnamefont {Wu}},
  \bibinfo {author} {\bibfnamefont {E.}~\bibnamefont {Schierle}}, \bibinfo
  {author} {\bibfnamefont {J.~C.}\ \bibnamefont {Lang}}, \bibinfo {author}
  {\bibfnamefont {G.}~\bibnamefont {Srajer}}, \bibinfo {author} {\bibfnamefont
  {S.~I.}\ \bibnamefont {Ikeda}}, \bibinfo {author} {\bibfnamefont
  {Y.}~\bibnamefont {Yoshida}}, \bibinfo {author} {\bibfnamefont
  {K.}~\bibnamefont {Iwata}}, \bibinfo {author} {\bibfnamefont
  {S.}~\bibnamefont {Katano}}, \bibinfo {author} {\bibfnamefont
  {N.}~\bibnamefont {Kikugawa}}, \ and\ \bibinfo {author} {\bibfnamefont
  {B.}~\bibnamefont {Keimer}},\ }\href {\doibase 10.1103/PhysRevB.77.224412}
  {\bibfield  {journal} {\bibinfo  {journal} {Phys. Rev. B}\ }\textbf {\bibinfo
  {volume} {77}},\ \bibinfo {pages} {224412} (\bibinfo {year}
  {2008})}\BibitemShut {NoStop}%
\bibitem [{\citenamefont {Maeno}\ \emph {et~al.}(1994)\citenamefont {Maeno},
  \citenamefont {Hashimoto}, \citenamefont {Yoshida}, \citenamefont
  {Nishizaki}, \citenamefont {Fujita}, \citenamefont {Bednorz},\ and\
  \citenamefont {Lichtenberg}}]{Maeno1994}%
  \BibitemOpen
  \bibfield  {author} {\bibinfo {author} {\bibfnamefont {Y.}~\bibnamefont
  {Maeno}}, \bibinfo {author} {\bibfnamefont {H.}~\bibnamefont {Hashimoto}},
  \bibinfo {author} {\bibfnamefont {K.}~\bibnamefont {Yoshida}}, \bibinfo
  {author} {\bibfnamefont {S.}~\bibnamefont {Nishizaki}}, \bibinfo {author}
  {\bibfnamefont {T.}~\bibnamefont {Fujita}}, \bibinfo {author} {\bibfnamefont
  {J.~G.}\ \bibnamefont {Bednorz}}, \ and\ \bibinfo {author} {\bibfnamefont
  {F.}~\bibnamefont {Lichtenberg}},\ }\href {\doibase 10.1038/372532a0}
  {\bibfield  {journal} {\bibinfo  {journal} {Nature}\ }\textbf {\bibinfo
  {volume} {372}},\ \bibinfo {pages} {532} (\bibinfo {year}
  {1994})}\BibitemShut {NoStop}%
\bibitem [{\citenamefont {Nakatsuji}\ and\ \citenamefont
  {Maeno}(2000)}]{Nakatsuji2000}%
  \BibitemOpen
  \bibfield  {author} {\bibinfo {author} {\bibfnamefont {S.}~\bibnamefont
  {Nakatsuji}}\ and\ \bibinfo {author} {\bibfnamefont {Y.}~\bibnamefont
  {Maeno}},\ }\href {\doibase 10.1103/PhysRevLett.84.2666} {\bibfield
  {journal} {\bibinfo  {journal} {Phys. Rev. Lett.}\ }\textbf {\bibinfo
  {volume} {84}},\ \bibinfo {pages} {2666} (\bibinfo {year}
  {2000})}\BibitemShut {NoStop}%
\bibitem [{\citenamefont {Ricc\`{o}}\ \emph {et~al.}(2018)\citenamefont
  {Ricc\`{o}}, \citenamefont {Kim}, \citenamefont {Tamai}, \citenamefont
  {McKeown~Walker}, \citenamefont {Bruno}, \citenamefont {Cucchi},
  \citenamefont {Cappelli}, \citenamefont {Besnard}, \citenamefont {Kim},
  \citenamefont {Dudin}, \citenamefont {Hoesch}, \citenamefont {Gutmann},
  \citenamefont {Georges}, \citenamefont {Perry},\ and\ \citenamefont
  {Baumberger}}]{Ricco2018}%
  \BibitemOpen
  \bibfield  {author} {\bibinfo {author} {\bibfnamefont {S.}~\bibnamefont
  {Ricc\`{o}}}, \bibinfo {author} {\bibfnamefont {M.}~\bibnamefont {Kim}},
  \bibinfo {author} {\bibfnamefont {A.}~\bibnamefont {Tamai}}, \bibinfo
  {author} {\bibfnamefont {S.}~\bibnamefont {McKeown~Walker}}, \bibinfo
  {author} {\bibfnamefont {F.~Y.}\ \bibnamefont {Bruno}}, \bibinfo {author}
  {\bibfnamefont {I.}~\bibnamefont {Cucchi}}, \bibinfo {author} {\bibfnamefont
  {E.}~\bibnamefont {Cappelli}}, \bibinfo {author} {\bibfnamefont
  {C.}~\bibnamefont {Besnard}}, \bibinfo {author} {\bibfnamefont {T.~K.}\
  \bibnamefont {Kim}}, \bibinfo {author} {\bibfnamefont {P.}~\bibnamefont
  {Dudin}}, \bibinfo {author} {\bibfnamefont {M.}~\bibnamefont {Hoesch}},
  \bibinfo {author} {\bibfnamefont {M.~J.}\ \bibnamefont {Gutmann}}, \bibinfo
  {author} {\bibfnamefont {A.}~\bibnamefont {Georges}}, \bibinfo {author}
  {\bibfnamefont {R.~S.}\ \bibnamefont {Perry}}, \ and\ \bibinfo {author}
  {\bibfnamefont {F.}~\bibnamefont {Baumberger}},\ }\href {\doibase
  10.1038/s41467-018-06945-0} {\bibfield  {journal} {\bibinfo  {journal} {Nat.
  Commun.}\ }\textbf {\bibinfo {volume} {9}},\ \bibinfo {pages} {4535}
  (\bibinfo {year} {2018})}\BibitemShut {NoStop}%
\bibitem [{\citenamefont {Yamase}\ and\ \citenamefont
  {Katanin}(2007)}]{Yamase2007}%
  \BibitemOpen
  \bibfield  {author} {\bibinfo {author} {\bibfnamefont {H.}~\bibnamefont
  {Yamase}}\ and\ \bibinfo {author} {\bibfnamefont {A.}~\bibnamefont
  {Katanin}},\ }\href {\doibase 10.1143/JPSJ.76.073706} {\bibfield  {journal}
  {\bibinfo  {journal} {J. Phys. Soc. Jpn.}\ }\textbf {\bibinfo {volume}
  {76}},\ \bibinfo {pages} {073706} (\bibinfo {year} {2007})}\BibitemShut
  {NoStop}%
\bibitem [{\citenamefont {Berridge}\ \emph {et~al.}(2009)\citenamefont
  {Berridge}, \citenamefont {Green}, \citenamefont {Grigera},\ and\
  \citenamefont {Simons}}]{Berridge2009}%
  \BibitemOpen
  \bibfield  {author} {\bibinfo {author} {\bibfnamefont {A.~M.}\ \bibnamefont
  {Berridge}}, \bibinfo {author} {\bibfnamefont {A.~G.}\ \bibnamefont {Green}},
  \bibinfo {author} {\bibfnamefont {S.~A.}\ \bibnamefont {Grigera}}, \ and\
  \bibinfo {author} {\bibfnamefont {B.~D.}\ \bibnamefont {Simons}},\ }\href
  {\doibase 10.1103/PhysRevLett.102.136404} {\bibfield  {journal} {\bibinfo
  {journal} {Phys. Rev. Lett.}\ }\textbf {\bibinfo {volume} {102}},\ \bibinfo
  {pages} {136404} (\bibinfo {year} {2009})}\BibitemShut {NoStop}%
\bibitem [{\citenamefont {Lester}\ \emph {et~al.}(2015)\citenamefont {Lester},
  \citenamefont {Ramos}, \citenamefont {Perry}, \citenamefont {Croft},
  \citenamefont {Bewley}, \citenamefont {Guidi}, \citenamefont {Manuel},
  \citenamefont {Khalyavin}, \citenamefont {Forgan},\ and\ \citenamefont
  {Hayden}}]{Lester2015}%
  \BibitemOpen
  \bibfield  {author} {\bibinfo {author} {\bibfnamefont {C.}~\bibnamefont
  {Lester}}, \bibinfo {author} {\bibfnamefont {S.}~\bibnamefont {Ramos}},
  \bibinfo {author} {\bibfnamefont {R.~S.}\ \bibnamefont {Perry}}, \bibinfo
  {author} {\bibfnamefont {T.~P.}\ \bibnamefont {Croft}}, \bibinfo {author}
  {\bibfnamefont {R.~I.}\ \bibnamefont {Bewley}}, \bibinfo {author}
  {\bibfnamefont {T.}~\bibnamefont {Guidi}}, \bibinfo {author} {\bibfnamefont
  {P.}~\bibnamefont {Manuel}}, \bibinfo {author} {\bibfnamefont {D.~D.}\
  \bibnamefont {Khalyavin}}, \bibinfo {author} {\bibfnamefont {E.}~\bibnamefont
  {Forgan}}, \ and\ \bibinfo {author} {\bibfnamefont {S.~M.}\ \bibnamefont
  {Hayden}},\ }\href {\doibase 10.1038/nmat4181} {\bibfield  {journal}
  {\bibinfo  {journal} {Nat. Mater.}\ }\textbf {\bibinfo {volume} {14}},\
  \bibinfo {pages} {373} (\bibinfo {year} {2015})}\BibitemShut {NoStop}%
\bibitem [{\citenamefont {Lindquist}\ \emph {et~al.}(2022)\citenamefont
  {Lindquist}, \citenamefont {Clepkens},\ and\ \citenamefont
  {Kee}}]{Lindquist2022}%
  \BibitemOpen
  \bibfield  {author} {\bibinfo {author} {\bibfnamefont {A.~W.}\ \bibnamefont
  {Lindquist}}, \bibinfo {author} {\bibfnamefont {J.}~\bibnamefont {Clepkens}},
  \ and\ \bibinfo {author} {\bibfnamefont {H.-Y.}\ \bibnamefont {Kee}},\ }\href
  {\doibase 10.1103/PhysRevResearch.4.023109} {\bibfield  {journal} {\bibinfo
  {journal} {Phys. Rev. Research}\ }\textbf {\bibinfo {volume} {4}},\ \bibinfo
  {pages} {023109} (\bibinfo {year} {2022})}\BibitemShut {NoStop}%
\bibitem [{\citenamefont {Lei}\ \emph {et~al.}(2018)\citenamefont {Lei},
  \citenamefont {Gu}, \citenamefont {Puggioni}, \citenamefont {Stone},
  \citenamefont {Peng}, \citenamefont {Ge}, \citenamefont {Wang}, \citenamefont
  {Wang}, \citenamefont {Yuan}, \citenamefont {Wang}, \citenamefont {Mao},
  \citenamefont {Rondinelli},\ and\ \citenamefont {Gopalan}}]{Lei2018}%
  \BibitemOpen
  \bibfield  {author} {\bibinfo {author} {\bibfnamefont {S.}~\bibnamefont
  {Lei}}, \bibinfo {author} {\bibfnamefont {M.}~\bibnamefont {Gu}}, \bibinfo
  {author} {\bibfnamefont {D.}~\bibnamefont {Puggioni}}, \bibinfo {author}
  {\bibfnamefont {G.}~\bibnamefont {Stone}}, \bibinfo {author} {\bibfnamefont
  {J.}~\bibnamefont {Peng}}, \bibinfo {author} {\bibfnamefont {J.}~\bibnamefont
  {Ge}}, \bibinfo {author} {\bibfnamefont {Y.}~\bibnamefont {Wang}}, \bibinfo
  {author} {\bibfnamefont {B.}~\bibnamefont {Wang}}, \bibinfo {author}
  {\bibfnamefont {Y.}~\bibnamefont {Yuan}}, \bibinfo {author} {\bibfnamefont
  {K.}~\bibnamefont {Wang}}, \bibinfo {author} {\bibfnamefont {Z.}~\bibnamefont
  {Mao}}, \bibinfo {author} {\bibfnamefont {J.~M.}\ \bibnamefont {Rondinelli}},
  \ and\ \bibinfo {author} {\bibfnamefont {V.}~\bibnamefont {Gopalan}},\ }\href
  {\doibase 10.1021/acs.nanolett.8b00633} {\bibfield  {journal} {\bibinfo
  {journal} {Nano Lett.}\ }\textbf {\bibinfo {volume} {18}},\ \bibinfo {pages}
  {3088} (\bibinfo {year} {2018})}\BibitemShut {NoStop}%
\bibitem [{\citenamefont {Dashwood}\ \emph {et~al.}(2020)\citenamefont
  {Dashwood}, \citenamefont {Veiga}, \citenamefont {Faure}, \citenamefont
  {Vale}, \citenamefont {Porter}, \citenamefont {Collins}, \citenamefont
  {Manuel}, \citenamefont {Khalyavin}, \citenamefont {Orlandi}, \citenamefont
  {Perry}, \citenamefont {Johnson},\ and\ \citenamefont
  {McMorrow}}]{Dashwood2020}%
  \BibitemOpen
  \bibfield  {author} {\bibinfo {author} {\bibfnamefont {C.~D.}\ \bibnamefont
  {Dashwood}}, \bibinfo {author} {\bibfnamefont {L.~S.~I.}\ \bibnamefont
  {Veiga}}, \bibinfo {author} {\bibfnamefont {Q.}~\bibnamefont {Faure}},
  \bibinfo {author} {\bibfnamefont {J.~G.}\ \bibnamefont {Vale}}, \bibinfo
  {author} {\bibfnamefont {D.~G.}\ \bibnamefont {Porter}}, \bibinfo {author}
  {\bibfnamefont {S.~P.}\ \bibnamefont {Collins}}, \bibinfo {author}
  {\bibfnamefont {P.}~\bibnamefont {Manuel}}, \bibinfo {author} {\bibfnamefont
  {D.~D.}\ \bibnamefont {Khalyavin}}, \bibinfo {author} {\bibfnamefont
  {F.}~\bibnamefont {Orlandi}}, \bibinfo {author} {\bibfnamefont {R.~S.}\
  \bibnamefont {Perry}}, \bibinfo {author} {\bibfnamefont {R.~D.}\ \bibnamefont
  {Johnson}}, \ and\ \bibinfo {author} {\bibfnamefont {D.~F.}\ \bibnamefont
  {McMorrow}},\ }\href {\doibase 10.1103/PhysRevB.102.180410} {\bibfield
  {journal} {\bibinfo  {journal} {Phys. Rev. B}\ }\textbf {\bibinfo {volume}
  {102}},\ \bibinfo {pages} {180410(R)} (\bibinfo {year} {2020})}\BibitemShut
  {NoStop}%
\bibitem [{\citenamefont {Faure}\ \emph {et~al.}(2023)\citenamefont {Faure},
  \citenamefont {Dashwood}, \citenamefont {Colin}, \citenamefont {Johnson},
  \citenamefont {Ressouche}, \citenamefont {Stenning}, \citenamefont {Spratt},
  \citenamefont {McMorrow},\ and\ \citenamefont {Perry}}]{Faure2023}%
  \BibitemOpen
  \bibfield  {author} {\bibinfo {author} {\bibfnamefont {Q.}~\bibnamefont
  {Faure}}, \bibinfo {author} {\bibfnamefont {C.~D.}\ \bibnamefont {Dashwood}},
  \bibinfo {author} {\bibfnamefont {C.~V.}\ \bibnamefont {Colin}}, \bibinfo
  {author} {\bibfnamefont {R.~D.}\ \bibnamefont {Johnson}}, \bibinfo {author}
  {\bibfnamefont {E.}~\bibnamefont {Ressouche}}, \bibinfo {author}
  {\bibfnamefont {G.~B.~G.}\ \bibnamefont {Stenning}}, \bibinfo {author}
  {\bibfnamefont {J.}~\bibnamefont {Spratt}}, \bibinfo {author} {\bibfnamefont
  {D.~F.}\ \bibnamefont {McMorrow}}, \ and\ \bibinfo {author} {\bibfnamefont
  {R.~S.}\ \bibnamefont {Perry}},\ }\href {\doibase
  10.1103/PhysRevResearch.5.013040} {\bibfield  {journal} {\bibinfo  {journal}
  {Phys. Rev. Res.}\ }\textbf {\bibinfo {volume} {5}},\ \bibinfo {pages}
  {013040} (\bibinfo {year} {2023})}\BibitemShut {NoStop}%
\bibitem [{\citenamefont {Yoshida}\ \emph {et~al.}(2005)\citenamefont
  {Yoshida}, \citenamefont {Ikeda}, \citenamefont {Matsuhata}, \citenamefont
  {Shirakawa}, \citenamefont {Lee},\ and\ \citenamefont
  {Katano}}]{Yoshida2005}%
  \BibitemOpen
  \bibfield  {author} {\bibinfo {author} {\bibfnamefont {Y.}~\bibnamefont
  {Yoshida}}, \bibinfo {author} {\bibfnamefont {S.-I.}\ \bibnamefont {Ikeda}},
  \bibinfo {author} {\bibfnamefont {H.}~\bibnamefont {Matsuhata}}, \bibinfo
  {author} {\bibfnamefont {N.}~\bibnamefont {Shirakawa}}, \bibinfo {author}
  {\bibfnamefont {C.~H.}\ \bibnamefont {Lee}}, \ and\ \bibinfo {author}
  {\bibfnamefont {S.}~\bibnamefont {Katano}},\ }\href {\doibase
  10.1103/PhysRevB.72.054412} {\bibfield  {journal} {\bibinfo  {journal} {Phys.
  Rev. B}\ }\textbf {\bibinfo {volume} {72}},\ \bibinfo {pages} {054412}
  (\bibinfo {year} {2005})}\BibitemShut {NoStop}%
\bibitem [{\citenamefont {Baumberger}\ \emph {et~al.}(2006)\citenamefont
  {Baumberger}, \citenamefont {Ingle}, \citenamefont {Kikugawa}, \citenamefont
  {Hossain}, \citenamefont {Meevasana}, \citenamefont {Perry}, \citenamefont
  {Shen}, \citenamefont {Lu}, \citenamefont {Damascelli}, \citenamefont {Rost},
  \citenamefont {Mackenzie}, \citenamefont {Hussain},\ and\ \citenamefont
  {Shen}}]{Baumberger2006}%
  \BibitemOpen
  \bibfield  {author} {\bibinfo {author} {\bibfnamefont {F.}~\bibnamefont
  {Baumberger}}, \bibinfo {author} {\bibfnamefont {N.~J.~C.}\ \bibnamefont
  {Ingle}}, \bibinfo {author} {\bibfnamefont {N.}~\bibnamefont {Kikugawa}},
  \bibinfo {author} {\bibfnamefont {M.~A.}\ \bibnamefont {Hossain}}, \bibinfo
  {author} {\bibfnamefont {W.}~\bibnamefont {Meevasana}}, \bibinfo {author}
  {\bibfnamefont {R.~S.}\ \bibnamefont {Perry}}, \bibinfo {author}
  {\bibfnamefont {K.~M.}\ \bibnamefont {Shen}}, \bibinfo {author}
  {\bibfnamefont {D.~H.}\ \bibnamefont {Lu}}, \bibinfo {author} {\bibfnamefont
  {A.}~\bibnamefont {Damascelli}}, \bibinfo {author} {\bibfnamefont
  {A.}~\bibnamefont {Rost}}, \bibinfo {author} {\bibfnamefont {A.~P.}\
  \bibnamefont {Mackenzie}}, \bibinfo {author} {\bibfnamefont {Z.}~\bibnamefont
  {Hussain}}, \ and\ \bibinfo {author} {\bibfnamefont {Z.-X.}\ \bibnamefont
  {Shen}},\ }\href {\doibase 10.1103/PhysRevLett.96.107601} {\bibfield
  {journal} {\bibinfo  {journal} {Phys. Rev. Lett.}\ }\textbf {\bibinfo
  {volume} {96}},\ \bibinfo {pages} {107601} (\bibinfo {year}
  {2006})}\BibitemShut {NoStop}%
\bibitem [{\citenamefont {Xing}\ \emph {et~al.}(2018)\citenamefont {Xing},
  \citenamefont {Wen}, \citenamefont {Shen}, \citenamefont {He}, \citenamefont
  {Cai}, \citenamefont {Peng}, \citenamefont {Wang}, \citenamefont {Tian},
  \citenamefont {Xu}, \citenamefont {Ku}, \citenamefont {Mao},\ and\
  \citenamefont {Liu}}]{Xing2018}%
  \BibitemOpen
  \bibfield  {author} {\bibinfo {author} {\bibfnamefont {H.}~\bibnamefont
  {Xing}}, \bibinfo {author} {\bibfnamefont {L.}~\bibnamefont {Wen}}, \bibinfo
  {author} {\bibfnamefont {C.}~\bibnamefont {Shen}}, \bibinfo {author}
  {\bibfnamefont {J.}~\bibnamefont {He}}, \bibinfo {author} {\bibfnamefont
  {X.}~\bibnamefont {Cai}}, \bibinfo {author} {\bibfnamefont {J.}~\bibnamefont
  {Peng}}, \bibinfo {author} {\bibfnamefont {S.}~\bibnamefont {Wang}}, \bibinfo
  {author} {\bibfnamefont {M.}~\bibnamefont {Tian}}, \bibinfo {author}
  {\bibfnamefont {Z.-A.}\ \bibnamefont {Xu}}, \bibinfo {author} {\bibfnamefont
  {W.}~\bibnamefont {Ku}}, \bibinfo {author} {\bibfnamefont {Z.}~\bibnamefont
  {Mao}}, \ and\ \bibinfo {author} {\bibfnamefont {Y.}~\bibnamefont {Liu}},\
  }\href {\doibase 10.1103/PhysRevB.97.041113} {\bibfield  {journal} {\bibinfo
  {journal} {Phys. Rev. B}\ }\textbf {\bibinfo {volume} {97}},\ \bibinfo
  {pages} {041113} (\bibinfo {year} {2018})}\BibitemShut {NoStop}%
\bibitem [{\citenamefont {Horio}\ \emph {et~al.}(2021)\citenamefont {Horio},
  \citenamefont {Wang}, \citenamefont {Granata}, \citenamefont {Kramer},
  \citenamefont {Sassa}, \citenamefont {J{\"{o}}hr}, \citenamefont {Sutter},
  \citenamefont {Bold}, \citenamefont {Das}, \citenamefont {Xu}, \citenamefont
  {Frison}, \citenamefont {Fittipaldi}, \citenamefont {Kim}, \citenamefont
  {Cacho}, \citenamefont {Rault}, \citenamefont {{Le F{\`{e}}vre}},
  \citenamefont {Bertran}, \citenamefont {Plumb}, \citenamefont {Shi},
  \citenamefont {Vecchione}, \citenamefont {Fischer},\ and\ \citenamefont
  {Chang}}]{Horio2019}%
  \BibitemOpen
  \bibfield  {author} {\bibinfo {author} {\bibfnamefont {M.}~\bibnamefont
  {Horio}}, \bibinfo {author} {\bibfnamefont {Q.}~\bibnamefont {Wang}},
  \bibinfo {author} {\bibfnamefont {V.}~\bibnamefont {Granata}}, \bibinfo
  {author} {\bibfnamefont {K.~P.}\ \bibnamefont {Kramer}}, \bibinfo {author}
  {\bibfnamefont {Y.}~\bibnamefont {Sassa}}, \bibinfo {author} {\bibfnamefont
  {S.}~\bibnamefont {J{\"{o}}hr}}, \bibinfo {author} {\bibfnamefont
  {D.}~\bibnamefont {Sutter}}, \bibinfo {author} {\bibfnamefont
  {A.}~\bibnamefont {Bold}}, \bibinfo {author} {\bibfnamefont {L.}~\bibnamefont
  {Das}}, \bibinfo {author} {\bibfnamefont {Y.}~\bibnamefont {Xu}}, \bibinfo
  {author} {\bibfnamefont {R.}~\bibnamefont {Frison}}, \bibinfo {author}
  {\bibfnamefont {R.}~\bibnamefont {Fittipaldi}}, \bibinfo {author}
  {\bibfnamefont {T.~K.}\ \bibnamefont {Kim}}, \bibinfo {author} {\bibfnamefont
  {C.}~\bibnamefont {Cacho}}, \bibinfo {author} {\bibfnamefont {J.~E.}\
  \bibnamefont {Rault}}, \bibinfo {author} {\bibfnamefont {P.}~\bibnamefont
  {{Le F{\`{e}}vre}}}, \bibinfo {author} {\bibfnamefont {F.}~\bibnamefont
  {Bertran}}, \bibinfo {author} {\bibfnamefont {N.~C.}\ \bibnamefont {Plumb}},
  \bibinfo {author} {\bibfnamefont {M.}~\bibnamefont {Shi}}, \bibinfo {author}
  {\bibfnamefont {A.}~\bibnamefont {Vecchione}}, \bibinfo {author}
  {\bibfnamefont {M.~H.}\ \bibnamefont {Fischer}}, \ and\ \bibinfo {author}
  {\bibfnamefont {J.}~\bibnamefont {Chang}},\ }\href {\doibase
  10.1038/s41535-021-00328-3} {\bibfield  {journal} {\bibinfo  {journal} {npj
  Quantum Mater.}\ }\textbf {\bibinfo {volume} {6}},\ \bibinfo {pages} {29}
  (\bibinfo {year} {2021})}\BibitemShut {NoStop}%
\bibitem [{\citenamefont {Markovi\'{c}}\ \emph {et~al.}(2020)\citenamefont
  {Markovi\'{c}}, \citenamefont {Watson}, \citenamefont {Clark}, \citenamefont
  {Mazzola}, \citenamefont {Abarca~Morales}, \citenamefont {Hooley},
  \citenamefont {Rosner}, \citenamefont {Polley}, \citenamefont
  {Balasubramanian}, \citenamefont {Mukherjee}, \citenamefont {Kikugawa},
  \citenamefont {Sokolov}, \citenamefont {Mackenzie},\ and\ \citenamefont
  {King}}]{Markovic2020}%
  \BibitemOpen
  \bibfield  {author} {\bibinfo {author} {\bibfnamefont {I.}~\bibnamefont
  {Markovi\'{c}}}, \bibinfo {author} {\bibfnamefont {M.~D.}\ \bibnamefont
  {Watson}}, \bibinfo {author} {\bibfnamefont {O.~J.}\ \bibnamefont {Clark}},
  \bibinfo {author} {\bibfnamefont {F.}~\bibnamefont {Mazzola}}, \bibinfo
  {author} {\bibfnamefont {E.}~\bibnamefont {Abarca~Morales}}, \bibinfo
  {author} {\bibfnamefont {C.~A.}\ \bibnamefont {Hooley}}, \bibinfo {author}
  {\bibfnamefont {H.}~\bibnamefont {Rosner}}, \bibinfo {author} {\bibfnamefont
  {C.~M.}\ \bibnamefont {Polley}}, \bibinfo {author} {\bibfnamefont
  {T.}~\bibnamefont {Balasubramanian}}, \bibinfo {author} {\bibfnamefont
  {S.}~\bibnamefont {Mukherjee}}, \bibinfo {author} {\bibfnamefont
  {N.}~\bibnamefont {Kikugawa}}, \bibinfo {author} {\bibfnamefont {D.~A.}\
  \bibnamefont {Sokolov}}, \bibinfo {author} {\bibfnamefont {A.~P.}\
  \bibnamefont {Mackenzie}}, \ and\ \bibinfo {author} {\bibfnamefont
  {P.~D.~C.}\ \bibnamefont {King}},\ }\href {\doibase 10.1073/pnas.2003671117}
  {\bibfield  {journal} {\bibinfo  {journal} {Proc. Natl. Acad. Sci. U.S.A.}\
  }\textbf {\bibinfo {volume} {117}},\ \bibinfo {pages} {15524} (\bibinfo
  {year} {2020})}\BibitemShut {NoStop}%
\bibitem [{Raz()}]{Razorbill}%
  \BibitemOpen
  \href@noop {} {}\bibinfo {note} {Razorbill Instruments,
  \url{https://razorbillinstruments.com}}\BibitemShut {NoStop}%
\bibitem [{\citenamefont {Chapon}\ \emph {et~al.}(2011)\citenamefont {Chapon},
  \citenamefont {Manuel}, \citenamefont {Radaelli}, \citenamefont {Benson},
  \citenamefont {Perrott}, \citenamefont {Ansell}, \citenamefont {Rhodes},
  \citenamefont {Raspino}, \citenamefont {Duxbury}, \citenamefont {Spill},\
  and\ \citenamefont {Norris}}]{Chapon2011}%
  \BibitemOpen
  \bibfield  {author} {\bibinfo {author} {\bibfnamefont {L.~C.}\ \bibnamefont
  {Chapon}}, \bibinfo {author} {\bibfnamefont {P.}~\bibnamefont {Manuel}},
  \bibinfo {author} {\bibfnamefont {P.~G.}\ \bibnamefont {Radaelli}}, \bibinfo
  {author} {\bibfnamefont {C.}~\bibnamefont {Benson}}, \bibinfo {author}
  {\bibfnamefont {L.}~\bibnamefont {Perrott}}, \bibinfo {author} {\bibfnamefont
  {S.}~\bibnamefont {Ansell}}, \bibinfo {author} {\bibfnamefont {N.~J.}\
  \bibnamefont {Rhodes}}, \bibinfo {author} {\bibfnamefont {D.}~\bibnamefont
  {Raspino}}, \bibinfo {author} {\bibfnamefont {D.}~\bibnamefont {Duxbury}},
  \bibinfo {author} {\bibfnamefont {E.}~\bibnamefont {Spill}}, \ and\ \bibinfo
  {author} {\bibfnamefont {J.}~\bibnamefont {Norris}},\ }\href {\doibase
  10.1080/10448632.2011.569650} {\bibfield  {journal} {\bibinfo  {journal}
  {Neutron News}\ }\textbf {\bibinfo {volume} {22}},\ \bibinfo {pages} {22}
  (\bibinfo {year} {2011})}\BibitemShut {NoStop}%
\bibitem [{\citenamefont {Cao}\ \emph {et~al.}(2003)\citenamefont {Cao},
  \citenamefont {Balicas}, \citenamefont {Xin}, \citenamefont {Crow},\ and\
  \citenamefont {Nelson}}]{Cao2003}%
  \BibitemOpen
  \bibfield  {author} {\bibinfo {author} {\bibfnamefont {G.}~\bibnamefont
  {Cao}}, \bibinfo {author} {\bibfnamefont {L.}~\bibnamefont {Balicas}},
  \bibinfo {author} {\bibfnamefont {Y.}~\bibnamefont {Xin}}, \bibinfo {author}
  {\bibfnamefont {J.~E.}\ \bibnamefont {Crow}}, \ and\ \bibinfo {author}
  {\bibfnamefont {C.~S.}\ \bibnamefont {Nelson}},\ }\href {\doibase
  10.1103/PhysRevB.67.184405} {\bibfield  {journal} {\bibinfo  {journal} {Phys.
  Rev. B}\ }\textbf {\bibinfo {volume} {67}},\ \bibinfo {pages} {184405}
  (\bibinfo {year} {2003})}\BibitemShut {NoStop}%
\bibitem [{\citenamefont {Yoshida}\ \emph {et~al.}(2008)\citenamefont
  {Yoshida}, \citenamefont {Ikeda}, \citenamefont {Shirakawa}, \citenamefont
  {Hedo},\ and\ \citenamefont {Uwatoko}}]{Yoshida2008}%
  \BibitemOpen
  \bibfield  {author} {\bibinfo {author} {\bibfnamefont {Y.}~\bibnamefont
  {Yoshida}}, \bibinfo {author} {\bibfnamefont {S.-I.}\ \bibnamefont {Ikeda}},
  \bibinfo {author} {\bibfnamefont {N.}~\bibnamefont {Shirakawa}}, \bibinfo
  {author} {\bibfnamefont {M.}~\bibnamefont {Hedo}}, \ and\ \bibinfo {author}
  {\bibfnamefont {Y.}~\bibnamefont {Uwatoko}},\ }\href {\doibase
  10.1143/jpsj.77.093702} {\bibfield  {journal} {\bibinfo  {journal} {J. Phys.
  Soc. Jpn.}\ }\textbf {\bibinfo {volume} {77}},\ \bibinfo {pages} {093702}
  (\bibinfo {year} {2008})}\BibitemShut {NoStop}%
\bibitem [{\citenamefont {Mohapatra}\ and\ \citenamefont
  {Singh}(2020)}]{Mohapatra2020}%
  \BibitemOpen
  \bibfield  {author} {\bibinfo {author} {\bibfnamefont {S.}~\bibnamefont
  {Mohapatra}}\ and\ \bibinfo {author} {\bibfnamefont {A.}~\bibnamefont
  {Singh}},\ }\href {\doibase 10.1088/1361-648X/abacad} {\bibfield  {journal}
  {\bibinfo  {journal} {J. Phys.: Condens. Matter}\ }\textbf {\bibinfo {volume}
  {32}},\ \bibinfo {pages} {485805} (\bibinfo {year} {2020})}\BibitemShut
  {NoStop}%
\bibitem [{\citenamefont {Puggioni}\ \emph {et~al.}(2020)\citenamefont
  {Puggioni}, \citenamefont {Horio}, \citenamefont {Chang},\ and\ \citenamefont
  {Rondinelli}}]{Puggioni2020}%
  \BibitemOpen
  \bibfield  {author} {\bibinfo {author} {\bibfnamefont {D.}~\bibnamefont
  {Puggioni}}, \bibinfo {author} {\bibfnamefont {M.}~\bibnamefont {Horio}},
  \bibinfo {author} {\bibfnamefont {J.}~\bibnamefont {Chang}}, \ and\ \bibinfo
  {author} {\bibfnamefont {J.~M.}\ \bibnamefont {Rondinelli}},\ }\href
  {\doibase 10.1103/PhysRevResearch.2.023141} {\bibfield  {journal} {\bibinfo
  {journal} {Phys. Rev. Res.}\ }\textbf {\bibinfo {volume} {2}},\ \bibinfo
  {pages} {023141} (\bibinfo {year} {2020})}\BibitemShut {NoStop}%
\bibitem [{\citenamefont {Slater}\ and\ \citenamefont
  {Koster}(1954)}]{Slater1954}%
  \BibitemOpen
  \bibfield  {author} {\bibinfo {author} {\bibfnamefont {J.~C.}\ \bibnamefont
  {Slater}}\ and\ \bibinfo {author} {\bibfnamefont {G.~F.}\ \bibnamefont
  {Koster}},\ }\href {\doibase 10.1103/PhysRev.94.1498} {\bibfield  {journal}
  {\bibinfo  {journal} {Phys. Rev.}\ }\textbf {\bibinfo {volume} {94}},\
  \bibinfo {pages} {1498} (\bibinfo {year} {1954})}\BibitemShut {NoStop}%
\bibitem [{\citenamefont {Peng}\ \emph {et~al.}(2010)\citenamefont {Peng},
  \citenamefont {Qu}, \citenamefont {Qian}, \citenamefont {Fobes},
  \citenamefont {Liu}, \citenamefont {Wu}, \citenamefont {Pham}, \citenamefont
  {Spinu},\ and\ \citenamefont {Mao}}]{Peng2010}%
  \BibitemOpen
  \bibfield  {author} {\bibinfo {author} {\bibfnamefont {J.}~\bibnamefont
  {Peng}}, \bibinfo {author} {\bibfnamefont {Z.}~\bibnamefont {Qu}}, \bibinfo
  {author} {\bibfnamefont {B.}~\bibnamefont {Qian}}, \bibinfo {author}
  {\bibfnamefont {D.}~\bibnamefont {Fobes}}, \bibinfo {author} {\bibfnamefont
  {T.}~\bibnamefont {Liu}}, \bibinfo {author} {\bibfnamefont {X.}~\bibnamefont
  {Wu}}, \bibinfo {author} {\bibfnamefont {H.~M.}\ \bibnamefont {Pham}},
  \bibinfo {author} {\bibfnamefont {L.}~\bibnamefont {Spinu}}, \ and\ \bibinfo
  {author} {\bibfnamefont {Z.~Q.}\ \bibnamefont {Mao}},\ }\href {\doibase
  10.1103/PhysRevB.82.024417} {\bibfield  {journal} {\bibinfo  {journal} {Phys.
  Rev. B}\ }\textbf {\bibinfo {volume} {82}},\ \bibinfo {pages} {024417}
  (\bibinfo {year} {2010})}\BibitemShut {NoStop}%
\bibitem [{\citenamefont {Rivero}\ \emph {et~al.}(2017)\citenamefont {Rivero},
  \citenamefont {Jin}, \citenamefont {Chen}, \citenamefont {Meunier},
  \citenamefont {Plummer},\ and\ \citenamefont {Shelton}}]{Rivero2017}%
  \BibitemOpen
  \bibfield  {author} {\bibinfo {author} {\bibfnamefont {P.}~\bibnamefont
  {Rivero}}, \bibinfo {author} {\bibfnamefont {R.}~\bibnamefont {Jin}},
  \bibinfo {author} {\bibfnamefont {C.}~\bibnamefont {Chen}}, \bibinfo {author}
  {\bibfnamefont {V.}~\bibnamefont {Meunier}}, \bibinfo {author} {\bibfnamefont
  {E.~W.}\ \bibnamefont {Plummer}}, \ and\ \bibinfo {author} {\bibfnamefont
  {W.}~\bibnamefont {Shelton}},\ }\href {\doibase 10.1038/s41598-017-10780-6}
  {\bibfield  {journal} {\bibinfo  {journal} {Sci. Rep.}\ }\textbf {\bibinfo
  {volume} {7}},\ \bibinfo {pages} {10265} (\bibinfo {year}
  {2017})}\BibitemShut {NoStop}%
\bibitem [{\citenamefont {Rivero}\ \emph {et~al.}(2018)\citenamefont {Rivero},
  \citenamefont {Meunier},\ and\ \citenamefont {Shelton}}]{Rivero2018}%
  \BibitemOpen
  \bibfield  {author} {\bibinfo {author} {\bibfnamefont {P.}~\bibnamefont
  {Rivero}}, \bibinfo {author} {\bibfnamefont {V.}~\bibnamefont {Meunier}}, \
  and\ \bibinfo {author} {\bibfnamefont {W.}~\bibnamefont {Shelton}},\ }\href
  {\doibase 10.1103/PhysRevB.97.134116} {\bibfield  {journal} {\bibinfo
  {journal} {Phys. Rev. B}\ }\textbf {\bibinfo {volume} {97}},\ \bibinfo
  {pages} {134116} (\bibinfo {year} {2018})}\BibitemShut {NoStop}%
\bibitem [{\citenamefont {Arnold}\ \emph {et~al.}(2014)\citenamefont {Arnold},
  \citenamefont {Bilheux}, \citenamefont {Borreguero}, \citenamefont {Buts},
  \citenamefont {Campbell}, \citenamefont {Chapon}, \citenamefont {Doucet},
  \citenamefont {Draper}, \citenamefont {{Ferraz Leal}}, \citenamefont {Gigg},
  \citenamefont {Lynch}, \citenamefont {Markvardsen}, \citenamefont
  {Mikkelson}, \citenamefont {Mikkelson}, \citenamefont {Miller}, \citenamefont
  {Palmen}, \citenamefont {Parker}, \citenamefont {Passos}, \citenamefont
  {Perring}, \citenamefont {Peterson}, \citenamefont {Ren}, \citenamefont
  {Reuter}, \citenamefont {Savici}, \citenamefont {Taylor}, \citenamefont
  {Taylor}, \citenamefont {Tolchenov}, \citenamefont {Zhou},\ and\
  \citenamefont {Zikovsky}}]{Arnold2014}%
  \BibitemOpen
  \bibfield  {author} {\bibinfo {author} {\bibfnamefont {O.}~\bibnamefont
  {Arnold}}, \bibinfo {author} {\bibfnamefont {J.~C.}\ \bibnamefont {Bilheux}},
  \bibinfo {author} {\bibfnamefont {J.~M.}\ \bibnamefont {Borreguero}},
  \bibinfo {author} {\bibfnamefont {A.}~\bibnamefont {Buts}}, \bibinfo {author}
  {\bibfnamefont {S.~I.}\ \bibnamefont {Campbell}}, \bibinfo {author}
  {\bibfnamefont {L.}~\bibnamefont {Chapon}}, \bibinfo {author} {\bibfnamefont
  {M.}~\bibnamefont {Doucet}}, \bibinfo {author} {\bibfnamefont
  {N.}~\bibnamefont {Draper}}, \bibinfo {author} {\bibfnamefont
  {R.}~\bibnamefont {{Ferraz Leal}}}, \bibinfo {author} {\bibfnamefont {M.~A.}\
  \bibnamefont {Gigg}}, \bibinfo {author} {\bibfnamefont {V.~E.}\ \bibnamefont
  {Lynch}}, \bibinfo {author} {\bibfnamefont {A.}~\bibnamefont {Markvardsen}},
  \bibinfo {author} {\bibfnamefont {D.~J.}\ \bibnamefont {Mikkelson}}, \bibinfo
  {author} {\bibfnamefont {R.~L.}\ \bibnamefont {Mikkelson}}, \bibinfo {author}
  {\bibfnamefont {R.}~\bibnamefont {Miller}}, \bibinfo {author} {\bibfnamefont
  {K.}~\bibnamefont {Palmen}}, \bibinfo {author} {\bibfnamefont
  {P.}~\bibnamefont {Parker}}, \bibinfo {author} {\bibfnamefont
  {G.}~\bibnamefont {Passos}}, \bibinfo {author} {\bibfnamefont {T.~G.}\
  \bibnamefont {Perring}}, \bibinfo {author} {\bibfnamefont {P.~F.}\
  \bibnamefont {Peterson}}, \bibinfo {author} {\bibfnamefont {S.}~\bibnamefont
  {Ren}}, \bibinfo {author} {\bibfnamefont {M.~A.}\ \bibnamefont {Reuter}},
  \bibinfo {author} {\bibfnamefont {A.~T.}\ \bibnamefont {Savici}}, \bibinfo
  {author} {\bibfnamefont {J.~W.}\ \bibnamefont {Taylor}}, \bibinfo {author}
  {\bibfnamefont {R.~J.}\ \bibnamefont {Taylor}}, \bibinfo {author}
  {\bibfnamefont {R.}~\bibnamefont {Tolchenov}}, \bibinfo {author}
  {\bibfnamefont {W.}~\bibnamefont {Zhou}}, \ and\ \bibinfo {author}
  {\bibfnamefont {J.}~\bibnamefont {Zikovsky}},\ }\href {\doibase
  10.1016/j.nima.2014.07.029} {\bibfield  {journal} {\bibinfo  {journal}
  {Nuclear Instruments and Methods in Physics Research A}\ }\textbf {\bibinfo
  {volume} {764}},\ \bibinfo {pages} {156} (\bibinfo {year}
  {2014})}\BibitemShut {NoStop}%
\bibitem [{\citenamefont {Porter}(2020)}]{Porter2020}%
  \BibitemOpen
  \bibfield  {author} {\bibinfo {author} {\bibfnamefont {D.~G.}\ \bibnamefont
  {Porter}},\ }\href {\doibase 10.5281/zenodo.3859719} {\enquote {\bibinfo
  {title} {Py16},}\ } (\bibinfo {year} {2020})\BibitemShut {NoStop}%
\bibitem [{\citenamefont {Basham}\ \emph {et~al.}(2015)\citenamefont {Basham},
  \citenamefont {Filik}, \citenamefont {Wharmby}, \citenamefont {Chang},
  \citenamefont {{El Kassaby}}, \citenamefont {Gerring}, \citenamefont
  {Aishima}, \citenamefont {Levik}, \citenamefont {Pulford}, \citenamefont
  {Sikharulidze}, \citenamefont {Sneddon}, \citenamefont {Webber},
  \citenamefont {Dhesi}, \citenamefont {Maccherozzi}, \citenamefont {Svensson},
  \citenamefont {Brockhauser}, \citenamefont {N\'{a}ray},\ and\ \citenamefont
  {Ashton}}]{Basham2015}%
  \BibitemOpen
  \bibfield  {author} {\bibinfo {author} {\bibfnamefont {M.}~\bibnamefont
  {Basham}}, \bibinfo {author} {\bibfnamefont {J.}~\bibnamefont {Filik}},
  \bibinfo {author} {\bibfnamefont {M.~T.}\ \bibnamefont {Wharmby}}, \bibinfo
  {author} {\bibfnamefont {P.~C.~Y.}\ \bibnamefont {Chang}}, \bibinfo {author}
  {\bibfnamefont {B.}~\bibnamefont {{El Kassaby}}}, \bibinfo {author}
  {\bibfnamefont {M.}~\bibnamefont {Gerring}}, \bibinfo {author} {\bibfnamefont
  {J.}~\bibnamefont {Aishima}}, \bibinfo {author} {\bibfnamefont
  {K.}~\bibnamefont {Levik}}, \bibinfo {author} {\bibfnamefont {B.~C.~A.}\
  \bibnamefont {Pulford}}, \bibinfo {author} {\bibfnamefont {I.}~\bibnamefont
  {Sikharulidze}}, \bibinfo {author} {\bibfnamefont {D.}~\bibnamefont
  {Sneddon}}, \bibinfo {author} {\bibfnamefont {M.}~\bibnamefont {Webber}},
  \bibinfo {author} {\bibfnamefont {S.~S.}\ \bibnamefont {Dhesi}}, \bibinfo
  {author} {\bibfnamefont {F.}~\bibnamefont {Maccherozzi}}, \bibinfo {author}
  {\bibfnamefont {O.}~\bibnamefont {Svensson}}, \bibinfo {author}
  {\bibfnamefont {S.}~\bibnamefont {Brockhauser}}, \bibinfo {author}
  {\bibfnamefont {G.}~\bibnamefont {N\'{a}ray}}, \ and\ \bibinfo {author}
  {\bibfnamefont {A.~W.}\ \bibnamefont {Ashton}},\ }\href {\doibase
  10.1107/S1600577515002283} {\bibfield  {journal} {\bibinfo  {journal} {J.
  Synchrotron Rad.}\ }\textbf {\bibinfo {volume} {22}},\ \bibinfo {pages} {853}
  (\bibinfo {year} {2015})}\BibitemShut {NoStop}%
\bibitem [{\citenamefont {Orobengoa}\ \emph {et~al.}(2009)\citenamefont
  {Orobengoa}, \citenamefont {Capillas}, \citenamefont {Aroyo},\ and\
  \citenamefont {Perez-Mato}}]{Orobengoa2009}%
  \BibitemOpen
  \bibfield  {author} {\bibinfo {author} {\bibfnamefont {D.}~\bibnamefont
  {Orobengoa}}, \bibinfo {author} {\bibfnamefont {C.}~\bibnamefont {Capillas}},
  \bibinfo {author} {\bibfnamefont {M.~I.}\ \bibnamefont {Aroyo}}, \ and\
  \bibinfo {author} {\bibfnamefont {J.~M.}\ \bibnamefont {Perez-Mato}},\ }\href
  {\doibase 10.1107/S0021889809028064} {\bibfield  {journal} {\bibinfo
  {journal} {J. App. Cryst.}\ }\textbf {\bibinfo {volume} {42}},\ \bibinfo
  {pages} {820} (\bibinfo {year} {2009})}\BibitemShut {NoStop}%
\bibitem [{\citenamefont
  {Rodr\'{i}guez-Carvajal}(1993)}]{Rodriguez-Carvajal1993}%
  \BibitemOpen
  \bibfield  {author} {\bibinfo {author} {\bibfnamefont {J.}~\bibnamefont
  {Rodr\'{i}guez-Carvajal}},\ }\href {\doibase 10.1016/0921-4526(93)90108-I}
  {\bibfield  {journal} {\bibinfo  {journal} {Physica B}\ }\textbf {\bibinfo
  {volume} {192}},\ \bibinfo {pages} {55} (\bibinfo {year} {1993})}\BibitemShut
  {NoStop}%
\bibitem [{\citenamefont {Li}\ \emph {et~al.}(2022)\citenamefont {Li},
  \citenamefont {Garst}, \citenamefont {Schmalian}, \citenamefont {Ghosh},
  \citenamefont {Kikugawa}, \citenamefont {Sokolov}, \citenamefont {Hicks},
  \citenamefont {Jerzembeck}, \citenamefont {Ikeda}, \citenamefont {Hu},
  \citenamefont {Ramshaw}, \citenamefont {Rost}, \citenamefont {Nicklas},\ and\
  \citenamefont {Mackenzie}}]{Li2022}%
  \BibitemOpen
  \bibfield  {author} {\bibinfo {author} {\bibfnamefont {Y.-S.}\ \bibnamefont
  {Li}}, \bibinfo {author} {\bibfnamefont {M.}~\bibnamefont {Garst}}, \bibinfo
  {author} {\bibfnamefont {J.}~\bibnamefont {Schmalian}}, \bibinfo {author}
  {\bibfnamefont {S.}~\bibnamefont {Ghosh}}, \bibinfo {author} {\bibfnamefont
  {N.}~\bibnamefont {Kikugawa}}, \bibinfo {author} {\bibfnamefont {D.~A.}\
  \bibnamefont {Sokolov}}, \bibinfo {author} {\bibfnamefont {C.~W.}\
  \bibnamefont {Hicks}}, \bibinfo {author} {\bibfnamefont {F.}~\bibnamefont
  {Jerzembeck}}, \bibinfo {author} {\bibfnamefont {M.~S.}\ \bibnamefont
  {Ikeda}}, \bibinfo {author} {\bibfnamefont {Z.}~\bibnamefont {Hu}}, \bibinfo
  {author} {\bibfnamefont {B.~J.}\ \bibnamefont {Ramshaw}}, \bibinfo {author}
  {\bibfnamefont {A.~W.}\ \bibnamefont {Rost}}, \bibinfo {author}
  {\bibfnamefont {M.}~\bibnamefont {Nicklas}}, \ and\ \bibinfo {author}
  {\bibfnamefont {A.~P.}\ \bibnamefont {Mackenzie}},\ }\href {\doibase
  10.1038/s41586-022-04820-z} {\bibfield  {journal} {\bibinfo  {journal}
  {Nature}\ }\textbf {\bibinfo {volume} {607}},\ \bibinfo {pages} {276}
  (\bibinfo {year} {2022})}\BibitemShut {NoStop}%
\end{thebibliography}%


\begin{thebibliography}{8}%
\makeatletter
\providecommand \@ifxundefined [1]{%
 \@ifx{#1\undefined}
}%
\providecommand \@ifnum [1]{%
 \ifnum #1\expandafter \@firstoftwo
 \else \expandafter \@secondoftwo
 \fi
}%
\providecommand \@ifx [1]{%
 \ifx #1\expandafter \@firstoftwo
 \else \expandafter \@secondoftwo
 \fi
}%
\providecommand \natexlab [1]{#1}%
\providecommand \enquote  [1]{``#1''}%
\providecommand \bibnamefont  [1]{#1}%
\providecommand \bibfnamefont [1]{#1}%
\providecommand \citenamefont [1]{#1}%
\providecommand \href@noop [0]{\@secondoftwo}%
\providecommand \href [0]{\begingroup \@sanitize@url \@href}%
\providecommand \@href[1]{\@@startlink{#1}\@@href}%
\providecommand \@@href[1]{\endgroup#1\@@endlink}%
\providecommand \@sanitize@url [0]{\catcode `\\12\catcode `\$12\catcode
  `\&12\catcode `\#12\catcode `\^12\catcode `\_12\catcode `\%12\relax}%
\providecommand \@@startlink[1]{}%
\providecommand \@@endlink[0]{}%
\providecommand \url  [0]{\begingroup\@sanitize@url \@url }%
\providecommand \@url [1]{\endgroup\@href {#1}{\urlprefix }}%
\providecommand \urlprefix  [0]{URL }%
\providecommand \Eprint [0]{\href }%
\providecommand \doibase [0]{http://dx.doi.org/}%
\providecommand \selectlanguage [0]{\@gobble}%
\providecommand \bibinfo  [0]{\@secondoftwo}%
\providecommand \bibfield  [0]{\@secondoftwo}%
\providecommand \translation [1]{[#1]}%
\providecommand \BibitemOpen [0]{}%
\providecommand \bibitemStop [0]{}%
\providecommand \bibitemNoStop [0]{.\EOS\space}%
\providecommand \EOS [0]{\spacefactor3000\relax}%
\providecommand \BibitemShut  [1]{\csname bibitem#1\endcsname}%
\let\auto@bib@innerbib\@empty
\bibitem [{\citenamefont {Bohnenbuck}\ \emph {et~al.}(2008)\citenamefont
  {Bohnenbuck}, \citenamefont {Zegkinoglou}, \citenamefont {Strempfer},
  \citenamefont {Sch\"{u}\ss{}ler-Langeheine}, \citenamefont {Nelson},
  \citenamefont {Leininger}, \citenamefont {Wu}, \citenamefont {Schierle},
  \citenamefont {Lang}, \citenamefont {Srajer}, \citenamefont {Ikeda},
  \citenamefont {Yoshida}, \citenamefont {Iwata}, \citenamefont {Katano},
  \citenamefont {Kikugawa},\ and\ \citenamefont {Keimer}}]{Bohnenbuck2008}%
  \BibitemOpen
  \bibfield  {author} {\bibinfo {author} {\bibfnamefont {B.}~\bibnamefont
  {Bohnenbuck}}, \bibinfo {author} {\bibfnamefont {I.}~\bibnamefont
  {Zegkinoglou}}, \bibinfo {author} {\bibfnamefont {J.}~\bibnamefont
  {Strempfer}}, \bibinfo {author} {\bibfnamefont {C.}~\bibnamefont
  {Sch\"{u}\ss{}ler-Langeheine}}, \bibinfo {author} {\bibfnamefont {C.~S.}\
  \bibnamefont {Nelson}}, \bibinfo {author} {\bibfnamefont {P.}~\bibnamefont
  {Leininger}}, \bibinfo {author} {\bibfnamefont {H.-H.}\ \bibnamefont {Wu}},
  \bibinfo {author} {\bibfnamefont {E.}~\bibnamefont {Schierle}}, \bibinfo
  {author} {\bibfnamefont {J.~C.}\ \bibnamefont {Lang}}, \bibinfo {author}
  {\bibfnamefont {G.}~\bibnamefont {Srajer}}, \bibinfo {author} {\bibfnamefont
  {S.~I.}\ \bibnamefont {Ikeda}}, \bibinfo {author} {\bibfnamefont
  {Y.}~\bibnamefont {Yoshida}}, \bibinfo {author} {\bibfnamefont
  {K.}~\bibnamefont {Iwata}}, \bibinfo {author} {\bibfnamefont
  {S.}~\bibnamefont {Katano}}, \bibinfo {author} {\bibfnamefont
  {N.}~\bibnamefont {Kikugawa}}, \ and\ \bibinfo {author} {\bibfnamefont
  {B.}~\bibnamefont {Keimer}},\ }\href {\doibase 10.1103/PhysRevB.77.224412}
  {\bibfield  {journal} {\bibinfo  {journal} {Phys. Rev. B}\ }\textbf {\bibinfo
  {volume} {77}},\ \bibinfo {pages} {224412} (\bibinfo {year}
  {2008})}\BibitemShut {NoStop}%
\bibitem [{\citenamefont {Dashwood}\ \emph {et~al.}(2020)\citenamefont
  {Dashwood}, \citenamefont {Veiga}, \citenamefont {Faure}, \citenamefont
  {Vale}, \citenamefont {Porter}, \citenamefont {Collins}, \citenamefont
  {Manuel}, \citenamefont {Khalyavin}, \citenamefont {Orlandi}, \citenamefont
  {Perry}, \citenamefont {Johnson},\ and\ \citenamefont
  {McMorrow}}]{Dashwood2020}%
  \BibitemOpen
  \bibfield  {author} {\bibinfo {author} {\bibfnamefont {C.~D.}\ \bibnamefont
  {Dashwood}}, \bibinfo {author} {\bibfnamefont {L.~S.~I.}\ \bibnamefont
  {Veiga}}, \bibinfo {author} {\bibfnamefont {Q.}~\bibnamefont {Faure}},
  \bibinfo {author} {\bibfnamefont {J.~G.}\ \bibnamefont {Vale}}, \bibinfo
  {author} {\bibfnamefont {D.~G.}\ \bibnamefont {Porter}}, \bibinfo {author}
  {\bibfnamefont {S.~P.}\ \bibnamefont {Collins}}, \bibinfo {author}
  {\bibfnamefont {P.}~\bibnamefont {Manuel}}, \bibinfo {author} {\bibfnamefont
  {D.~D.}\ \bibnamefont {Khalyavin}}, \bibinfo {author} {\bibfnamefont
  {F.}~\bibnamefont {Orlandi}}, \bibinfo {author} {\bibfnamefont {R.~S.}\
  \bibnamefont {Perry}}, \bibinfo {author} {\bibfnamefont {R.~D.}\ \bibnamefont
  {Johnson}}, \ and\ \bibinfo {author} {\bibfnamefont {D.~F.}\ \bibnamefont
  {McMorrow}},\ }\href {\doibase 10.1103/PhysRevB.102.180410} {\bibfield
  {journal} {\bibinfo  {journal} {Phys. Rev. B}\ }\textbf {\bibinfo {volume}
  {102}},\ \bibinfo {pages} {180410(R)} (\bibinfo {year} {2020})}\BibitemShut
  {NoStop}%
\bibitem [{\citenamefont {Hicks}\ \emph {et~al.}(2014)\citenamefont {Hicks},
  \citenamefont {Brodsky}, \citenamefont {Yelland}, \citenamefont {Gibbs},
  \citenamefont {Bruin}, \citenamefont {Barber}, \citenamefont {Edkins},
  \citenamefont {Nishimura}, \citenamefont {Yonezawa}, \citenamefont {Maeno},\
  and\ \citenamefont {Mackenzie}}]{Hicks2014}%
  \BibitemOpen
  \bibfield  {author} {\bibinfo {author} {\bibfnamefont {C.~W.}\ \bibnamefont
  {Hicks}}, \bibinfo {author} {\bibfnamefont {D.~O.}\ \bibnamefont {Brodsky}},
  \bibinfo {author} {\bibfnamefont {E.~A.}\ \bibnamefont {Yelland}}, \bibinfo
  {author} {\bibfnamefont {A.~S.}\ \bibnamefont {Gibbs}}, \bibinfo {author}
  {\bibfnamefont {J.~A.~N.}\ \bibnamefont {Bruin}}, \bibinfo {author}
  {\bibfnamefont {M.~E.}\ \bibnamefont {Barber}}, \bibinfo {author}
  {\bibfnamefont {S.~D.}\ \bibnamefont {Edkins}}, \bibinfo {author}
  {\bibfnamefont {K.}~\bibnamefont {Nishimura}}, \bibinfo {author}
  {\bibfnamefont {S.}~\bibnamefont {Yonezawa}}, \bibinfo {author}
  {\bibfnamefont {Y.}~\bibnamefont {Maeno}}, \ and\ \bibinfo {author}
  {\bibfnamefont {A.~P.}\ \bibnamefont {Mackenzie}},\ }\href {\doibase
  10.1126/science.1248292} {\bibfield  {journal} {\bibinfo  {journal}
  {Science}\ }\textbf {\bibinfo {volume} {344}},\ \bibinfo {pages} {283}
  (\bibinfo {year} {2014})}\BibitemShut {NoStop}%
\bibitem [{\citenamefont {Steppke}\ \emph {et~al.}(2017)\citenamefont
  {Steppke}, \citenamefont {Zhao}, \citenamefont {Barber}, \citenamefont
  {Scaffidi}, \citenamefont {Jerzembeck}, \citenamefont {Rosner}, \citenamefont
  {Gibbs}, \citenamefont {Maeno}, \citenamefont {Simon}, \citenamefont
  {Mackenzie},\ and\ \citenamefont {Hicks}}]{Steppke2017}%
  \BibitemOpen
  \bibfield  {author} {\bibinfo {author} {\bibfnamefont {A.}~\bibnamefont
  {Steppke}}, \bibinfo {author} {\bibfnamefont {L.}~\bibnamefont {Zhao}},
  \bibinfo {author} {\bibfnamefont {M.~E.}\ \bibnamefont {Barber}}, \bibinfo
  {author} {\bibfnamefont {T.}~\bibnamefont {Scaffidi}}, \bibinfo {author}
  {\bibfnamefont {F.}~\bibnamefont {Jerzembeck}}, \bibinfo {author}
  {\bibfnamefont {H.}~\bibnamefont {Rosner}}, \bibinfo {author} {\bibfnamefont
  {A.~S.}\ \bibnamefont {Gibbs}}, \bibinfo {author} {\bibfnamefont
  {Y.}~\bibnamefont {Maeno}}, \bibinfo {author} {\bibfnamefont {S.~H.}\
  \bibnamefont {Simon}}, \bibinfo {author} {\bibfnamefont {A.~P.}\ \bibnamefont
  {Mackenzie}}, \ and\ \bibinfo {author} {\bibfnamefont {C.~W.}\ \bibnamefont
  {Hicks}},\ }\href {\doibase 10.1126/science.aaf9398} {\bibfield  {journal}
  {\bibinfo  {journal} {Science}\ }\textbf {\bibinfo {volume} {355}},\ \bibinfo
  {pages} {eaaf9398} (\bibinfo {year} {2017})}\BibitemShut {NoStop}%
\bibitem [{\citenamefont {Malinowski}\ \emph {et~al.}(2020)\citenamefont
  {Malinowski}, \citenamefont {Jiang}, \citenamefont {Sanchez}, \citenamefont
  {Mutch}, \citenamefont {Liu}, \citenamefont {Went}, \citenamefont {Liu},
  \citenamefont {Ryan}, \citenamefont {Kim},\ and\ \citenamefont
  {Chu}}]{Malinowski2020}%
  \BibitemOpen
  \bibfield  {author} {\bibinfo {author} {\bibfnamefont {P.}~\bibnamefont
  {Malinowski}}, \bibinfo {author} {\bibfnamefont {Q.}~\bibnamefont {Jiang}},
  \bibinfo {author} {\bibfnamefont {J.~J.}\ \bibnamefont {Sanchez}}, \bibinfo
  {author} {\bibfnamefont {J.}~\bibnamefont {Mutch}}, \bibinfo {author}
  {\bibfnamefont {Z.}~\bibnamefont {Liu}}, \bibinfo {author} {\bibfnamefont
  {P.}~\bibnamefont {Went}}, \bibinfo {author} {\bibfnamefont {J.}~\bibnamefont
  {Liu}}, \bibinfo {author} {\bibfnamefont {P.~J.}\ \bibnamefont {Ryan}},
  \bibinfo {author} {\bibfnamefont {J.-W.}\ \bibnamefont {Kim}}, \ and\
  \bibinfo {author} {\bibfnamefont {J.-H.}\ \bibnamefont {Chu}},\ }\href
  {\doibase 10.1038/s41567-020-0983-9} {\bibfield  {journal} {\bibinfo
  {journal} {Nat. Phys.}\ }\textbf {\bibinfo {volume} {16}},\ \bibinfo {pages}
  {1189} (\bibinfo {year} {2020})}\BibitemShut {NoStop}%
\bibitem [{\citenamefont {Kim}\ \emph {et~al.}(2021)\citenamefont {Kim},
  \citenamefont {Lefran\c{c}ois}, \citenamefont {Kummer}, \citenamefont
  {Fumagalli}, \citenamefont {Brookes}, \citenamefont {Betto}, \citenamefont
  {Nakata}, \citenamefont {Tortora}, \citenamefont {Porras}, \citenamefont
  {Loew}, \citenamefont {Barber}, \citenamefont {Braicovich}, \citenamefont
  {Mackenzie}, \citenamefont {Hicks}, \citenamefont {Keimer}, \citenamefont
  {Minola},\ and\ \citenamefont {Le~Tacon}}]{Kim2021}%
  \BibitemOpen
  \bibfield  {author} {\bibinfo {author} {\bibfnamefont {H.-H.}\ \bibnamefont
  {Kim}}, \bibinfo {author} {\bibfnamefont {E.}~\bibnamefont {Lefran\c{c}ois}},
  \bibinfo {author} {\bibfnamefont {K.}~\bibnamefont {Kummer}}, \bibinfo
  {author} {\bibfnamefont {R.}~\bibnamefont {Fumagalli}}, \bibinfo {author}
  {\bibfnamefont {N.~B.}\ \bibnamefont {Brookes}}, \bibinfo {author}
  {\bibfnamefont {D.}~\bibnamefont {Betto}}, \bibinfo {author} {\bibfnamefont
  {S.}~\bibnamefont {Nakata}}, \bibinfo {author} {\bibfnamefont
  {M.}~\bibnamefont {Tortora}}, \bibinfo {author} {\bibfnamefont
  {J.}~\bibnamefont {Porras}}, \bibinfo {author} {\bibfnamefont
  {T.}~\bibnamefont {Loew}}, \bibinfo {author} {\bibfnamefont {M.~E.}\
  \bibnamefont {Barber}}, \bibinfo {author} {\bibfnamefont {L.}~\bibnamefont
  {Braicovich}}, \bibinfo {author} {\bibfnamefont {A.~P.}\ \bibnamefont
  {Mackenzie}}, \bibinfo {author} {\bibfnamefont {C.~W.}\ \bibnamefont
  {Hicks}}, \bibinfo {author} {\bibfnamefont {B.}~\bibnamefont {Keimer}},
  \bibinfo {author} {\bibfnamefont {M.}~\bibnamefont {Minola}}, \ and\ \bibinfo
  {author} {\bibfnamefont {M.}~\bibnamefont {Le~Tacon}},\ }\href {\doibase
  10.1103/PhysRevLett.126.037002} {\bibfield  {journal} {\bibinfo  {journal}
  {Phys. Rev. Lett.}\ }\textbf {\bibinfo {volume} {126}},\ \bibinfo {pages}
  {037002} (\bibinfo {year} {2021})}\BibitemShut {NoStop}%
\bibitem [{\citenamefont {Worasaran}\ \emph {et~al.}(2021)\citenamefont
  {Worasaran}, \citenamefont {Ikeda}, \citenamefont {Palmstrom}, \citenamefont
  {Straquadine}, \citenamefont {Kivelson},\ and\ \citenamefont
  {Fisher}}]{Worasaran2021}%
  \BibitemOpen
  \bibfield  {author} {\bibinfo {author} {\bibfnamefont {T.}~\bibnamefont
  {Worasaran}}, \bibinfo {author} {\bibfnamefont {M.~S.}\ \bibnamefont
  {Ikeda}}, \bibinfo {author} {\bibfnamefont {J.~C.}\ \bibnamefont
  {Palmstrom}}, \bibinfo {author} {\bibfnamefont {J.~A.~W.}\ \bibnamefont
  {Straquadine}}, \bibinfo {author} {\bibfnamefont {S.~A.}\ \bibnamefont
  {Kivelson}}, \ and\ \bibinfo {author} {\bibfnamefont {I.~R.}\ \bibnamefont
  {Fisher}},\ }\href {\doibase 10.1126/science.abb9280} {\bibfield  {journal}
  {\bibinfo  {journal} {Science}\ }\textbf {\bibinfo {volume} {372}},\ \bibinfo
  {pages} {973} (\bibinfo {year} {2021})}\BibitemShut {NoStop}%
\bibitem [{\citenamefont {Bartlett}\ \emph {et~al.}(2021)\citenamefont
  {Bartlett}, \citenamefont {Steppke}, \citenamefont {Hosoi}, \citenamefont
  {Noad}, \citenamefont {Park}, \citenamefont {Timm}, \citenamefont
  {Shibauchi}, \citenamefont {Mackenzie},\ and\ \citenamefont
  {Hicks}}]{Bartlett2021}%
  \BibitemOpen
  \bibfield  {author} {\bibinfo {author} {\bibfnamefont {J.~M.}\ \bibnamefont
  {Bartlett}}, \bibinfo {author} {\bibfnamefont {A.}~\bibnamefont {Steppke}},
  \bibinfo {author} {\bibfnamefont {S.}~\bibnamefont {Hosoi}}, \bibinfo
  {author} {\bibfnamefont {H.}~\bibnamefont {Noad}}, \bibinfo {author}
  {\bibfnamefont {J.}~\bibnamefont {Park}}, \bibinfo {author} {\bibfnamefont
  {C.}~\bibnamefont {Timm}}, \bibinfo {author} {\bibfnamefont {T.}~\bibnamefont
  {Shibauchi}}, \bibinfo {author} {\bibfnamefont {A.~P.}\ \bibnamefont
  {Mackenzie}}, \ and\ \bibinfo {author} {\bibfnamefont {C.~W.}\ \bibnamefont
  {Hicks}},\ }\href {\doibase 10.1103/PhysRevX.11.021038} {\bibfield  {journal}
  {\bibinfo  {journal} {Phys. Rev. X}\ }\textbf {\bibinfo {volume} {11}},\
  \bibinfo {pages} {021038} (\bibinfo {year} {2021})}\BibitemShut {NoStop}%
\end{thebibliography}%

\section*{Acknowledgements}

We thank Richard Thorogate for assistance with the resistivity measurements, Daniel Nye and Gavin Stenning for assistance with the powder x-ray diffraction and Laue alignment in the Materials Characterisation Laboratory at the ISIS Neutron and Muon Source, Mike Matthews for technical support at I16, and Jacob Simms, Katherine Mordecai, Jon Bones and David Keymer for technical support at WISH. C.D.D.~was supported by the Engineering and Physical Sciences Research Council (EPSRC) Centre for Doctoral Training in the Advanced Characterisation of Materials under Grant No.~EP/L015277/1. A.H.W.~was supported by the EPSRC under Grant No.~EP/N509577/1. D.D.K.~was supported by the EPSRC under Grant No.~EP/W00562X/1. Work at UCL was supported by the EPSRC under Grants No.~EP/W005786/1, EP/N027671/1, EP/P013449/1 and EP/N509577/1. Experiments at the ISIS Neutron and Muon Source were supported by beamtime allocation RB1920210 from the Science and Technology Facilities Council. We acknowledge the Diamond Light Source for time on beamline I16 under proposals MM23580 and MM25554. We thank Institut Laue Langevin for access to the neutron diffractometer D9 under proposal EASY-951.

\section*{Author Contributions}

R.S.P.~and C.D.D.~grew and prepared the single crystal samples respectively. C.D.D., L.S.I.V., Q.F., R.S.P., R.D.J., P.M., D.D.K.~and F.O.~performed the neutron scattering measurements under stress, and C.D.D., P.M., D.D.K.~and F.O.~analysed the data. C.D.D., L.S.I.V., Q.F., R.S.P., R.D.J.~and D.G.P.~performed the x-ray scattering measurements under stress, and C.D.D.~and D.G.P.~analysed the data. Q.F., C.V.C.~and O.F.~performed the neutron diffraction measurements and all analysed the data. A.H.W., M.P.K., F.K.~and A.G.G.~developed the theoretical model. C.D.D.~and A.H.W.~wrote the paper, with input from all authors. A.G.G.~and D.F.M.~oversaw the study.

\section*{Competing Interests}

The authors declare no competing interests.

\end{document}


\title{Supplementary Information for ``Strain control of a bandwidth-driven spin reorientation in \CaRuO{}''}

\author{C.~D.~Dashwood}
\author{A.~H.~Walker}
\affiliation{London Centre for Nanotechnology and Department of Physics and Astronomy, University College London, London, WC1E 6BT, United Kingdom}

\author{M.~P.~Kwasigroch}
\affiliation{Department of Mathematics, University College London, London, WC1H 0AY, United Kingdom}
\affiliation{Trinity College, Cambridge, CB2 1TQ, United Kingdom}

\author{L.~S.~I.~Veiga}
\affiliation{London Centre for Nanotechnology and Department of Physics and Astronomy, University College London, London, WC1E 6BT, United Kingdom}
\affiliation{Diamond Light Source, Harwell Science and Innovation Campus, Didcot, Oxfordshire, OX11 0DE, United Kingdom}

\author{Q.~Faure}
\affiliation{London Centre for Nanotechnology and Department of Physics and Astronomy, University College London, London, WC1E 6BT, United Kingdom}
\affiliation{Laboratoire Le\'{o}n Brillouin, CEA, CNRS, Universit\'{e} Paris-Saclay, CEA-Saclay, 91191 Gif-sur-Yvette, France}

\author{J.~G.~Vale}
\affiliation{London Centre for Nanotechnology and Department of Physics and Astronomy, University College London, London, WC1E 6BT, United Kingdom}

\author{D.~G.~Porter}
\affiliation{Diamond Light Source, Harwell Science and Innovation Campus, Didcot, Oxfordshire, OX11 0DE, United Kingdom}

\author{P.~Manuel}
\author{D.~D.~Khalyavin}
\author{F.~Orlandi}
\affiliation{ISIS Neutron and Muon Source, STFC Rutherford Appleton Laboratory, Didcot, Oxfordshire, OX11 0QX, United Kingdom}

\author{C.~V.~Colin}
\affiliation{Universit\'{e} Grenoble Alpes, CNRS, Institut N\'{e}el, 38000 Grenoble, France}

\author{O.~Fabelo}
\affiliation{Institut Laue-Langevin, 71 Avenue des Martyrs, CS 20156, 38042 Grenoble, France}

\author{F.~Kr\"{u}ger}
\affiliation{London Centre for Nanotechnology and Department of Physics and Astronomy, University College London, London, WC1E 6BT, United Kingdom}
\affiliation{ISIS Neutron and Muon Source, STFC Rutherford Appleton Laboratory, Didcot, Oxfordshire, OX11 0QX, United Kingdom}

\author{R.~S.~Perry}
\affiliation{London Centre for Nanotechnology and Department of Physics and Astronomy, University College London, London, WC1E 6BT, United Kingdom}

\author{R.~D.~Johnson}
\affiliation{Department of Physics and Astronomy, University College London, London, WC1E 6BT, United Kingdom}

\author{A.~G.~Green}
\author{D.~F.~McMorrow}
\affiliation{London Centre for Nanotechnology and Department of Physics and Astronomy, University College London, London, WC1E 6BT, United Kingdom}

\maketitle

\section*{Supplementary Note 1: Additional neutron scattering data}

\begin{figure*}
	\includegraphics[width=\linewidth]{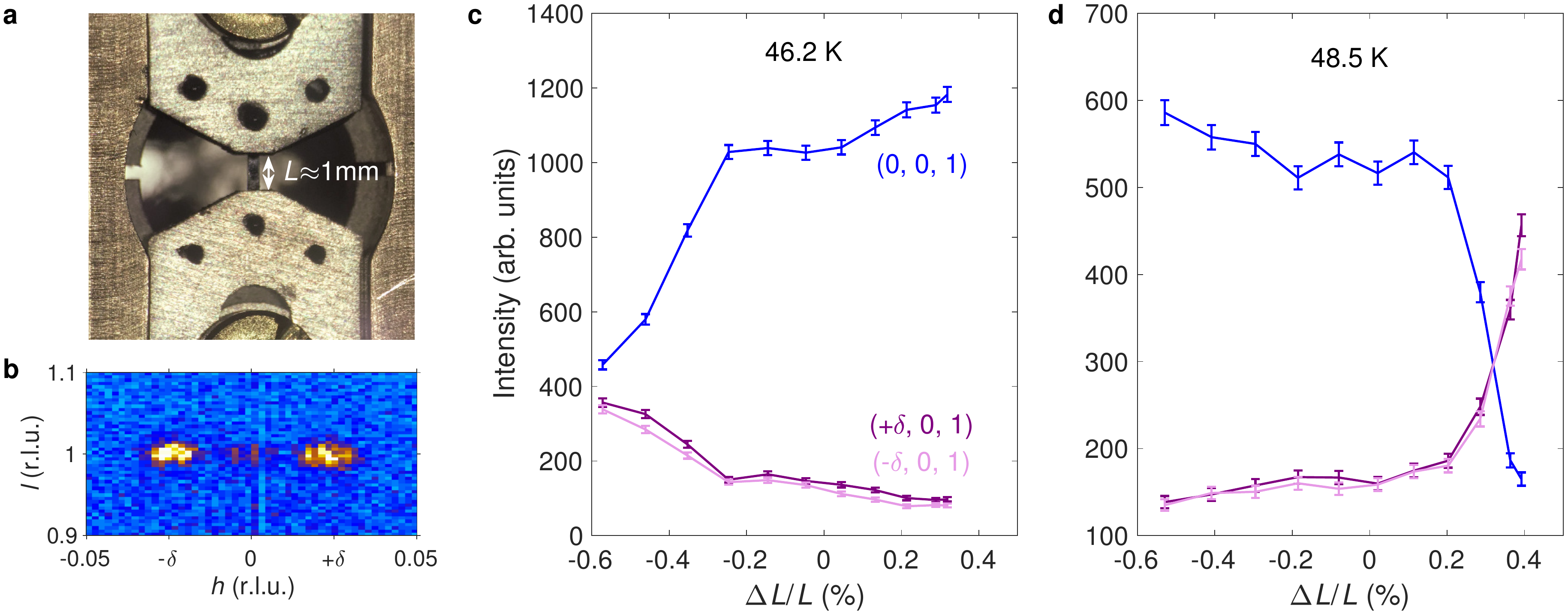}
	\caption{\label{Ca327_strain_SM_neutron}\textbf{Neutron scattering under stress. a} Image of the bar-shaped \CaRuO{} sample mounted over a distance $L \approx$ \SI{1}{\milli\metre} between the sample plates of the CS200T cell. \textbf{b} $(h, 0, l)$ reciprocal space map at \SI{47.7}{\kelvin} and zero strain showing satellite peaks at $(\pm\delta, 0, 1)$ around the commensurate $(0, 0, 1)$ position. \textbf{c--d} Integrated intensity of the commensurate $(0, 0, 1)$ and incommensurate $(\delta, 0, 1)$ peaks as a function of applied strain at \SI{46.2}{\kelvin} and \SI{48.5}{\kelvin}. Errors are standard deviations.}
\end{figure*}

In the main text, we present neutron scattering data taken on a strained \CaRuO{} crystal. Supplementary Fig.~\ref{Ca327_strain_SM_neutron}a shows an image of the sample used for these measurements mounted on the CS200T strain cell. The black, bar-shaped sample, seen in the centre of the image, is mounted between titanium sample plates over a distance of around \SI{1}{\milli\metre}. At zero strain and \SI{47.7}{\kelvin}, in the centre of the ICC phase, this sample produces the diffraction pattern shown in Supplementary Fig.~\ref{Ca327_strain_SM_neutron}b. A remnant central peak at $(0, 0, 1)$, which arises from the collinear \AFMa{} and \AFMb{} phases \cite{Bohnenbuck2008}, can be seen due to slight temperature and strain gradients through the probed region of the sample. Either side of this, satellite peaks are visible at $(\pm\delta, 0, 1)$ with $\delta \approx 0.023$ due to the ICC phase \cite{Dashwood2020}.

We tracked the intensity of these peaks under strain. In Fig.~1c of the main text, we present the results obtained at \SI{47.7}{\kelvin}, which show a suppression of the satellite peaks and concomitant increase in intensity of the commensurate peak under both positive and negative strain. In Supplementary Fig.~\ref{Ca327_strain_SM_neutron}c we show the results of similar measurements performed at \SI{46.2}{\kelvin} where we are in the \AFMb{} phase at zero strain. We see a relative insensitivity of the peaks to tensile strain, whereas compressive strains beyond $\Delta L / L <$ -0.3\% lead to a strong decrease of the $(0, 0, 1)$ intensity and simultaneous increase of the $(\pm\delta, 0, 1)$ intensities. Supplementary Fig.~\ref{Ca327_strain_SM_neutron}d shows results at \SI{48.5}{\kelvin}, where we are in the \AFMa{} phase at zero strain. Here, we see the opposite behaviour, with little change under compression and an exchange of intensity above 0.2\% tensile strain. These data confirm the conclusion in the main text that increasing strain (i.e.~from compressive to tensile) drives transitions from \AFMa{} to ICC, and then from ICC to \AFMb{}.

\newpage

\section*{Supplementary Note 2: Structural analysis}

\begin{figure*}
	\includegraphics[width=\linewidth]{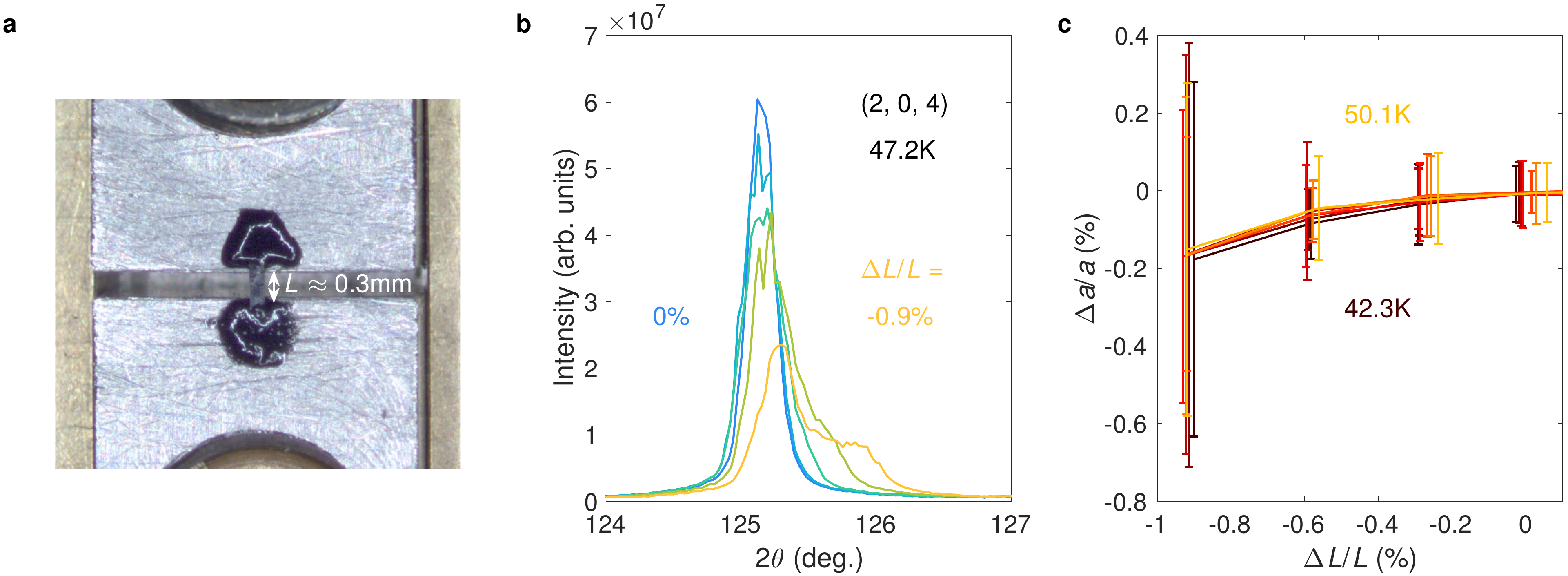}
	\caption{\label{Ca327_strain_SM_xray}\textbf{X-ray scattering with stress along the $a$-axis. a} Image of the \CaRuO{} sample mounted over a distance $L \approx$ \SI{0.3}{\milli\metre} between the sample plates of the CS100 cell. \textbf{b} $2\theta$ scans of the structural $(2, 0, 4)$ Bragg peak as a function of applied strain with stress along the $a$-axis. \textbf{c} True strain along the $a$-axis as a function of applied strain at a range of temperatures through the SRT. Errors are standard deviations.}
\end{figure*}

Our x-ray scattering setup enabled a detailed study of how the crystal structure of \CaRuO{} responds to applied stress. We measured samples with stress applied along both the $\mathbf{a}$ and $\mathbf{b}$ directions, with data from the latter shown in Fig.~2 of the main text. Supplementary Fig.~\ref{Ca327_strain_SM_xray}a shows an image of a sample stressed along $\mathbf{a}$ mounted on the CS100 cell, from which the phase diagram in Fig.~3a of the main text was constructed. To maximise beam access, the sample is mounted on top of raised sample plates (across a gap of around \SI{0.3}{\milli\metre}), which causes a slight bending of the sample under strain and appears to impair our ability to transmit tensile strain, as discussed below.

$2\theta$ scans of the structural $(2, 0, 4)$ Bragg peak of this sample are shown in Supplementary Fig.~\ref{Ca327_strain_SM_xray}b at various applied strains. We see a shifting of the peak under compressive strain due to the expected change in the lattice parameters. Unlike for the sample stressed along $\mathbf{b}$ (see Fig.~2b of the main text), however, we also see a significant broadening of the peak under strain, with a shoulder forming on the high-$2\theta$ side. This indicates that the magnitude of the strain is inhomogeneous in the probed region. We attribute this to preexisting defects in the sample, as a number of Bragg peaks were asymmetric before any strain was applied. Given the strain inhomogeneity, we used the centre of mass of the peaks to determine the true strain, shown in Supplementary Fig.~\ref{Ca327_strain_SM_xray}c.

\begin{figure*}
	\includegraphics[width=\linewidth]{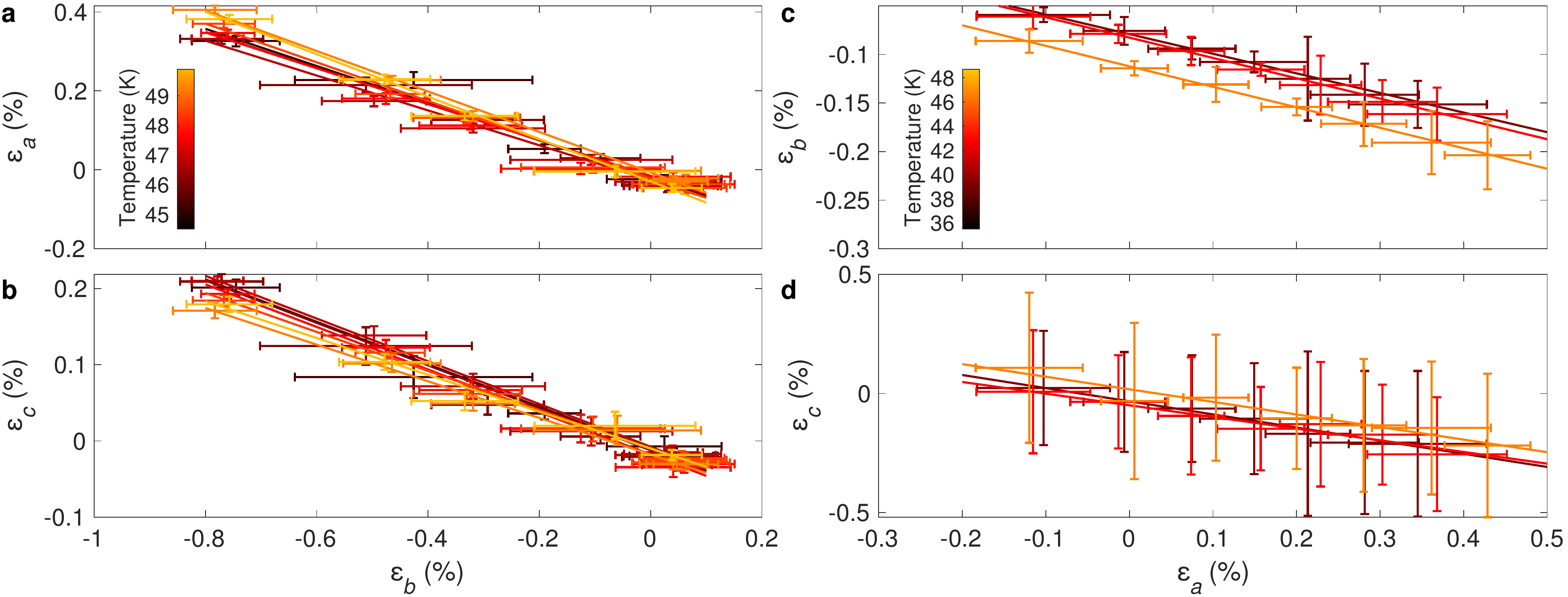}
	\caption{\label{Ca327_strain_SM_Poisson_ratios}\textbf{Poisson ratios. a--b} Induced strain along the $a$- and $c$-axes as a function of strain along the $b$-axis at a range of temperatures. \textbf{c--d} Induced strain along the $b$- and $c$-axes as a function of strain along the $a$-axis at a range of temperatures. The solid lines are linear fits at each temperature used to determine the Poisson ratios, and all errors are standard deviations.}
\end{figure*}

Measurement of multiple Bragg peaks as a function of strain allows us to determine the strain along all three crystallographic directions. Supplementary Fig.~\ref{Ca327_strain_SM_Poisson_ratios} shows the relationship between the orthogonal strains, which are found to be linear at all temperatures (note that the data in Supplementary Fig.~\ref{Ca327_strain_SM_Poisson_ratios}c--d are obtained from a different sample than that in Supplementary Fig.~\ref{Ca327_strain_SM_xray}). The Poisson ratios, shown in Fig.~2d of the main text, are determined from the slope of linear fits to these dependences.

\begin{figure*}
	\includegraphics[width=\linewidth]{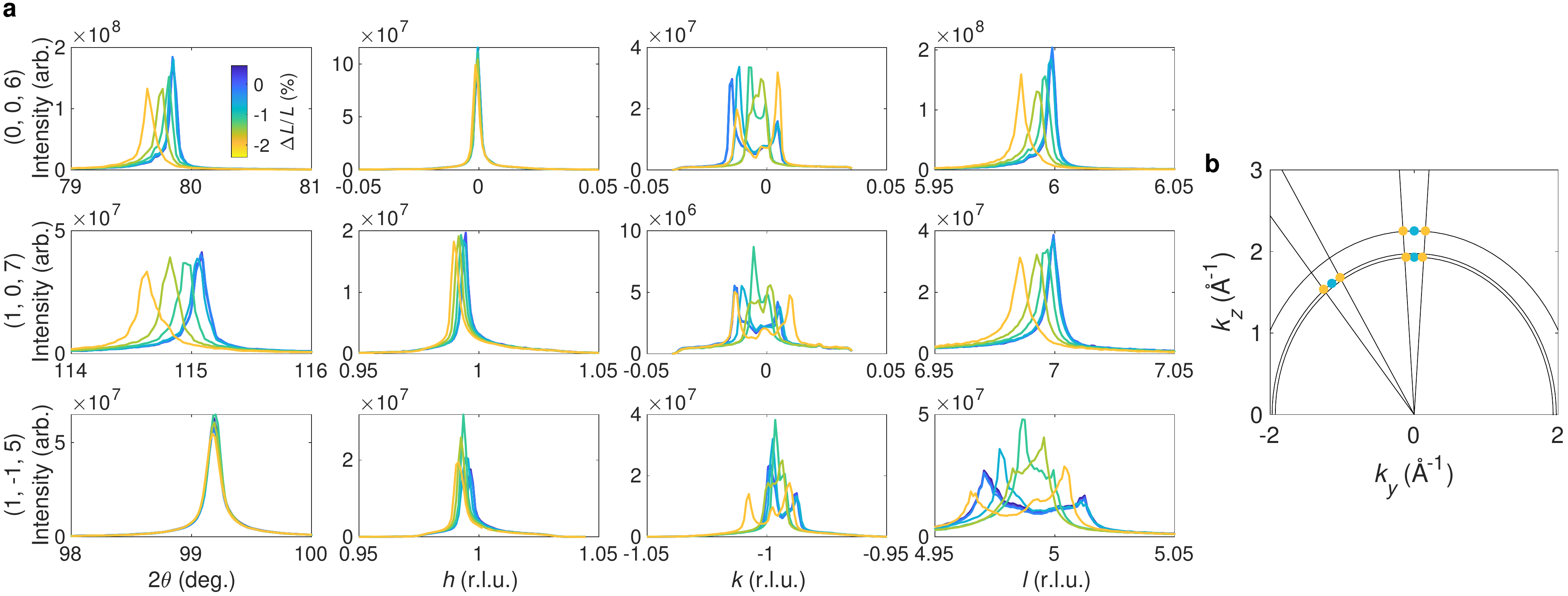}
	\caption{\label{Ca327_strain_SM_sample_bending}\textbf{Sample bending. a} Various reciprocal-space cuts through the structural $(0, 0, 6)$ (top row), $(1, 0, 7)$ (central row), and $(1, -1, 5)$ (lower row) peaks at a range of applied strains with stress along the $b$-axis at \SI{44.5}{\kelvin}. The left-most column shows $2\theta$ scans, the next column $h$ scans, then $k$ scans, and finally $l$ scans. \textbf{b} Projected $k$--$l$ reciprocal space map in inverse \r{A}ngstroms, showing the position of the split peaks at zero (blue) and maximum compressive (yellow) applied strain. The black lines depict the arcs and radial lines on which the split peaks lie.}
\end{figure*}

As well as probing average changes in the lattice parameters, we can leverage the large area detector at I16 to efficiently map the peaks in full 3D reciprocal space, giving insight into stress-induced variations of the lattice through the probed region. The left panels of Supplementary Fig.~\ref{Ca327_strain_SM_sample_bending} show $2\theta$, $h$, $k$ and $l$ cuts through three structural Bragg peaks at a range of applied strains (for the sample stressed along $\mathbf{b}$ at \SI{44.5}{\kelvin}). Alongside the shifting of the peaks, a splitting can be seen in the $k$--$l$ plane, with a distribution of intensity between the extremal split components. The positions of the main components, converted into reciprocal \r{A}ngstroms, are plotted in the right panel for zero and maximum compressive strain. Under compression, the split peaks lie on arcs in reciprocal space and are separated by approximately the same angle. The splitting can therefore be naturally explained by a bending of the sample under stress, which occurs due to the asymmetric mounting that transmits the strain mostly through the lower face of the sample. Despite the bending, the peaks do not broaden in $2\theta$ beyond our angular resolution, showing that the magnitude of the strain remains fairly homogeneous through the probed region. Interestingly, the peak splitting occurs for both tensile and compressive applied strain. This indicates that our inability to transmit tensile strain to the sample is not due to it fully fracturing (which would result in no response to tensile strain), but must be due to some other mechanism of strain relaxation in the sample and/or epoxy. We saw this same behaviour (peak splitting but no shift in $2\theta$ under tensile strain) in all samples mounted asymmetrically (including in the resistivity measurements described below), but in the neutron measurements with the sample sandwiched symmetrically between sample plates we see no bending and are able to transmit tensile strain.

These results highlight the importance of proper structural analysis when performing strain experiments. The majority of strain studies to date have relied on the applied strain (i.e.~the separation of the sample plates) as a proxy for the true strain (often with a constant factor introduced to account for deformation of the epoxy on the basis of finite element analysis) \cite{Hicks2014, Steppke2017, Malinowski2020, Kim2021, Worasaran2021}. Had we done the same, we would have significantly overestimated the compressive strain, been unsure why the magnetic phases did not respond to tensile strain, had no ability to determine the Poisson ratios, and been unaware of the sample bending. 

\newpage

\section*{Supplementary Note 3: Temperature-strain phase diagrams}

\begin{figure*}
	\includegraphics[width=\linewidth]{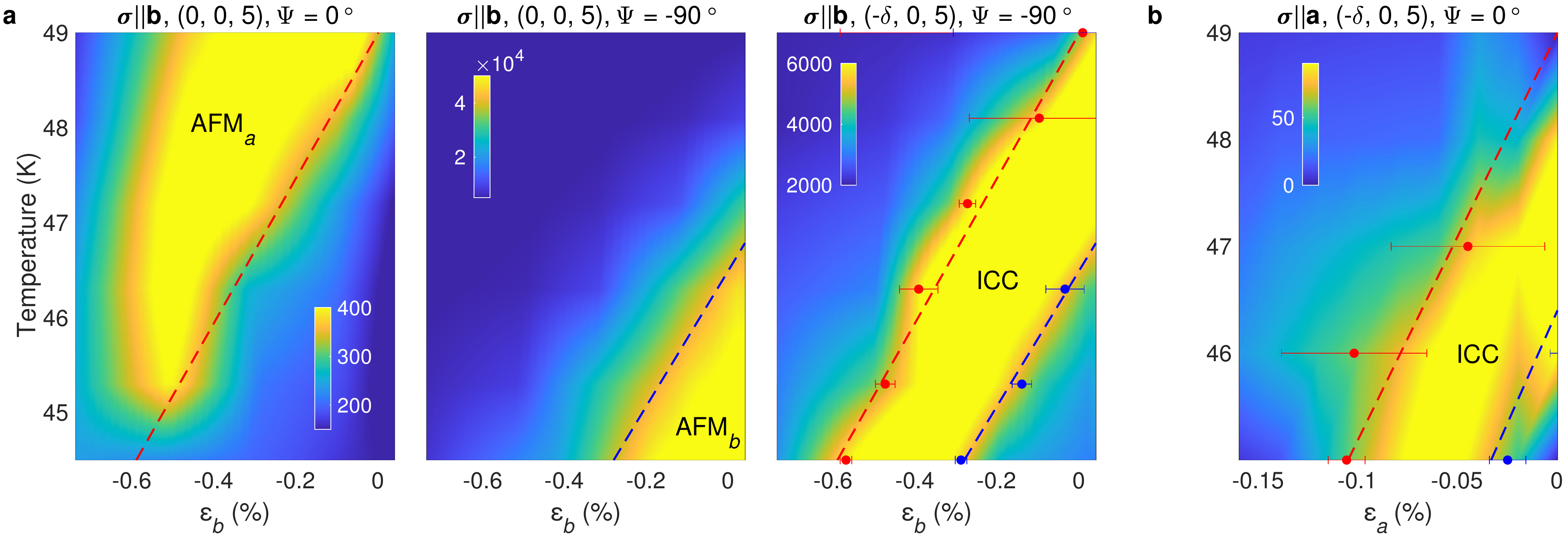}
	\caption{\label{Ca327_strain_SM_phase_diagrams}\textbf{Phase diagrams. a} Phase diagrams for stress applied along the $b$-axis. In the left panel the colourmap shows the integrated intensity of the commensurate $(0, 0, 5)$ peak at an azimuth of \ang{0} (sensitive to the component of the moment along the $a$-axis), the central panel shows the intensity of the $(0, 0, 5)$ peak at an azimuth of \ang{-90} (sensitive to the moment along the $b$-axis), and the right panel shows the intensity of the satellite $(-\delta, 0, 5)$ peak. \textbf{b} Phase diagram for stress applied along the $a$-axis, with the colourmap showing the intensity of the satellite $(-\delta, 0, 5)$ peak. The points with error bars (standard deviations) are fits to the ICC phase boundaries defined by the positions of half-maximum intensity, and the dashed red (blue) lines are linear fits through these points.}
\end{figure*}

To construct the magnetic phase diagrams shown in Fig.~3 of the main text, we tracked the intensities of the $(0, 0, 5)$ (at two azimuths) and $(\delta, 0, 5)$ peaks during strain sweeps at multiple temperatures. These intensities are plotted as colourmaps in Supplementary Fig.~\ref{Ca327_strain_SM_phase_diagrams}a--b for stress applied along $\mathbf{b}$ and $\mathbf{a}$ respectively. From left to right, the three panels in Supplementary Fig.~\ref{Ca327_strain_SM_phase_diagrams}a show the intensity of the commensurate $(0, 0, 5)$ peak at an azimuth of $\Psi =$ \ang{0}, sensitive to the \AFMa{} phase, the $(0, 0, 5)$ at an azimuth of $\Psi =$ \ang{-90}, sensitive to \AFMb{}, and the $(-\delta, 0, 5)$ satellite peak, sensitive to the ICC phase \cite{Dashwood2020}. The \AFMb{} phase can be seen in the lower right corner of the phase diagrams, disappearing with increasing temperature or compressive strain to be replaced by the ICC and then the \AFMa{} phases. One unexpected feature in the leftmost panel is the suppression of commensurate intensity above $|\varepsilon_b| \approx$ 0.6\% compressive strain at all temperatures. The reason for this loss of intensity requires further investigation, but it may be connected to the bending of the sample at high strain, which could disrupt the long-range order of the already-weak \AFMa{} phase [the (0, 0, 5) intensity is two orders of magnitude lower in the \AFMa{} phase than the \AFMb{}]. We note that there is no evidence of a similar suppression in our neutron results, where the bending does not occur.

Supplementary Fig.~\ref{Ca327_strain_SM_phase_diagrams}b similarly shows the intensity of the $(-\delta, 0, 5)$ satellite for stress along $\mathbf{a}$. As discussed in the previous section, the data quality for this sample was significantly worse than that with stress along $\mathbf{b}$, but despite this approximately linear phase boundaries can still be seen. The phase boundaries plotted in Fig.~3 of the main text are determined by finding the strain values at each temperature at which the satellite intensity falls to half its maximum value. These points are shown in red/blue in Supplementary Fig.~\ref{Ca327_strain_SM_phase_diagrams}, with error bars purely from fitting of the intensity profile and not accounting for the errors in the strain values. The dashed lines show linear fits through these points.

\newpage

\section*{Supplementary Note 4: Transport measurements under stress}

\begin{figure*}
	\includegraphics[width=\linewidth]{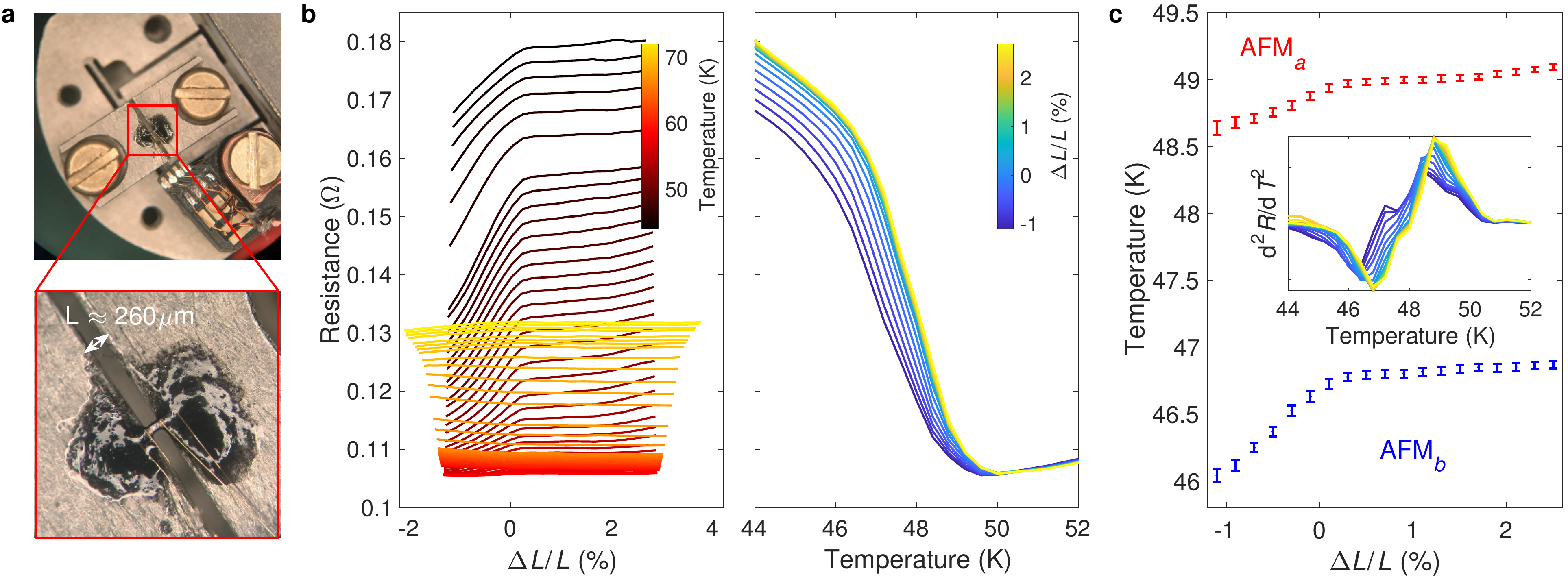}
	\caption{\label{Ca327_strain_SM_resistivity}\textbf{Transport measurements. a} The upper panel shows an image of the CS100 cell set up for resistivity measurements, with the WP100 wiring platform visible on the bottom right. The lower panel shows a close-up of a \CaRuO{} sample mounted on the cell, with gold wires contacted to the top surface in a standard four-probe configuration and stress applied along the $a$-axis. \textbf{b} The left panel shows the measured resistance as a function of applied strain at a range of temperatures. The right panel shows the same data plotted as a function of temperature for a range of applied strains. \textbf{c} Strain dependence of the transition temperatures, determined from the maxima and minima of $d^2R / dT^2$ shown in the inset.}
\end{figure*}

To correlate the changes in the crystal and magnetic structures with the Fermi surface reconstruction, we also performed four-probe resistance measurements under strain in a Quantum Design PPMS. For this we used the same CS100 cell as the x-ray measurements, with the addition of a wiring platform to allow electrical contacts to be made to the sample (see the upper panel in Supplementary Fig.~\ref{Ca327_strain_SM_resistivity}a). Four \SI{25}{\micro\metre}-diameter gold wires were contacted along the length of the sample with Dupont 6838 silver paint, and cured at \SI{900}{\celsius} for \SI{5}{\minute}. This results in mechanically robust contacts that can survive the repeated deformation caused by the applied stress, and it was confirmed that the brief exposure to high temperature does not change the transport properties of the sample. The other ends of the wires were connected with silver epoxy to pads on the wiring platform, which was in turn connected to a standard PPMS puck. The sample was mounted only with sample plates below as in the x-ray measurements, and \SI{10}{\micro\metre} nylon threads were placed in the epoxy below the sample to prevent electrical contact with the titanium cell (which is grounded during the measurement). The sample was mounted with the current contacts in the unstrained regions covered by epoxy and the voltage contacts in the strained region suspended between the sample plates (see the lower panel in Supplementary Fig.~\ref{Ca327_strain_SM_resistivity}a). As the contacts are on the top face of the sample, the current will spread downwards over a length scale $t\sqrt{\rho_c / \rho_{ab}} \sim$ \SI{100}{\micro\metre} \cite{Bartlett2021}. This is of the same magnitude as the length between the voltage contacts, so there will be a significant contribution to the measured resistivity from $\rho_c$. Thankfully, $\rho_{ab}$ and $\rho_c$ follow the same trend above \SI{45}{\kelvin}, so this does not affect the determination of the transition temperatures that we are interested in. The CS100 was mounted on a custom PPMS probe supplied by Razorbill Instruments. No temperature sensor was mounted on the cell, but through comparison with the neutron and x-ray data a constant \SI{2}{\kelvin} offset was found and corrected between the cryostat and sample temperatures.

The left panel of Supplementary Fig.~\ref{Ca327_strain_SM_resistivity}b shows strain dependences of the in-plane resistance at a range of temperatures for the sample with stress applied along $\mathbf{a}$. As we have no access to the true strain in the transport measurements, we plot these dependences as a function of applied strain, $\Delta L / L$. At the lowest temperatures the resistance is highest and we see a clear decrease under compressive applied strain. There is little response to tensile strain, which is expected from the asymmetric mounting as in the x-ray experiments. On warming into the ICC phase, the slope under compression decreases and the resistance appears to increase slightly under high tensile strain $\Delta L / L >$ 2\%. This again suggests that the sample is not broken, but that some other mechanism relieves tensile strain. On further warming into the \AFMa{} phase, the resistance instead begins to increase with compressive applied strain, before again falling under compression above $T_N$. The same data are plotted in the right panel of Supplementary Fig.~\ref{Ca327_strain_SM_resistivity}b as a function of temperature in the vicinity of the ICC phase, showing the resistance curve shifting down in temperature under compression. At zero strain, the boundaries of the ICC phase coincide with changes of slope of the resistance \cite{Dashwood2020}. To check whether this correspondence is maintained under strain, we plot the maxima and minima of $d^2 R / dT^2$ as a function of applied strain in Supplementary Fig.~\ref{Ca327_strain_SM_resistivity}c. We again find linear dependences under compression, with gradients that are comparable to those of the magnetic phase boundaries in Supplementary Fig.~\ref{Ca327_strain_SM_phase_diagrams} (after converting true strain to applied strain). This confirms that the electronic and magnetic transitions are locked together under strain.

\bibliography{Ca327_strain_SM}